\documentclass[reqno,12pt]{article}
\usepackage{enumerate}
\usepackage{ifpdf}
\usepackage{ae}
\usepackage[T1]{fontenc}
\usepackage[ansinew]{inputenc}
\usepackage{mathrsfs}
\usepackage{amsmath}
\usepackage{amssymb}
\usepackage{graphicx}
\usepackage{color}
\definecolor{darkblue}{cmyk}{0.9,0.9,0,0}
\definecolor{darkgreen}{rgb}{0,0.55,0}
\usepackage[colorlinks=true,linkcolor=darkblue,citecolor=darkblue,urlcolor=darkblue]{hyperref}
\usepackage{epsfig}
\usepackage{epstopdf}

\newcommand{\captionfonts}{\small}

\makeatletter
\long\def\@makecaption#1#2{
  \vskip\abovecaptionskip
  \sbox\@tempboxa{{\captionfonts #1: #2}}
  \ifdim \wd\@tempboxa >\hsize
    {\captionfonts #1: #2\par}
  \else
    \hbox to\hsize{\hfil\box\@tempboxa\hfil}
  \fi
  \vskip\belowcaptionskip}
\makeatother

\newcommand{\beq}{\begin{equation}}
\newcommand{\eeq}{\end{equation}}
\newcommand{\beqq}{\begin{equation*}}
\newcommand{\eeqq}{\end{equation*}}
\newcommand\beqa{\begin{eqnarray}}
\newcommand\eeqa{\end{eqnarray}}
\newcommand\beqaa{\begin{eqnarray*}}
\newcommand\eeqaa{\end{eqnarray*}}
\newcommand\bea{\begin{array}}
\newcommand\eea{\end{array}}

\def\XXint#1#2#3{{\setbox0=\hbox{$#1{#2#3}{\int}$}
\vcenter{\hbox{$#2#3$}}\kern-.5\wd0}}

\definecolor{mygreen}{RGB}{113,200,55}

\newcommand{\nn}{\nonumber}

\newcommand{\neqa}{\nonumber\end{eqnarray}}

\def\tr{{\rm tr~}}

\renewcommand{\d}{\partial}

\newcommand{\<}{{\langle}}
\renewcommand{\>}{{\rangle}}

\newcommand{\re}{\relax{\rm I\kern-.18em R}}

\renewcommand{\sp}{p\hspace{-.40em}/}

\newcommand{\beqy} {\begin{eqnarray}}
\newcommand{\eeqy} {\end{eqnarray}}
\newcommand{\bsmat}{\begin{smallmatrix}}
\newcommand{\esmat}{\end{smallmatrix}}
\newcommand{\bmat}{\begin{matrix}}
\newcommand{\emat}{\end{matrix}}

\def\({\left(}
\def\){\right)}
\def\[{\left[}
\def\]{\right]}

\def\<{\langle}
\def\>{\rangle}
\def\pd{\partial}
\def\ra{\rightarrow}

\newcommand{\mc}{\mathcal}

\def\a{\alpha}

\def\c{\chi}
\def\g{\gamma}

\def\d{\delta}
\def\D{\Delta}
\def\e{\epsilon}

\def\th{\theta}

\def\i{\iota}

\def\l{\lambda}

\def\o{\omega}

\def\s{\sigma}

\def\mc{\mathcal}

\def\pd{\partial}
\def\pdb{\bar{\partial}}

\def\su2{{SU(2)}}

\def\o{{\omega}}
\def\a{{\alpha}}

\def\[{\left[}
\def\]{\right]}

\def\l{\lambda}
\def\e{\epsilon}

\def\s{\sigma}
\def\a{\alpha}

\def\th{\theta}

\def\({\left(}
\def\){\right)}
\def\[{\left[}
\def\]{\right]}

\def\<{\langle}
\def\>{\rangle}

\def\i2{\frac{i}{2}}

\def\spi{\relax{\rm \pi\kern-0.5em /}}
\def\sA{\relax{\rm A\kern-0.5em /}}
\def\sp{\relax{\rm p\kern-0.5em /}}
\def\sd{\relax{\rm \d\kern-0.5em /}}
\def\sk{\relax{\rm k\kern-0.5em /}}
\def\sn{\relax{\rm n\kern-0.5em /}}
\def\sl{\relax{\rm l\kern-0.5em /}}
\def\sP{\relax{\rm P\kern-0.7em /}}
\def\sBethe{\relax{\rm \Bethe\kern-0.5em /}}

\usepackage{varioref}
\usepackage{makeidx}
\makeindex

\usepackage[english]{babel}
\begin{document}

\thispagestyle{empty}

\renewcommand{\thefootnote}{\fnsymbol{footnote}}
\setcounter{page}{1}
\setcounter{footnote}{0}
\setcounter{figure}{0}
\begin{center}
$$$$

{\Large\textbf{\mathversion{bold}
$\chi-$Systems for Correlation Functions}\par}
\vspace{1.0cm}

\textrm{J. Caetano$^{a,b,c}$, \ J. Toledo$^{a,b}$}
\\ \vspace{1.2cm}
\footnotesize{

\textit{$^a$Perimeter Institute for Theoretical Physics\\ Waterloo,
Ontario N2L 2Y5, Canada}  \\
\vspace{4mm}
\textit{$^b$Department of Physics and Astronomy \& Guelph-Waterloo Physics Institute,\\
University of Waterloo, Waterloo, Ontario N2L 3G1, Canada} \\
\vspace{4mm}
\textit{$^c$Centro de F\'isica do Porto e Departamento de F\'isica e Astronomia,\\
Faculdade de Ci\^encias da Universidade do Porto,\\
Rua do Campo Alegre, 687, 4169-007 Porto, Portugal} \\
\vspace{5mm}
\small\texttt{jd.caetano.s,jonathan.campbell.toledo@gmail.com} \\
}

\par\vspace{1.5cm}

\textbf{Abstract}\vspace{2mm}
\end{center}

\noindent
We consider the strong coupling limit of 4-point functions of heavy operators in $\mathcal{N}=4$ SYM dual to strings with no spin in AdS. We restrict our discussion for operators inserted on a line. The string computation factorizes into a state-dependent sphere part and a universal AdS contribution which depends only on the dimensions of the operators and the cross ratios.  We use the integrability of the AdS string equations to compute the AdS part for operators of arbitrary conformal dimensions.  The solution takes the form of TBA-like integral equations with the minimal $AdS$ string-action computed by a corresponding free-energy-like functional.  These TBA-like equations stem from a peculiar system of functional equations which we call a $\chi$-system.  In principle one could use the same method to solve for the $AdS$ contribution in the $N$-point function.   An interesting feature of the solution is that it encodes multiple string configurations corresponding to different classical saddle-points.  The discrete data that parameterizes these solutions enters through the analog of the chemical-potentials in the TBA-like equations.  Finally, for operators dual to strings spinning in the same equator in $S^5$ (i.e. BPS operators of the same type) the sphere part is simple to compute.  In this case (which is generically neither extremal nor protected) we can construct the complete, strong-coupling 4-point function. 

\vspace*{\fill}

\setcounter{page}{1}
\renewcommand{\thefootnote}{\arabic{footnote}}
\setcounter{footnote}{0}

\newpage

 \def\nref#1{{(\ref{#1})}}

\tableofcontents


\section{Introduction}
One of the most interesting objects to study in an interacting quantum field theory is the four-point function.  In conformal field theories these are the first $N$-point functions whose spacetime dependence is not explicitly fixed by conformal symmetry.  The computation of this correlator is generally highly nontrivial and obtaining explicit expressions outside the perturbative regime is typically impossible. However, the advent of the AdS/CFT correspondence has made it possible to access the strong coupling limits of special QFT's.  In particular the correspondence \cite{adscft1} maps the strong coupling limit of $\mathcal{N}=4$ SYM to a theory of classical strings moving in an $AdS_5 \times S^5$ background.  This allows the computation of leading strong coupling behavior of certain 4-point functions in $\mathcal{N}=4$ SYM, which is the main purpose of this paper.  While we have mostly focused on the 4-point computation, we note that the method used below could in principle be applied for the corresponding $N$-point computation. \\
\indent  There are also interesting indirect applications.  One of our main motivations for this work is the possibility of exploring the spectrum and the structure constants of $\mathcal{N}=4$ SYM through the operator product expansion (OPE) of the four point function. Another potentially interesting application of this calculation is in the weak-strong coupling connections reported in \cite{tailor2,CE,tailor3,quantumint,KOSTOV} that suggests the existence of an underlying common structure in both regimes.\\
\indent At strong coupling, the problem of computing the correlation function is that of finding the area of the minimal surface in $AdS_5 \times S^5$ that goes to the $AdS$ boundary at the operator insertion points $x_a$. In this paper, we compute the $AdS$ part of the correlation function for arbitrary heavy scalar operators inserted along a line. The method used here is inspired by the integrability techniques originally developed for the Null Polygonal Wilson loop problem \cite{AMSV} and later applied to the computation of three-point functions \cite{JW2,KK1,KK2}.  As in these previous applications, integrability allows one to compute the minimal $AdS$ action without knowing the explicit classical solution. For the four point correlation function the connection with Hitchin systems and the formalism developed in \cite{GMN} is used intensely. As in \cite{AMSV, JW2, KK1} the starting point of the method is the map of the string equations of motion in $AdS$ to a certain the auxilliary linear problem by Pohlmeyer reduction.  Ultimately the solution takes the form of a set of functional equations that we call a \emph{$\chi$-system}.  These functional equations are similar in spirit to the $Y$-system appearing in \cite{AMSV} and which naturally arise in the solutions of integrable QFT's.  \\
\indent For some specific BPS operators dual to strings spinning on the same great circle of $S^5$ the sphere contribution is well know. In this case we can construct the full strong-coupling correlation function.  We emphasize that these 4-point functions are generically neither extremal nor protected.  Complete, non-protected results for correlation functions of heavy operators at strong coupling are quite rare.  For example, in \cite{JW2}, the $AdS$ part of the three-point function is computed, but the only case for which the sphere is known (BPS operators) is protected.   The only complete, non-protected result that we know of is \cite{KK2} where the strong-coupling three-point functions of GKP strings is computed.\footnote{Using the results of this paper it may be possible to extend the results of \cite{KK2} to the complete $N$-point functions of GKP strings at strong coupling since the mathematical problem is similar to the one treated here.}   \\
\indent The layout of the paper is as follows.  In the section \ref{4ptgen} we start by giving the general strategy of this work and discuss some physically relevant aspects of the problem. Then, in section \ref{PR} we write the $AdS$ part of the correlation function in terms of objects which are naturally computed using the integrability of the string equations of motion. In section \ref{Linear Problem} we introduce a formalism that will lead to the $\chi$-system which is the set of functional equations that allows one to compute the minimal $AdS$ action. In section \ref{AdS action} we compute explicitly the $AdS$ part of the correlation function using the $\chi$-system and explain the mechanism by which the dependence on the spacetime points emerges. We also present some numerical tests of the results. In section \ref{sphere}, we compute the sphere part for the specific case of BPS operators with large charges. In section \ref{saddle point}, we discuss the saddle points of the fourth insertion point of the correlation function and perform some numerical studies on this issue. In the same section, we study the extremal limit of the correlation function which is an analytical test of the result. In section \ref{conclusion} we conclude and discuss some open problems.\\
\indent Finally, because the computation involves many intermediate steps in the flow-chart below we summarize for the reader the basic components of the method.  The flow-chart provides a map of the paper and summarizes the main steps of the computation which are shown in the individual boxes.  Each box contains a reference to the relevant section where that part of the computation is discussed.
\begin{figure}[h!]
\begin{center}
\includegraphics[width=1.0\linewidth]{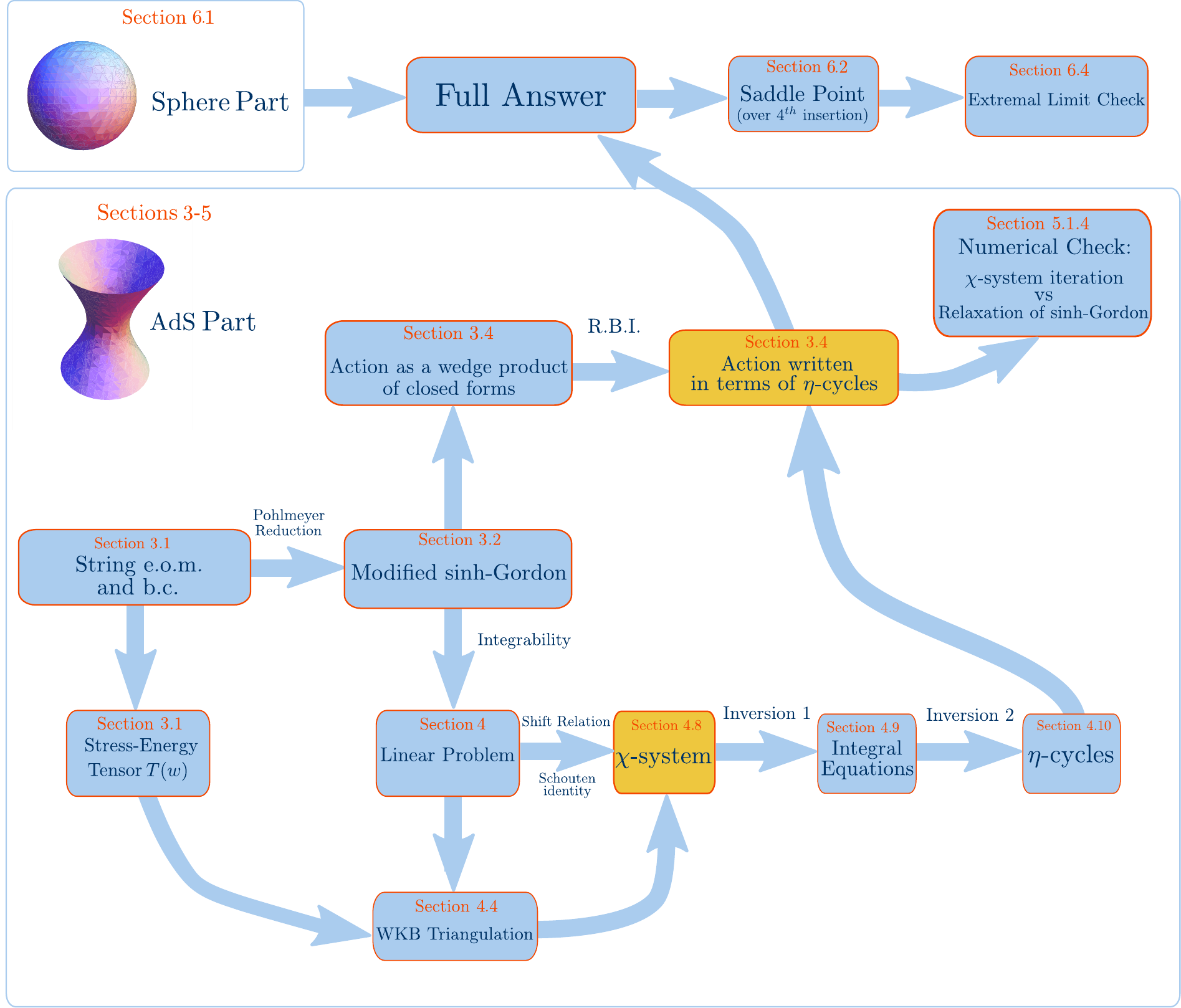}
\end{center}
\end{figure}
\section{Four point function generalities}\label{4ptgen}
For large 't Hooft coupling $\lambda$, the semi-classical computation of correlation functions corresponds to the evaluation of the $AdS_5$ and $S^5$ actions for classical solutions with the topology of a four punctured sphere.   The boundary conditions are that the solution close to each puncture $P_a$, which is associated with the gauge theory operator $\mathcal{O}_a \! \(x_a\)$, approaches the $AdS$ boundary at the point $x_a$ in the same way as a 2-point function involving $\mathcal{O}_a \!\(x_a\)$ and some other heavy, scalar operator.  In this paper, we study the simplest case where the operators are inserted on a line in $\mathbb{R}^4$. This implies that the string solution is contained in a Euclidean $AdS_2$ subspace of $AdS_5 $. Moreover, there is only one independent cross-ratio.\\
The conformal symmetry of $\mathcal{N}=4$ constrains the four-point correlation function to take the form
\beq\label{can4pt}
\langle \mathcal{O}_1(x_1)\mathcal{O}_2(x_2)\mathcal{O}_3(x_3)\mathcal{O}_4(x_4) \rangle = f(u)\prod_{a>b}^4 (x_{ab})^{\D_{ab}}\,,
\eeq
where $x_{ab}=x_a-x_b$, $\Delta_a$ is the dimension of operator $\mathcal{O}_a$, $\D_{ab}=\(\sum_c \D_c\)/3-\D_a-\D_b$ and $u$ is the conformal cross-ratio
\beq\label{crat}
u=\frac{x_{14} x_{23}}{x_{12} x_{34}}
\eeq\\
\indent Both the $AdS$ and sphere contributions contain divergences as the string approaches the position of the operators at the boundary of $AdS$, which requires a cut-off $z=\mathcal{E}$. 
To describe the world-sheet we use complex variables $w, \bar{w}$.  On the 4-punctured sphere, the physical cut-off $\mathcal{E}$ corresponds to cutting out small disks of radius $\epsilon_a$ around each puncture $P_a$ at $w_a$. Ultimately, we will need to establish a precise relation between the cut-off's $\e_a$ and $\mathcal{E}$.  As we will review later, this is possible given the data accessible from integrability \cite{JW2}. \\
\indent In this paper, we will consider operators with charges scaling as $\sqrt{\lambda}$, and without spin in $AdS$. Following the prescription developed in \cite{JW1, JW2}, we account for the states in the sphere by introducing an extra contribution of wave-functions. Therefore, the semi-classical four-point function is given schematically by
\beq\label{semi4pt}
\int{dw_4}\,e^{-\frac{\sqrt{\lambda}}{\pi} \int_{\Sigma \backslash \{ \epsilon_a\}}\mathcal{L}_{AdS_2}} e^{-\frac{\sqrt{\lambda}}{\pi} \int_{\Sigma \backslash \{ \epsilon_a\}}\mathcal{L}_{S^5}} \Psi_1 \Psi_2 \Psi_3 \Psi_4 
\eeq
where the actions are evaluated on a classical (Euclidean) string solution approaching the boundary of $AdS$ at the positions of the insertion points $x_a$.  \\
\begin{figure}[t!]
\begin{center}
\includegraphics[width=1\linewidth]{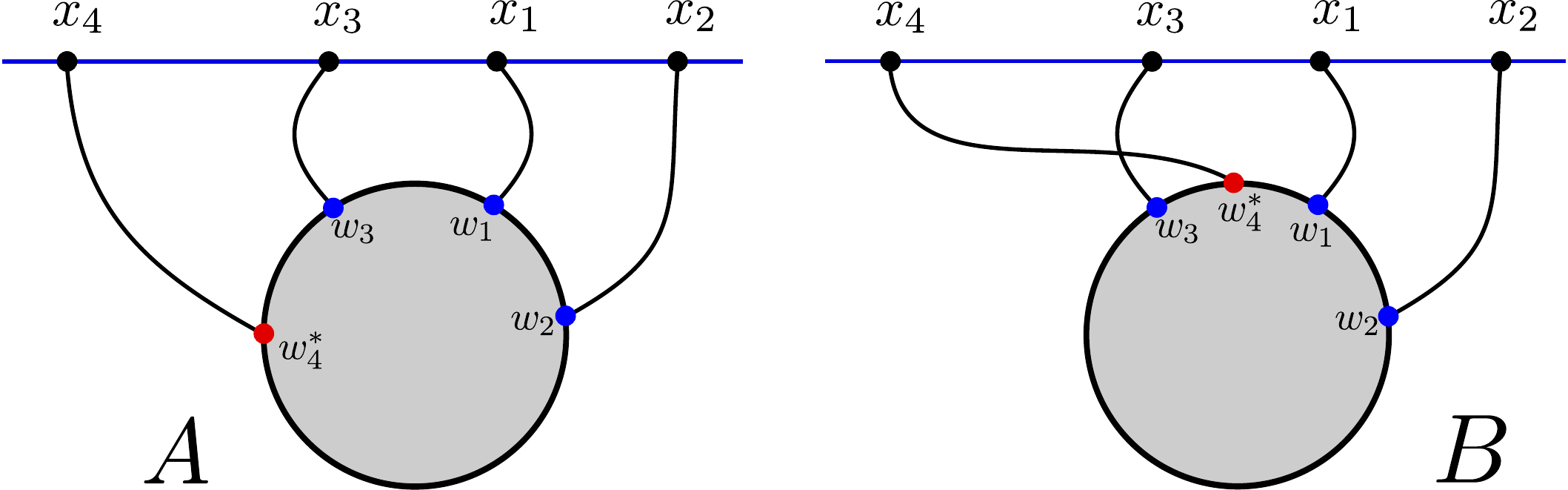}
\caption{Insertions on the 4-punctured sphere. The gray ball represents the world-sheet (the complex plane plus the point at infinity, or simply `the sphere') and the black boundary of the ball represents the equator of the sphere.  The points $w_a$ are the punctures on the sphere corresponding to the operators inserted at the positions $x_a$ at the boundary of $AdS_2$, which is represented by the strait line.  We fix the points $w_1$, $w_2$, $w_3$ and $x_1$, $x_2$, $x_3$ using the world-sheet and target-space conformal symmetry respectively.  The position of the fourth insertion $w_4$ should be fixed at the dominant saddle point $w_4^*$ of the integrand of \eqref{semi4pt}.  By symmetry we expect this saddle point to also be along the real axis, and thus we have a notion of an ordering of the 4 punctures (but see footnote \ref{foot: saddle caveat}).  In particular, there is three distinct ranges for the location of $w_4^*$.  Consider the ordering of the $x_a$ shown in this figure.  If the dominant saddle point is located between $w_2$ and $w_3$ (as in panel $A$) then the insertions will not cross and the string embedding will look schematically like the one shown in figure \ref{stringconfigs2}A.  If the dominant saddle-point is located between $w_3$ and $w_1$ (as in panel $B$) or between $w_1$ and $w_2$ then the insertions cross each-other and we expect the string embedding to look like the one shown in figure \ref{stringconfigs2}B.}\label{insertions}
\end{center}
\end{figure}
\indent In principle, there is an integral over all four insertion-points on the worldsheet.  In \eqref{semi4pt} we only integrate over the insertion $w_4$ since the position of the other punctures can be fixed by conformal transformations.  Since we are considering the $\l \ra \infty$ limit, one can evaluate the integral over $w_4$ by saddle point and the end result is the integrand of \eqref{semi4pt} evaluated at the dominant saddle point.   \\
\indent Let us consider the issue of the saddle point in some detail since it will be an interesting aspect of our computation.  There are two issues here:  the positions of the operators on the boundary and the positions of the insertion points on the sphere.  We can use the target-space conformal symmetry to place three of the operators at $x_1 = 1$, $x_2 = \infty$, $x_3 = -1$ and the world-sheet conformal symmetry to fix $w_1 = 1$, $w_2 = \infty$, $w_3 = -1$.  
The position $x_4$ is an input since we can put $\mathcal{O}_4$ anywhere along the line that contains $\mathcal{O}_{1,2,3}$.  On the other hand, once we choose $x_4$ the position of the fourth puncture is fixed at $w_4=w_4^*$ by the saddle-point condition.  By symmetry we expect the dominant saddle-point to be located on the real axis and in this case we have a notion of an ordering of the punctures.\footnote{We have confirmed numerically that there are saddle points along the real axis.  There may also be the possibility saddle points off of the real axis and occurring in complex-conjugate pairs, but we have not investigated this possibility.  This issue certainly requires further investigation. \label{foot: saddle caveat}}  In particular, there are three possible in-equivalent orderings depending on the position of $w_4$.  Figure \ref{insertions} shows two of these possibilities.  If the ordering of the $x_a$ is the same as the $w_a$ then the insertions do not cross each other, as in figure \ref{insertions}$A$. If the ordering of the $x_a$ is different from that of the $w_a$, then the insertions will cross as in figure \ref{insertions}$B$.  These two possibilities lead to two types of string embeddings with distinctly different properties as is shown in figure \ref{stringconfigs2}.   We will see that two types of solutions arise naturally in our construction.  We are able to characterize the qualitative features of the spacetime embeddings  and compute the minimal $AdS$ action of both types of solutions.  We will return to this topic below. 
\begin{figure}[t!]
\begin{center}
\includegraphics[width=1\linewidth]{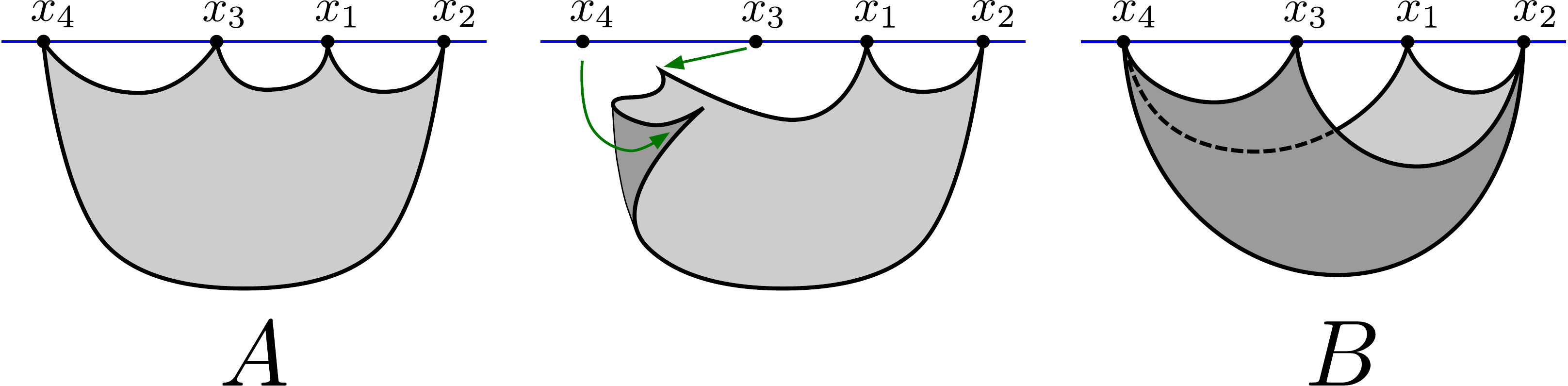}
\caption{Two different possible string embeddings in $AdS_2$ which obey the required boundary conditions. These two solutions are shown in panels $A$ and $B$.  The center panel shows how to generate the configuration of panel $B$ from that of panel $A$ by interchanging the order in which the insertions on the sphere attach to the boundary; this interchange results in the characteristic folding shown in the embedding of panel $B$. These two types of solutions arise from the possibility that for a given choice of operator insertion points $x_a$ the insertion point $w_4^*$ (see figure \ref{insertions}) can be located in any of the three possible intervals $\(w_2, w_3\)$, $\(w_3, w_1\)$, $\(w_1, w_2\)$.}\label{stringconfigs2}   
\end{center}
\end{figure}
\section{$AdS_2$ Pohlmeyer reduction}\label{PR}
In this section we briefly review the Pohlmeyer-reduction process.  We begin with a discussion of the string equations of motion and the stress-energy tensor, which is the starting point of the reduction.  We then introduce the function $\g$ in terms of which the $AdS$ Lagrangian can be written.  It turns out that $\g$ satisfies a non-linear but \emph{scalar} equation of motion that is a modified version of the well-know sinh-Gordon equation.  Next we show how the different types of string embeddings discussed in section \ref{4ptgen} are encoded though the boundary conditions imposed on $\g$.  Finally we use the function $\g$ to  write the $AdS$ action in a form where integrability is more readily applied.
\subsection{Equations of motion and stress-energy tensor}\label{eqmotion}
Recall that we can consider (euclidean) $AdS_2$ as a surface in $R^{1,2}$ obeying the constraint
\beq
Y \cdot Y=\(Y_1\)^2-\(Y_2\)^2+\(Y_3\)^2 = -1.\label{AdS constraint}
\eeq
 We write the action for a string in $AdS_2$ as 
\beq
S= \frac{1}{2}\int d^2 \s \[\pd_{\a}Y \cdot \pd^{\a} Y + \l \(Y\cdot Y +1\)\]
\eeq
and the resulting equations of motion as 
\beq
\square Y = \( \pd Y \cdot \bar{\pd} Y\) Y \label{AdS2 EOM}
\eeq
The first term in the action is just the free string action in $R^{1,2}$; the second term is a Lagrange multiplier term that imposes $(\ref{AdS constraint})$.  
\\
\begin{figure}[t!]
\begin{center}
\includegraphics[width=0.85\linewidth]{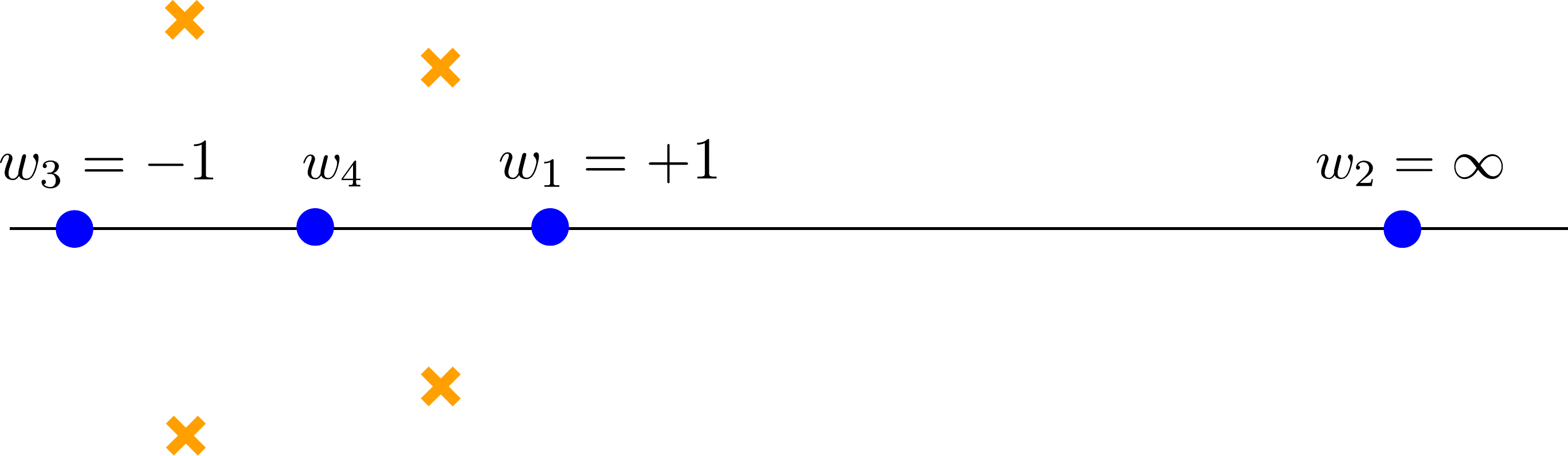}
\caption{Schematic analytic structure of $T$.  The blue dots represent the (double) poles of $T$ at locations $w_a$ and corresponding to the operator insertions $\mathcal{O}_a \! \(x_a\)$.  The yellow crosses indicated zeros of $T$. We have fixed the positions of $w_1$, $w_2$ and $w_3$ using the world-sheet conformal symmetry.  We have arbitrarily placed $w_4$ in the interval $\(w_3,w_1\)$ although generically the saddle-point $w_4^*$ can be located in any of the three intervals along the real axis.   }\label{schematicT} 
\end{center}
\end{figure}
\indent The equations of motion ($\ref{AdS2 EOM}$) must be supplemented by the Virasoro constraints and boundary conditions.  The Virasoro constraint requires $T_{AdS}+T_{S} = 0$.  In particular, the $AdS$ contribution to the stress-energy tensor does not vanish.  Fortunately the boundary conditions allow us to completely fix the form of  $T_{AdS}=-T_{S}$.  Here we are interested in solutions with the topology of a four-punctured sphere where the punctures are at the position of the operator insertions and thus the boundary conditions give the behavior of the string solutions near the insertion points.  The correct prescription is to demand that the string goes to the boundary at the insertion points.  Furthermore, it should approach the boundary in a specific way as dictated by the vertex operators.  The behavior of the solution near the boundary will be dominated by the operator inserted there, independent of the properties or number of other operators inserted at different points.  This means that the behavior near the insertion points can be determined from the 2-point function, where the string solution is know explicitly.  From the explicit solution for the 2-point function one finds that the desired property of the solution near insertion point $w_a$ is \cite{JW2}
\beqy
(\pd Y)^2 \equiv T\(w\) \sim \frac{\D_a^2}{4(w-w_a)^2} \;\;\;\;\; \(w \ra w_a \)\label{T near P}
\eeqy
where $T\(w\)$ is the holomorphic component of $T_{AdS}$.  The corresponding property also is required for the anti-holomorphic component $\bar{T}\(\bar{w}\)$.  Thus we know that $T$ should be an analytic function on the (4-punctured) Riemann sphere with double-pole singularities at the punctures.  This fixes $T$ to be a specific rational function.  \\
\indent First consider the denominator of the rational function $T$.  The polynomial in the denominator is determined by the positions of the insertions. Three of the insertions can be fixed by conformal symmetry, leaving one final insertion.  The integrand of \eqref{semi4pt} will be a function of this final insertion point.  In the limit $\sqrt{\l}\ra \infty$ the integral localizes at the saddle point $w_4=w_4^*$, thus fixing completely the denominator of $T$.  \\
\indent Now consider the numerator of $T$.  Without loss of generality we can consider the case where there is no insertion at infinity since we can perform a transformation that maps any arbitrary point to infinity.  Then the polynomial in the numerator can be at most of degree 4 (otherwise $T$ would not be regular at infinity) and therefore it is characterized by 5 parameters.  Four of these parameters are fixed by the condition \eqref{T near P}.  The final unfixed parameter, which we will call $U$, parameterizes the single cross-ratio of the four operators (recall that four points in a line have only one independent cross-ratio).  The precise map between the parameter $U$ and the cross-ratio $u$ is quite involved but fortunately we will not need it since the cross-ratio will be encoded in the $\chi$-system in a simple way.  The analytic structure of $T$ is shown schematically in figure 
\ref{schematicT}.  We will use this sort of figure to represent $T$ throughout this paper.   
\subsection{The function $\g$}\label{action formula}
Our objective is to evaluate the $AdS$ part of the string action.  In Poincar\'{e} coordinates the on-shell action becomes\footnote{The $AdS_2$ Poincar\'{e} coordinates are given by
\beq
Y^1 = -\frac{1}{2z}\(1-x^2-z^2\), \;\;\;\;\;
Y^2 = \frac{1}{2z}\(1+x^2+z^2\),\;\;\;\;\; 
Y^3 = \frac{x}{z}.
\eeq}
\beq\label{L}
\pd Y \cdot \pdb Y = \frac{\pd x\pdb x+\pd z\pdb z}{z^2}=\sqrt{T\bar{T}} \cosh \g
\eeq
where the above formula defines the function $\g(w,\bar{w})$.  It follows from the equations of motion that $\g$ satisfies the modified sinh-Gordon equation
\beq
\pd \pdb \g = \sqrt{T \bar{T}}\sinh \g.\label{SGE}
\eeq
It is well known that this equation is classically integrable, and in what follows we exploit this integrability to compute the $AdS$ action.  \\ 
\indent Now let us determine what boundary conditions should be imposed on $\g$. For the 2-point function $\g=0$.  Recall that the string solution should approach that of the 2-point function as the string approaches the boundary at the operator insertion points $x_a$.  Therefore we should require that $\g\ra 0$ as $w \ra w_a$ \cite{JW2}.  Furthermore,  in order to have a non-singular world-sheet metric the right-hand side of $(\ref{L})$ should never vanish.  Thus when $T$ has a zero $\g$ must have a logarithmic singularity to cancel it.  In summary, the boundary conditions on $\g$ are
\beqy
\g &\ra& \pm \frac{1}{2} \log T\bar{T}\;\;\;\;\; (w \ra z_a )\label{g at zeros} \\
\g &\ra& 0 \;\;\;\;\;\;\;\;\;\;\;\;\;\;\;\;\;\;\;\; (w \ra w_a)\label{g at punctures} 
\eeqy
where $z_a$ denotes a zero of $T$ and $w_a$ a pole of $T$.   Notice that the regularity of the world-sheet metric does not fix the sign of the logarithmic `spike' in ($\ref{g at zeros}$) and, in principle, different choices are possible at each zero (recall that generically $T$ will have 4 zeros for the 4-point function, as follows from the discussion of the previous section).  These different choices correspond to different string solutions having differing properties, and generically different total action.   We will refer to the spikes with the $+$ $\(-\)$ sign as $u$-spikes ($d$-spikes).  We will now describe how the choice of these signs is related to the string embeddings shown in figure \ref{stringconfigs2}.
\begin{figure}[t!]
\begin{center}
\includegraphics[width=0.65\linewidth]{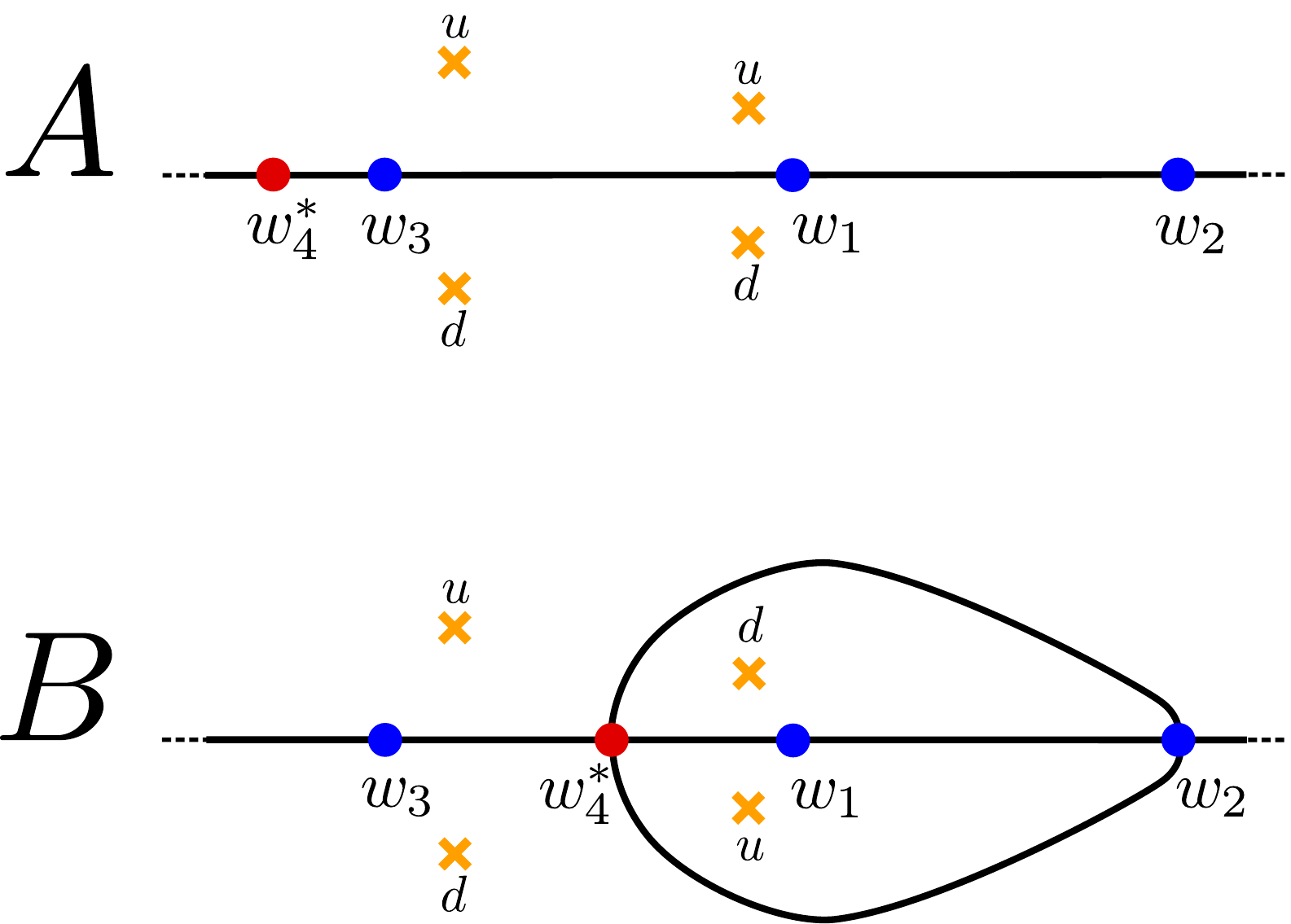} 
\caption{Contours where $\g=0$ based on the choice of signs in equation \eqref{g at zeros}.  These contours are shown schematically by the black curves.  The label $u$ ($d$) at a zero indicates the choice of sign $+$ ($-$) in equation \eqref{g at zeros}. We give a detailed discussion of why these are the only possible structures for these contours in appendix \ref{gamma props}.  The key in relating these figures to the embeddings in figure \ref{stringconfigs2} is that contours on the world-sheet where $\g=0$ map onto folds of the embedding.}\label{SpikeConfigsMainText}
\end{center}
\end{figure}
\subsection{Spikes, fold-lines and string embeddings}\label{folds and spikes}
\indent As mentioned in the previous section there are $4$ zeros of $T$ and at each zero we have a $\mathbb{Z}_2$ ambiguity (see equation \eqref{g at zeros}) in the choice of spikes of $\g$.  A priori there are $2^4$ different choices for the spikes.   However, it turns out that there are only $2$ distinct choices that correspond to target-space solutions with the desired properties.  These two possibilities are shown in figure \ref{SpikeConfigsMainText}.  A discussion of why these are the only two possible choices is given in appendix \ref{gamma props}.\footnote{The main ideas are:  first,  configurations related by $\g \ra -\g$ are not distinct since this is a symmetry of \eqref{SGE}, and second, one should choose the spikes such that $\g \ra - \g$ under reflection about the real axis.  See appendix \ref{gamma props}.}   These two different possibilities correspond precisely to the two different possible string solutions shown in figure \ref{stringconfigs2}.  The key ingredient in making this correspondence is the observation that contours on the world-sheet where $\g=0$ correspond to fold-lines in the string embedding (see appendix \ref{gamma props}).  The location of these contours is directly connected with the choice of spikes.  For example, between a $u$-spike and a $d$-spike we know that there must be at least one such contour.  In figure \ref{SpikeConfigsMainText} the $\g=0$ contours are indicated by the black curves.  In appendix \ref{gamma props} we discuss in detail how the structure of these contours is inferred from the choice of spikes.  \\
\indent Let us describe in more detail how we relate the two spike configurations in figure \ref{SpikeConfigsMainText} to the two string embeddings in figure \ref{stringconfigs2}.  As mentioned above, the key ingredient is to study the fold lines in the two figures.  First consider the target-space solution.  In figure \ref{stringconfigs2}A it is clear that there is a single fold-line that runs through each of the punctures in sequence.  That is, there is a fold-line directly connecting $x_4$ with $x_3$ then $x_3$ with $x_1$, etc.   This is in agreement with the fold-structure implied by figure \ref{SpikeConfigsMainText}A since for this choice of spikes we can deduce that there is a single contour where $\g=0$ running along the real axis connecting $w_4$ to $w_3$ then $w_3$ to $w_1$, etc.  Thus the spike configuration of figure \ref{SpikeConfigsMainText} describes a string embedding of the type shown in figure \ref{stringconfigs2}A.\\
\indent Now consider the folds of the embedding shown in figure \ref{stringconfigs2}B.  Insertions $x_1$ and $x_3$ are both connected by fold lines directly to the insertions $x_2$ and $x_4$.  Furthermore, $x_2$ and $x_4$ are connected to each other by \emph{two} fold-lines.   This is because this configuration is double-folded along that line, as one can see from the construction shown in the center panel of figure \ref{stringconfigs2}.  All of this is in perfect agreement with the fold-structure implied by figure \ref{SpikeConfigsMainText}B.  In particular, for this choice of spikes both $w_1$ and $w_3$ are directly connected to $w_2$ and $w_4^*$ by contours where $\g=0$.  Moreover, $w_2$ and $w_4^*$ are connected by \emph{two} contours where $\g=0$, precisely corresponding to the double-fold line connecting $x_2$ and $x_4$ in figure \ref{stringconfigs2}B. \\
\indent Let us comment on a subtle point regarding figure \ref{SpikeConfigsMainText}.  Note that we have placed the saddle point $w_4^*$ in different intervals in the two figures.  On one hand, we should do this in order to be in agreement with figures \ref{insertions} and \ref{stringconfigs2}. 
However, as we will see, given a cross-ratio $u$ and the saddle point $w^*_4$, the fold structure is fixed. So, to compare two different fold structures for a given cross ratio we are forced to place the saddle point $w_4^*$ in different intervals. This is in perfect agreement with the intuitive perspective of figures \ref{insertions} and \ref{stringconfigs2}.   We will return to this point in section \ref{saddle and configs}.

\subsection{The action as a wedge product}\label{sec:  action as wedge}
We will now return to the computation of the minimal $AdS$ action (see equation \eqref{semi4pt}).  Explicitly, the quantity we want to evaluate is
\beq
- \frac{\sqrt{\l}}{\pi} \int_{\Sigma \backslash \{\e_a \}}   \frac{\pd x\pdb x+\pd z\pdb z}{z^2}
\eeq
where $\Sigma \backslash \{\e_a \}$ denotes the sphere with small disks of radius $\e_a$ cut out at each puncture.  These cut-offs are not independent and are all fixed in terms of the single target-space cut-off $z=\mathcal{E}$; this is important in recovering the spacetime dependence of the correlation function and we will return to this point below \cite{JW2}.  It is convenient to separate the action into a piece that is independent of the cut-offs, and a piece where the dependence can be explicitly evaluated.  This can be done because the solution near the punctures is know.  In particular, we may write \cite{AMSV, JW2}
\beq
\mathcal{A}= - \frac{\sqrt{\l}}{\pi} \int_{\Sigma} \sqrt{T \bar{T}}\(\cosh \g-1\) -  \frac{\sqrt{\l}}{\pi}  \int_{\Sigma \backslash \{\e_a \}} \sqrt{T \bar{T}}\label{action}
\eeq
To extend the integration to the full sphere in the first term we have used the fact that the action ($\ref{L}$) goes like $\sqrt{T \bar{T}}$ near the punctures as follows from $(\ref{g at punctures})$.   We will refer to the first and second term in ($\ref{action}$) as $A_{fin}$ and $A_{div}$ respectively.  Since $T$ is known $A_{div}$ can be evaluated explicitly in terms of the $\e_a$, but to eliminate the $\e_a$ in terms of $\mathcal{E}$ requires detailed information about the string solution itself.  Fortunately, the tools necessary for computing $A_{fin}$ will also provide the necessary information to complete the calculation of $A_{div}$.  Thus, let us focus for the time being on the computation of $A_{fin}$ and return to $A_{div}$ afterwards.  \\
\indent We would like to write $A_{fin}$ in a form where the integrability of ($\ref{SGE}$) is more readily usable.  Following \cite{AMSV, JW2} we introduce the forms
\beqy
\o &=& \sqrt{T}dw\label{omega} \\
\eta &=& \frac{1}{2} \sqrt{\bar{T}}\(\cosh \g -1\)d\bar{w}+\frac{1}{4}\frac{1}{\sqrt{T}}\(\pd \g\)^2 dw \label{eta}
\eeqy
and then from a direct computation it follows that
\beq
A_{fin}= \frac{i}{2} \int_{\widetilde{\Sigma}} \o \wedge \eta\label{action as wedge}
\eeq
where $\widetilde{\Sigma}$ denotes the double cover of the sphere defined by $y^2=T(w)$.   Extending the integration from $\Sigma$ to $\widetilde{\Sigma}$ simply involves a factor of 2 since each form is odd under sheet-exchange.  An important property of these forms is that they are both closed.  The form $\o$ is clearly closed since it is holomorphic, and the closure of $\eta$ follows from the equations of motion for $\g$.  Notice that ($\ref{action as wedge}$) would be true for any choice of the $dw$ component of $\eta$.  The specific coefficient appearing in $(\ref{eta})$ is necessary for the closure of the form. \\
\indent Now we would like to apply the Riemann bilinear identity (RBI) to reduce ($\ref{action as wedge}$) to one-dimensional integrals over cycles on $\widetilde{\Sigma}$. There are two caveats in doing this -- the singularities in $\o$ and the singularities in $\eta$.  These issues where resolved in \cite{JW2}, and we follow the approach used there (see \cite{JW2} for a more detailed treatment and also \cite{KK1} for a different approach). The basic idea of the RBI is to write one of the forms of the wedge product as an exact form, $\o = dF$ where $F=\int^{P}_{P_0}\o$, which is always possible on a Riemann surface minus some contour, $L$.  In the present case $\o$ has single poles and thus $F$ will have logarithmic cuts which need to be accounted for.  A way to side-step this complication is to spread the single poles in $\o$ into a small square-root cuts such that $F$ has only square-root cuts and no singularities.  The cost of doing this is that the genus of $\widetilde{\Sigma}$ increases, but the upside is that the application of the RBI is simplified.  This takes care of the singularites in $\o$.  Now consider the form $\eta$ which behaves as $\eta \sim (w-z_a)^{-5/2}$ near the zeros of $T$.  The prescription of \cite{JW2} is to remove the points $z_a$ from the domain by taking $L$ to be the sum of the standard contour for a Riemann surface of genus $g$ and small contours $C_a$ encircling the points $z_a$.   The integrand of ($\ref{action as wedge}$) can then be written as $d \(F \eta\)$ (since $d\eta=0$ on the domain) and then Stokes theorem can be used to reduce the surface integral to a line integral over the usual boundary of the genus $g$ Riemann surface and the contours $C_a$.  The end result is that each boundary $C_a$ contributes a correction of $\pi/12$ to $A_{fin}$ in $(\ref{action as wedge})$ while the integral over the boundary of $\widetilde{\Sigma}$ gives the usual sum over cycles on $\widetilde{\Sigma}$ and thus we have\cite{JW2}
\beq\label{action as cycles}
A_{fin}=(\mbox{number of zeros}) \frac{\pi}{12} -\frac{i}{2} \(\oint_{\g_a} \!\! \o \) I^{-1}_{ab} \(\oint_{\g_b} \!\! \eta \)
\eeq
where $\{\g_{a}\}$ is a complete basis of cycles on $\widetilde{\Sigma}$ and $I_{ab}$ is their intersection matrix.  For the four-point function there is generically 4 zeros and 4 poles.  When we spread the four poles we introduce an additional 4 cuts and thus the surface is genus 5 and there will be 5 a-cycles and 5 b-cycles; that is $\{\g_{a}\}=\{\g_{a_1}, \g_{b_1}, \g_{a_2},...,\g_{a_5},\g_{b_5}\}$.  The main point is that we have reduced the computation of the surface integral ($\ref{action as wedge}$) into a sum of 1-dimensional cycle integrals of a closed form.  Such integrals are precisely what integrability is good at computing.  In the following section we will see how to compute the cycles $\oint_{\g_a}\!\! \eta$ by exploiting the integrability of $(\ref{SGE})$.
\section{The linear problem}\label{Linear Problem}
\indent To compute the $\eta$-cycles appearing in $(\ref{action as cycles})$ it is useful to consider the linear problem associated with equation $(\ref{SGE})$.  Consider a function $\psi$ obeying
\beq
\(\pd+J_w\)\psi=0, \;\;\;\;\; \(\pdb+J_{\bar{w}}\)\psi=0\label{LP}
\eeq
where the components of the connection $J=J_w dw + J_{\bar{w}}d\bar{w}$ are given by
\beq\label{connection1}
J_{w} = A_w + \frac{1}{\xi} \Phi_w, \;\;\;\;\;\; J_{\bar{w}} = A_{\bar{w}} + \xi \Phi_{\bar{w}}
\eeq
where $A$ and $\Phi$ are independent of the spectral parameter $\xi$ and given in terms of $\g$ and $T$, $\bar{T}$.   We give the explicit forms of $A$ and $\Phi$ in appendix \ref{App LP}.   Note that we will frequently write the spectral parameter as $\xi = e^{\th}$.  \\
\indent Compatibility of equations \eqref{LP} for all $\xi$ is equivalent to the flatness of $J$, which is satisfied if $\g$ obeys the equation of motion ($\ref{SGE}$) and $T$ ($\bar{T}$) is purely holomorphic (anti-holomorphic).  In the following section we will discuss the relation between the solutions of the \eqref{LP} and the $\eta$-cycles appearing in \eqref{action as cycles}.
\subsection{Basic properties}\label{basic properties}
There are a few aspects of the linear problem which will be essential for the following analysis.  Let us comment on each of them in turn.
\begin{itemize}
\item \emph{Solutions near punctures.} Using $(\ref{T near P})$ and $(\ref{g at punctures})$ one can show that near the punctures $P_a$ there are two linear-independent solutions of the form (see Appendix $\ref{App LP}$)
\beqy
\hat{\psi}^{\pm}\(w\) &\equiv& \(T/\bar{T}\)^{1/8} e^{\pm \frac{1}{2} \int^w \xi^{-1}\o+\xi \bar{\o} } |\pm\rangle\label{psi near p} \\
                                   & \sim   & \(w-w_a\)^{\pm \frac{1}{4}\D_a \xi^{-1}-\frac{1}{4}}\(\bar{w}-\bar{w}_a\)^{\pm \frac{1}{4}\bar{\D}_a \xi +\frac{1}{4}}|\pm \rangle\label{psi near p 2}
\eeqy
where $|\pm\>$ are the eigenvectors of $\sigma^3$.  Notice that there is a solution that is exponentially big and one that is exponentially small as one approaches the puncture $P_a$.\footnote{In going from \eqref{psi near p} and \eqref{psi near p 2} we have been careless about the branches of $\o$.  In particular, we may choose a particular branch at some $P_a$ such that the near-puncture solutions take the form \eqref{psi near p 2} but then if we smoothly continue $\sqrt{T}$ to some other puncture $P_b$ it is possible that the small and big solutions correspond to the opposite components from the small and big solutions at $P_a$.  This will be very important below, since it will usually be the case in the construction we will use.}
\item \emph{`Small' solutions.} Demanding that a function is both a solution of the linear problem and also small at some puncture $P$ uniquely defines that solution (up to overall normalization).  Thus there is a family of `small' solutions $s_a$ each of which is small at puncture $P_a$.  On the other hand, specifying that a solution has the big asymptotic near $P$ does not uniquely determine the solution since one could create another solution obeying the same boundary conditions by adding an arbitrary multiple of $s_P$.
\item \emph{$\mathbb{Z}_2$ symmetry and `big' solutions.}  Even though one cannot uniquely specify a solution by demanding that it has the big asymptotic near $P$, there is nevertheless a special solution big near $P$ that is uniquely defined.  This follows from the $\mathbb{Z}_2$ symmetry of the connection $(\ref{connection1})$ which is given by
\beqy
U J(\xi) U^{-1}=J(e^{i\pi}\xi)
\eeqy
where $U=i \sigma^3$.  This symmetry implies that if $s_P(\xi)$ (we are suppressing the $w$, $\bar{w}$ dependence) is the solution to $(\ref{LP1})$ small at $P$ then 
\beq
\tilde{s}_P \equiv \sigma^3  s_P(e^{-i \pi}\xi)\label{big solutions}
\eeq
 is another solution of the linear problem.  Moreover, from $(\ref{psi near p})$ it follows that $\tilde{s}_P$ is \emph{big} at $P$.  Thus we have a second uniquely defined family of solutions $\tilde{s}_{a}$, each of which is big at puncture $P_a$.
\item \emph{Products of solutions.}  Given two solutions of the linear problem $\psi_1$ and $\psi_2$, there is a natural $SL_2$ invariant inner product 
\beq
\(\psi_1 \wedge \psi_2\) \equiv \mbox{Det}\[\{\psi_1,\psi_2\}\]
\eeq
This inner product is equivalent to the Wronskian of the two solutions.  Important properties of this Wronskian are that it is independent of $w$ and $\bar{w}$, and thus only depends on the spectral parameter $\xi$.  Further, the product will vanish if the two solutions are linearly \emph{dependent}.  
\end{itemize}
\indent Now that we have introduced these basic facts of the linear problem, we can state what is perhaps the key ingredient in the whole computation.\footnote{To our knowledge the following fact first appear in \cite{AMSV}.  Later it was used in \cite{JW2,KK1} for 3-point function computations.  We give a derivation in appendix \ref{WKB};  we thank  Pedro Vieira and Amit Sever for explaining the basic components of the derivation used for \cite{AMSV}.  A different derivation appears in \cite{KK1}.  }  We claim that the $\xi\ra 0$ expansion of the inner product of two small solutions is the following\cite{AMSV,JW2}
\beq
\(s_a \wedge s_b \) \sim \mbox{exp}\[\frac{1}{2} \xi^{-1}\varpi_{ab} + \frac{1}{2}\xi \bar{\varpi}_{ab} + \xi \int_{a}^{b}\eta +\mathcal{O}\(\xi^{2}\)  \]\label{prod WKB}
\eeq 
where $\eta$ is precisely the same form $(\ref{eta})$ that appears in the action formula $(\ref{action as cycles})$ and $\varpi_{ab}$, $\bar{\varpi}_{ab}$ are explicitly known in terms of integrals of $\o$ and $\bar{\o}$.\footnote{To be more precise, this expansion will be true for certain $s_a$ and $s_b$ depending on certain conditions stemming from the form of $T$ and also depending on the value of $\mbox{Arg}\(\xi\)$.  Furthermore, the contour of integration will be precisely defined by these conditions.  We will discuss these conditions in detail below.}  A derivation of \eqref{prod WKB} is given in appendix \ref{WKB}. The point is that by computing the inner products $\(s_a \wedge s_b\)\(\xi\)$ we can extract the ``puncture-puncture"  integrals $\int_{a}^{b}\eta$ by extracting the $\mathcal{O}\(\xi\)$ coefficient of this inner product.  All of the $\eta$-cycles appearing in $(\ref{action as cycles})$  can be written in terms of linear combinations of these puncture-puncture integrals.  Thus, we can compute area $(\ref{action as cycles})$  by computing the inner products $\(s_a \wedge s_b\)$.  The rest of this section is devoted to explaining how we compute such inner products using techniques from integrability.  
\subsection{Defining solutions globally}\label{defining globally} 
\begin{figure}[ht!]
\begin{center}
\includegraphics[width=0.4\linewidth]{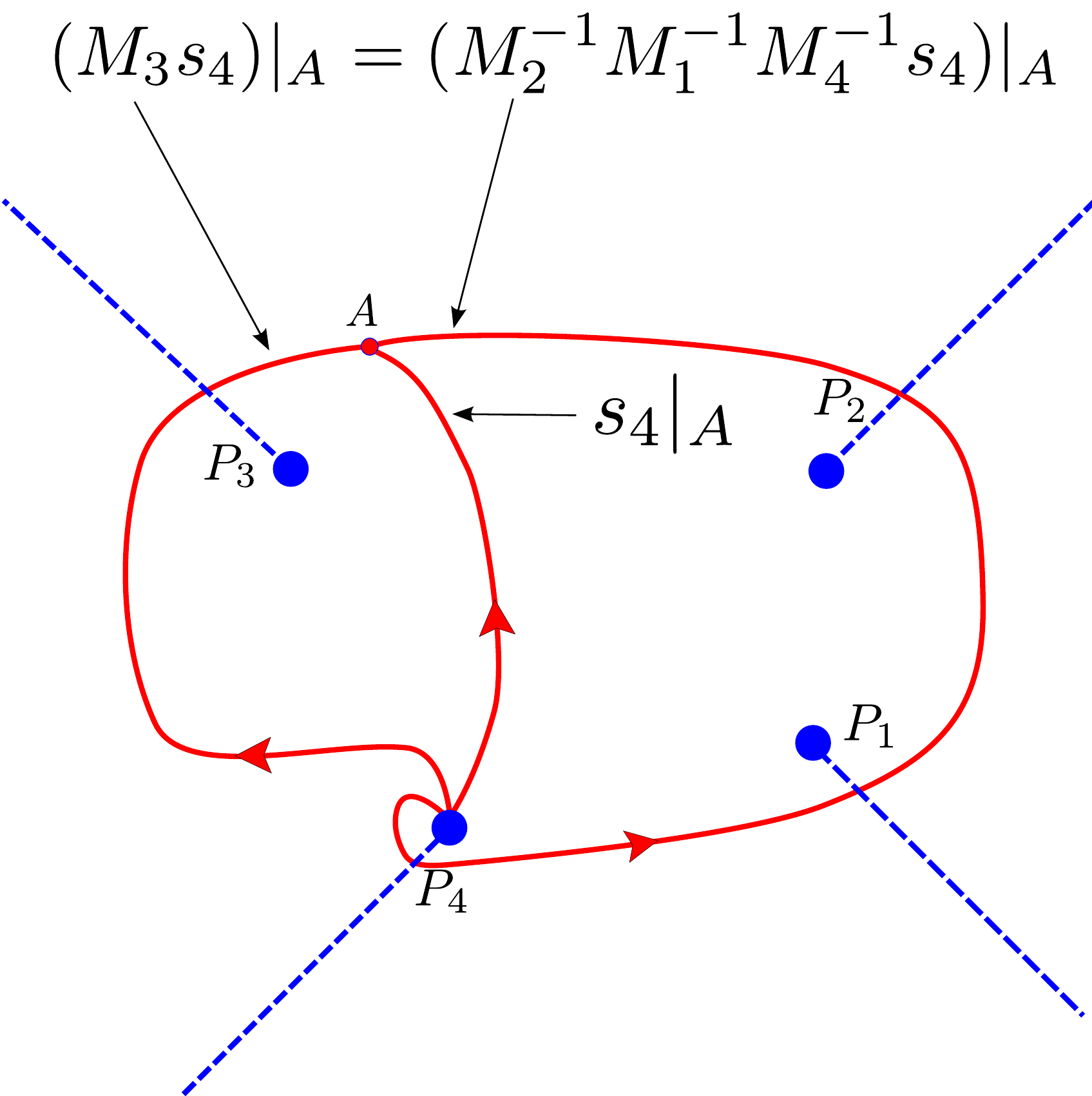}
\caption{Our conventions for defining the solutions globally.  The dashed blue lines emanating from the punctures indicate the conventions for `cutting' the full Riemann surface, thus defining the sheets.  The red lines indicate the parallel transport of a solution from $P_4$ to the point $A$ along three paths.  Two of the paths are homotopically equivalent due to the triviality of the total monodromy $M_4 M_3 M_2 M_1 =1$ (which follows from the fact that any path encircling all the punctures with the same orientation is contractable on the sphere).  The third path is homotopically distinct from the other two, and thus the value of the solution at $A$ will differ by monodromy factors. }\label{DefiningMonodromies}
\end{center}
\end{figure}

Let us now comment on how to globally define the small solutions. Suppose that we want to construct the small solution $s_P$ away from puncture $P$, say at some generic point $A$.  To do this we need to use the connection to transport the solution along some path from the neighborhood of $P$ to the point $A$.  However, it is clear from  $(\ref{psi near p})$  that the solutions of the linear problem have non-trivial monodromies around the punctures and therefore homotopically different paths on the 4-punctured sphere will result in different values of the small solution at $A$. In other words, solutions of the linear problem live on a (generically infinite-sheeted) Riemann surface with branch points at the punctures.  For the purposes of calculating it is convenient to fix some conventions for dealing with the multivaluedness of the solutions.   We first define the sheets by cutting the Riemann surface as shown in figure \ref{DefiningMonodromies}. 
The cuts all join at a common point and the monodromy about that point is the identity since a path passing through all the cuts is contractable on the sphere.  We then define the value of the small solution associated with puncture $P$ at some point $A$ as follows.  Draw any curve from the neighborhood of $P$ to $A$.  In the neighborhood of $P$ one starts with $s_P$.  For every time the path crosses a cut emanating from some puncture $Q$  in the \emph{clockwise} (\emph{counterclockwise}) sense attach a factor $M_Q$ ($M_Q^{-1}$).\footnote{Note that the result of a monodromy can be expressed as the linear map $M$ since both $s$ and $Ms$ are solutions of the linear problem.  Therefore they can both be expanded in terms of two linearly independent solutions of the linear problem, and thus they are related to each other simply by a linear map, or in other words simply by multiplication by some matrix, $M$.}  In this way, if we transport along a path that is homotopically equivalent to a path that does not cross any cuts then the small solution at $A$ will be $s_P|_A$.  If the path crosses the cut emanating from puncture $Q$ once in the clockwise sense, then the value of the small solution at $A$ will be $(M_Q s_P)|_A$, and so on (see figure \ref{DefiningMonodromies}).  In the case when $s_P$ is transported around the puncture $P$ one can see from the explicit form ($\ref{psi near p}$) of $s_P$ near $P$ that the result will be multiplication by a constant.  That is 
\beqy
M_P s_P &=& \mu_P s_P\label{mu} \\
M_P \tilde{s}_P &=& \tilde{\mu}_P \tilde{s}_P\label{mut} 
\eeqy
so that $s_P$ and $\tilde{s}_P$ are eigenvectors of $M_P$ with eigenvalues $\mu_P$ and $\tilde{\mu}_P = 1/\mu_P$ respectively.  One cannot repeat such an analysis to evaluate $M_Q s_P$ since generically one does not know the explicit form of $s_P$ in the neighborhood of $Q$.
\subsection{WKB approximation and WKB Curves}\label{WKB curves}  
\begin{figure}[ht!]
\begin{center}
\includegraphics[width=0.350\linewidth]{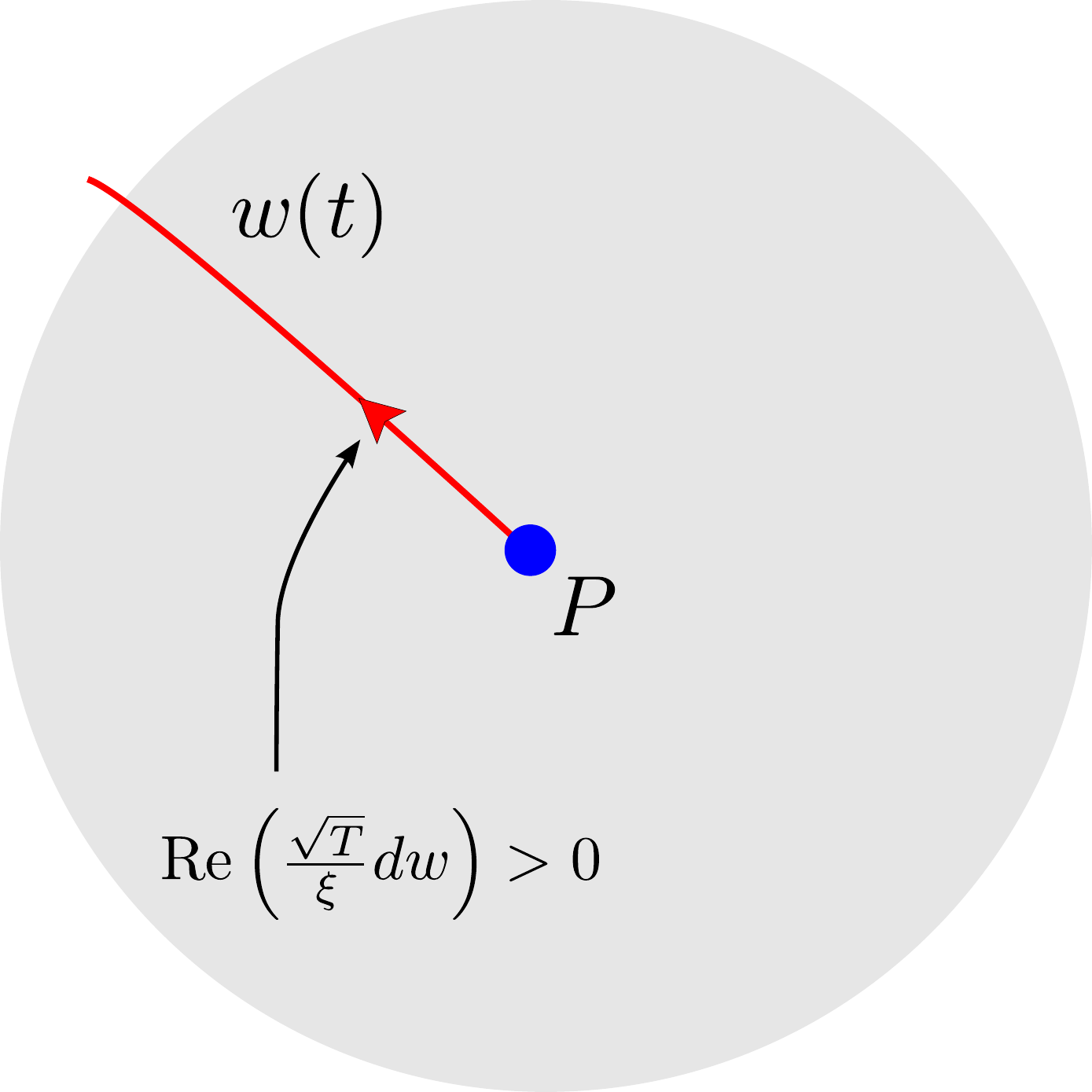}
\caption{Transporting $s_P$ away from $P$ along $w(t)$.  We have chosen the branch of $\Phi$ in $(\ref{Diag Phi 1})$ such that $s_P \propto |+\>$ near $w_P$.  In other words, we have chosen the branch of $\Phi$ such that $\mbox{Re}\(\<+|- \Phi_w/\xi dw |+\> \) = \mbox{Re}\(dw \sqrt{T}/\xi\) > 0$ for $dw$ pointing along $w(t)$ \emph{away} from $P$ and thus $\mbox{exp}\(\int^{w(t)}_{w'_P} dw \sqrt{T}/\xi \)$ is exponentially diverging as $\xi \ra 0$.  }\label{ExplainingWKB}
\end{center}
\end{figure}

As we will discuss shortly, it will be essential to have control over the $\xi \ra 0,\infty$ asymptotics of the inner products $\(s_P \wedge s_Q\)\(\xi\)$.  It is clear from $(\ref{LP})-(\ref{connection1})$ that these are both singular limits, and the basic idea of extracting this singularity -- which is called the \emph{WKB approximation} --  is as follows.\footnote{See appendix $\ref{WKB}$ or \cite{GMN} for a more detailed treatment.}   As discussed above, we have good control over the solutions in the neighborhood of the punctures.   Thus we would like to study, in the limits  $\xi \ra 0,\infty$, the transport of small $s_P$ along a curve $w(t)$ which connects a neighborhood of a puncture $P$ with a neighborhood of another puncture $Q$.  Let us consider the transport away from $P$ (see figure $\ref{ExplainingWKB}$).
We will discuss the $\xi \ra 0$ limit since the $\xi \ra \infty$ limit is similar.   \\
\indent At any point in $\Sigma$ the matrix $\Phi$ has the two eigenvalues $\mp \o/2 = \mp \sqrt{T}/2 \; dw$ (which are single valued on the double cover $\widetilde{\Sigma}$), and thus we can choose a gauge along $w(t)$ where $\Phi$ is diagonal and given by 
\beqy
\Phi = 
\frac{1}{2} \left(
\begin{matrix}
-\o  & 0   \\
0    & \o   
\end{matrix}
\right)
=
\frac{1}{2} \left(
\begin{matrix}
-\sqrt{T}dw  & 0   \\
0    & \sqrt{T}dw   
\end{matrix}
\right) 
\label{Diag Phi 1}
\eeqy
In the limit $\xi \ra 0$ some component of $\Phi/\xi$ will dominate and thus the leading contribution to $s_a$ at some point $w$ along $w(t)$ will be given by
\beq 
e^{-\int_{w'_a}^w \Phi/\xi} |\s\>\label{WKB singularity}
\eeq
where the value of $\s=\pm$ depends on the branch of  $\Phi$ we have chosen (recall that $|\pm\>$ are the eigenvectors of $\sigma^3$).   This is the singular contribution in the limit $\xi \ra 0$ for the same reason that it is the small solution -- namely, because
\beq 
\mbox{Re}\(\<\s|\(\!-\Phi/\xi\)|\s\> \) > 0\label{WKB condition}
\eeq 
along a path traveling \emph{away} from $P_a$.  The basic statement of the WKB approximation is that so long as we transport along paths such that $(\ref{WKB condition})$ is true along the whole path then the leading contribution to $s_P$ in the $\xi \ra 0$ limit is indeed given by ($\ref{WKB singularity}$).  In other words, as long as we transport along curves satisfying $(\ref{WKB condition})$ everywhere, then we can reliably extract the singularity as $\xi \ra 0$ as it is simply given by $(\ref{WKB singularity})$.  The curves along which $(\ref{WKB condition})$ is satisfied most strongly are those for which
\beq 
\mbox{Im}\(\<\s|\(\!-\Phi/\xi\)|\s\> \) = 0\label{WKB lines}
\eeq 
Curves satisfying this condition are called \emph{WKB curves}.  If we transport along some curve satisfying $(\ref{WKB lines})$ for $\mbox{Arg}\(\xi\) = \phi$, then the condition $(\ref{WKB condition})$ will be satisfied for $\mbox{Arg}\(\xi\) \in \(\phi-\pi/2,\phi+\pi/2\)$. In fact, we will need to control the asymptotics of $s_P$ in precisely such a wedge of the $\xi$-plane, and thus we should always transport along WKB lines.  We will give the a very brief overview of the properties of these lines in the next subsection.  For a detailed treatment see \cite{GMN}.
\subsection{WKB triangulation}\label{WKB Triangulation}
\begin{figure}[t!]
\begin{center}
\includegraphics[width=1\linewidth]{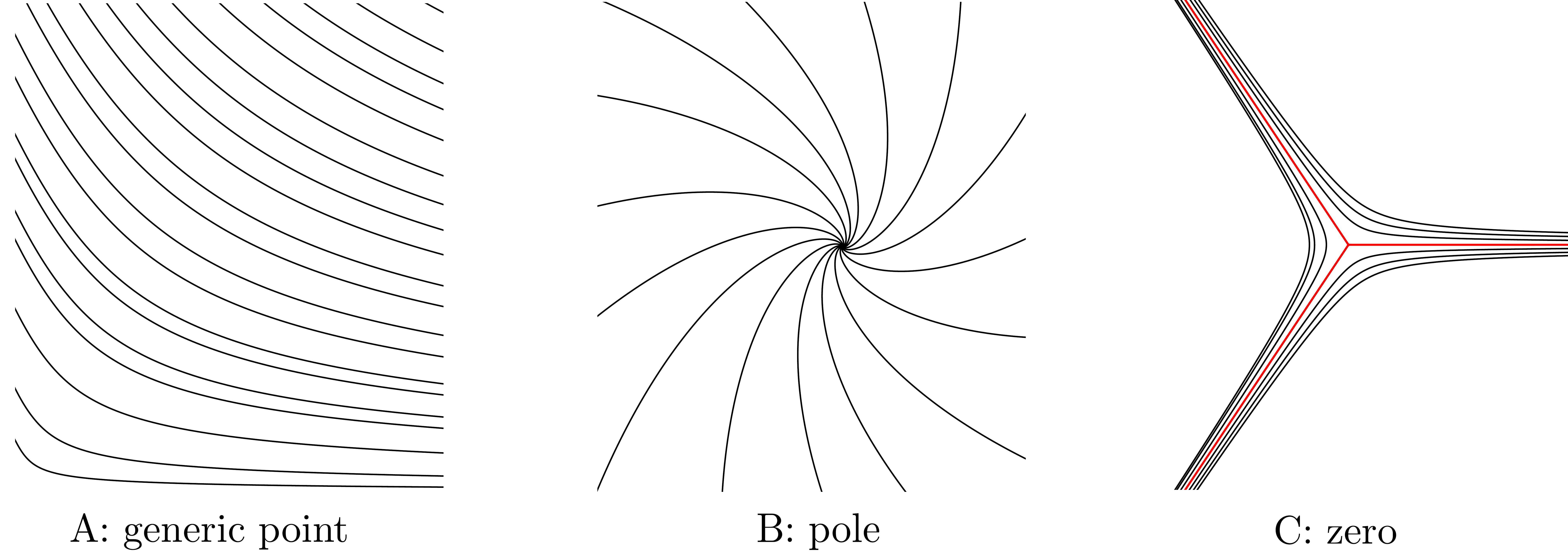}
\caption{Local structure of WKB lines in the neighborhood of, A: a generic point; B:  a double pole of $T$; C: a simple zero of $T$.  In the case of a generic point the WKB curves form continuous non-intersecting lines.  In the case of a singular point they form logarithmic spirals for generic values of $\mbox{Arg}(\xi)$.  The exact nature of these spirals will not be important.  What is important is that the singular points act as sources/sinks of WKB curves.  In the case of a zero, there are three special WKB curves that asymptote to the zero which are the red curves in panel $C$.  These special curves, called separating curves, determine the global structure of the WKB foliation.  }\label{LocalWKB}
\end{center}
\end{figure}

As we discussed in section $\ref{defining globally}$ we define the solutions of $(\ref{LP})$ globally by transporting along specific paths.  Transport of solutions along homotopically equivalent paths will lead to the same result, whereas transport along homotopically inequivalent paths generically will give different results.  For this reason it is useful to set up a system of fiducial paths between the punctures which we will use to globally define the solutions.   Because we will need to control the large/small $\xi$ asymptotics of the Wronskians, it is best to choose these paths to be WKB curves -- i.e. curves satisfying $(\ref{WKB lines})$.  \\
\indent We will first consider the local structure of WKB curves.  In the neighborhood of a generic point on the punctured sphere the WKB curves are smooth and non-intersecting (see figure \ref{LocalWKB}A).  In the neighborhood of a (double) pole of $T$ the WKB curves follow logarithmic spirals that asymptote to the singular point (see figure \ref{LocalWKB}B).  All that will be important here is that the poles act as sources/sinks of WKB curves but the exact nature of these spirals will not be important.  Finally, working in the neighborhood of a simple zero of $T$ one can show that there are three special WKB curves that asymptote to the zero and which govern the WKB lines near the zero (see figure \ref{LocalWKB}C). 
\begin{figure}[t!]
\begin{center}
\includegraphics[width=0.6\linewidth]{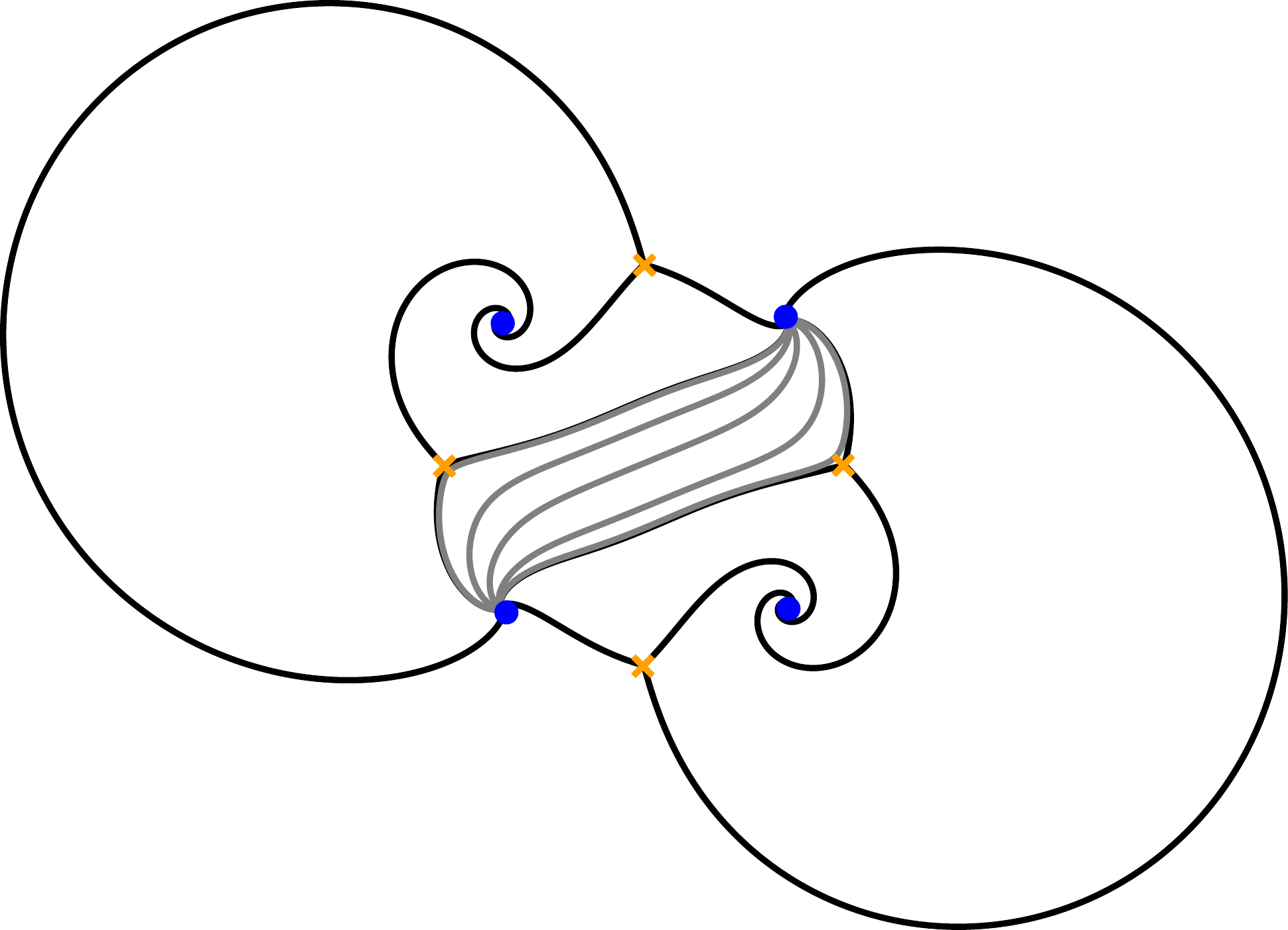}
\caption{Global WKB structure for an example with 4 punctures.  The separating curves are shown in black.  In one cell we show several examples of homotopically equivalent curves (shown in gray) that sweep the cell.  Each cell defined by the separating curves has a 1-parameter family of such curves.  By choosing a representative curve from each family we obtain the triangulation shown in figure \ref{ExampleTriangulation}.   Notice that near each puncture (the blue dots) we see the spiral structure shown in panel B of figure \ref{LocalWKB} and near each zero (yellow $\times$) we see the local structure shown in panel C of figure \ref{LocalWKB}.}\label{ExCellConstruc2}
\end{center}
\end{figure}\\
\indent Now consider the global structure of the WKB curves.  All WKB curves fall into one of the following types \cite{GMN}
\begin{itemize}
\item \emph{Generic WKB curves} which are those that asymptote in both directions to a pole of $T$ (potentially the same one);
\item \emph{Separating WKB curves} which asymptote to a zero of $T$ in one direction and to a pole of $T$ in the other;
\item \emph{Finite WKB curves} which are closed or asymptote in both directions to a zero of $T$ (potentially the same one).
\end{itemize}
\begin{figure}[t!]
\begin{center}
\includegraphics[width=0.4\linewidth]{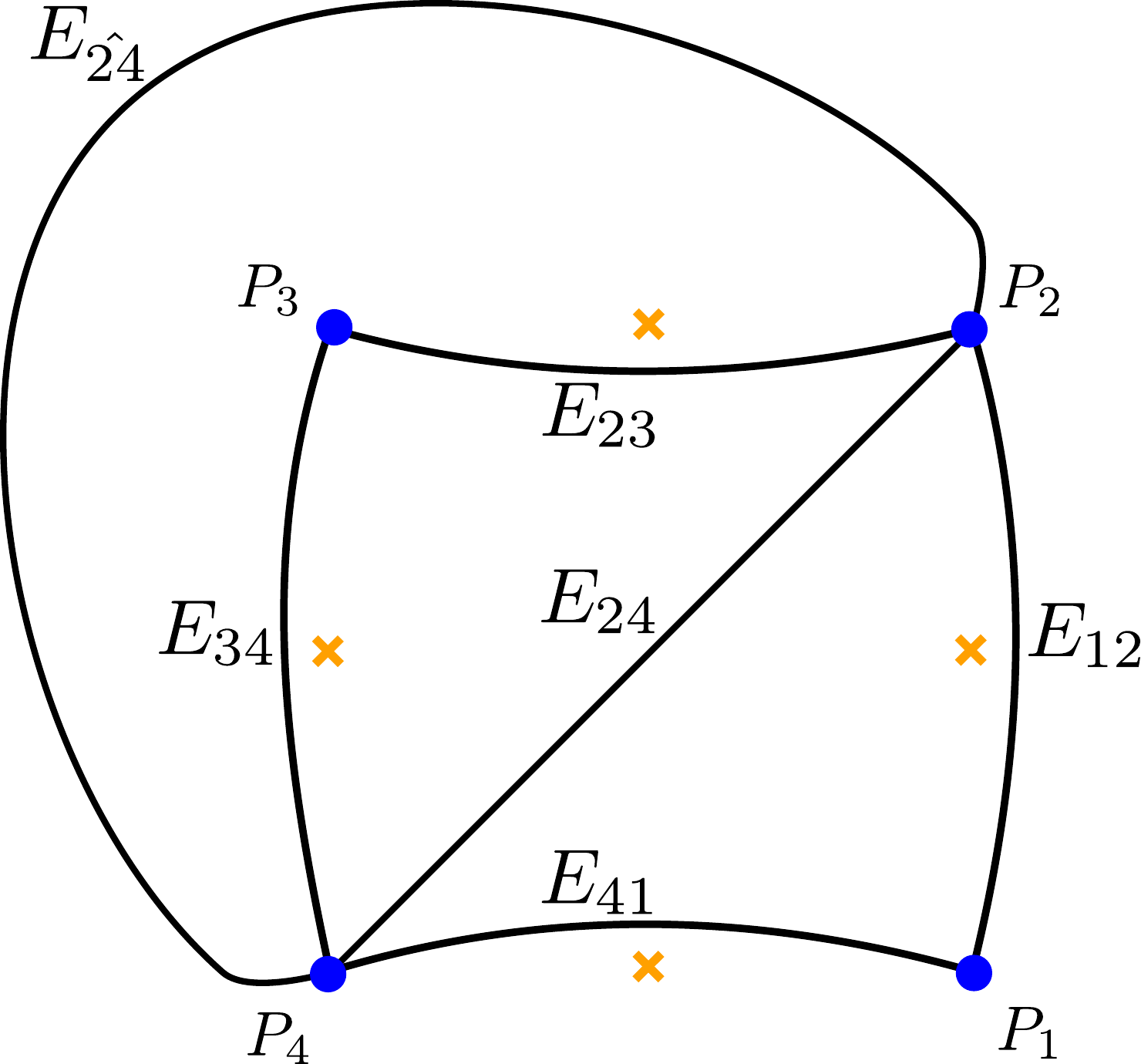}
\caption{The WKB triangulation of the 4-punctured sphere following from the WKB foliation shown in figure \ref{ExCellConstruc2}.  Each edge $E_{ab}$ of the triangulation is a representative from one of the families of homotopotically equivalent lines in each cell of figure \ref{ExCellConstruc2}.  This triangulation will be of central interest in the 4-point function computation.}\label{ExampleTriangulation}
\end{center}
\end{figure}
We will now describe how we use the WKB curves to set up a system of fiducial paths, or more specifically, a \emph{triangulation}.  By triangulation we mean a triangulation of the punctured sphere with all vertices at the punctures and at least one edge incident on each vertex.  Consider fixed $T$ and $\mbox{Arg}(\xi)$ such that there are no finite WKB curves (this can always be done since such curves only appear at special, discrete values of $\mbox{Arg}(\xi)$).  First draw all of the separating WKB curves -- there will be $3N_Z$ of these, where $N_Z$ is the number of zeros of $T$ (since for the moment we are not allowing finite WKB curves).  These curves will divide the punctured sphere up into cells with each cell defining a family of homotopically equivalent \emph{generic} WKB curves as shown in figure \ref{ExCellConstruc2} for an example of the 4-punctured sphere.  To construct the triangulation,  choose a representative curve from each cell, e.g. any one of the silver curves shown in figure \ref{ExCellConstruc2}.  The claim is that the collection of these representative curves, which we will call \emph{edges}, gives the desired triangulation \cite{GMN}.\footnote{To see this in general consider a single zero of $T$ as shown in figure \ref{LocalWKB}.   The zero is on the boundary of three cells.  Choosing edges from the family of curves in each cell we see that they form a triangle.  Thus the edges form a triangulation of the punctured sphere with each triangle containing a zero of $T$.}   As a concrete example, the triangulation associated with the cell-construction of figure \ref{ExCellConstruc2} is shown in figure \ref{ExampleTriangulation}.   This same triangulation will play an important role in the 4-point function computation below. \\
\indent We have now finished the discussion of how to construct the WKB triangulation for a given $T$ and $\mbox{Arg}\(\xi\)$.  Before moving on the the next section let us discuss one final point.  In the following it will be useful to lift edges of the triangulation to the double cover $\widetilde{\Sigma}$ and to endow the lifted edges with an orientation.  Recall that $\o = \sqrt{T} dw$ is a single valued form on $\widetilde{\Sigma}$.  Let $\pd_t$ be a tangent vector of the lifted edge $E$ at a point on $\widetilde{\Sigma}$.  There are of course two possible orientations for $\pd_t$.  Note that by virtue of \eqref{WKB lines} we have $e^{-i \phi}\o \cdot \pd_t \in \mathbb{R}$.  We define the orientation of the lifted edge $E$ by the condition $e^{-i \phi}\o \cdot \pd_t >0$.  Notice that each edge on the punctured sphere will lift to two edges -- one on each sheet of $\widetilde{\Sigma}$ and that these two edges will have opposite relative orientation.  Picking a particular orientation of some edge is equivalent to picking a particular sheet of $\widetilde{\Sigma}$.
\subsection{Coordinates}\label{coordinates}
\begin{figure}[t!]
\begin{center}
\includegraphics[width=0.4\linewidth]{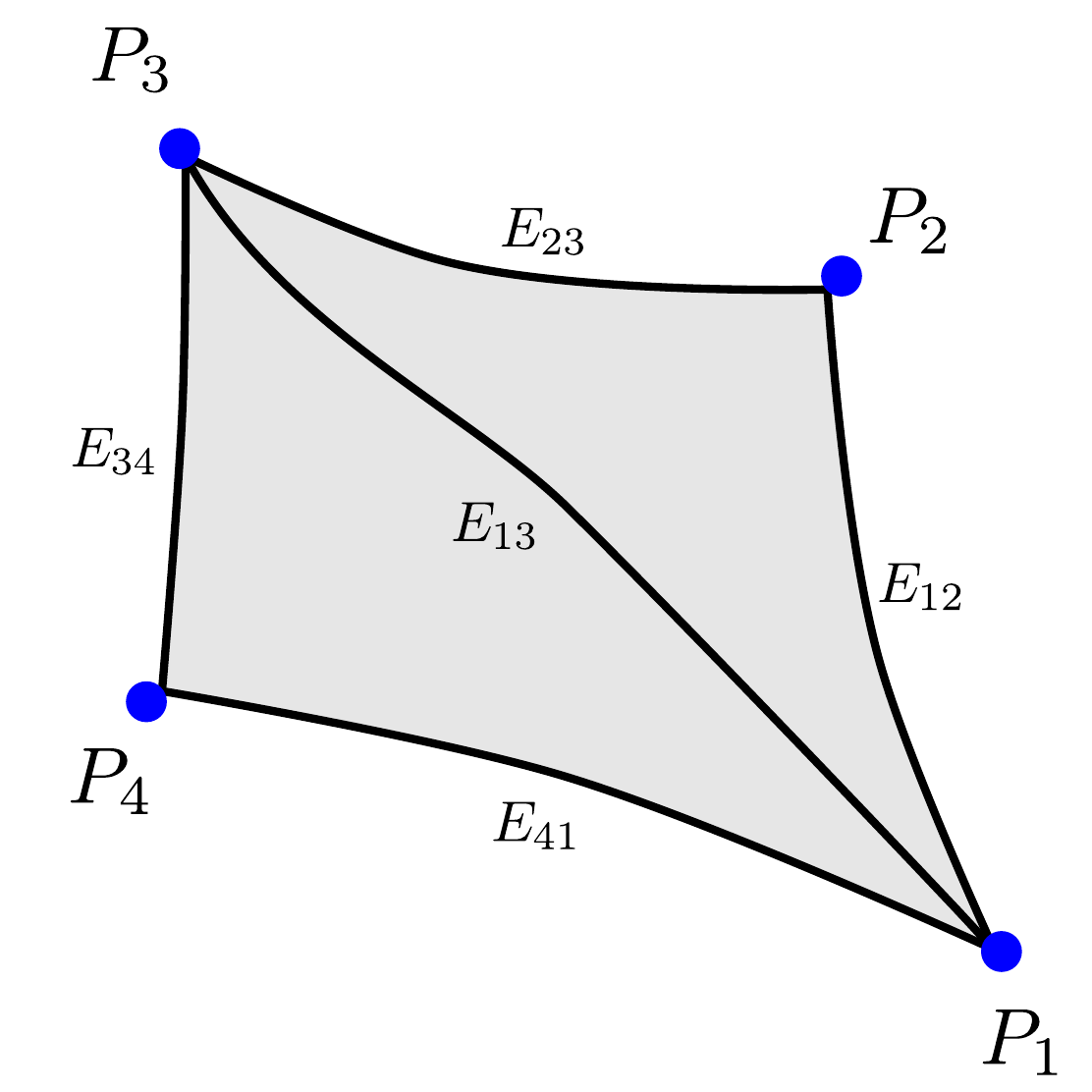}
\caption{The two triangles sharing the edge $E_{13}$.  These two triangles define the quadrilateral $Q_{E_{13}}$, which is shown in gray shading.  Each blue dot represents a puncture, which are the verticies of the triangulation and each black line and is an edge.}\label{ChiConstruction0}
\end{center}
\end{figure}
From the WKB triangulation we will now construct the so-called Fock-Goncharov coordinates \cite{GMN}.  These are natural objects to work with because they are gauge invariant and independent of the normalization of the small solutions.  From the coordinates we will be able to extract the $\eta$-cycles that we need to compute the action \eqref{action as cycles}. \\
\indent Consider some edge $E$ of the triangulation.  This edge is shared by precisely two triangles, and these triangles form the quadrilateral $Q_E$ (see figure $\ref{ChiConstruction0}$).  Number the vertices of $Q_E$ as shown in figure  $\ref{ChiConstruction0}$ with $E$ going between $P_1$ and $P_3$.  As we mentioned in section \ref{basic properties}, associated with each puncture $P_a$ there is a small solution $s_a$.  The solutions  cannot be made globally smooth and single valued on the punctured sphere due to the monodromy around each puncture.  However, we can define them such that they are single valued and smooth throughout $Q_E$.\footnote{We will show this in some concrete examples momentarily.}  We then define the Fock-Goncharov coordinate as \cite{GMN}
\beq
\chi_{E}=(-1)\frac{(s_1 \wedge s_2)(s_3 \wedge s_4)}{(s_2\wedge s_3)(s_4 \wedge s_1)}\label{FG coord}
\eeq  
where all the $s_a$ are evaluated at a common point in $Q_E$.   \\
\begin{figure}[t!]
\begin{center}
\includegraphics[width=1\linewidth]{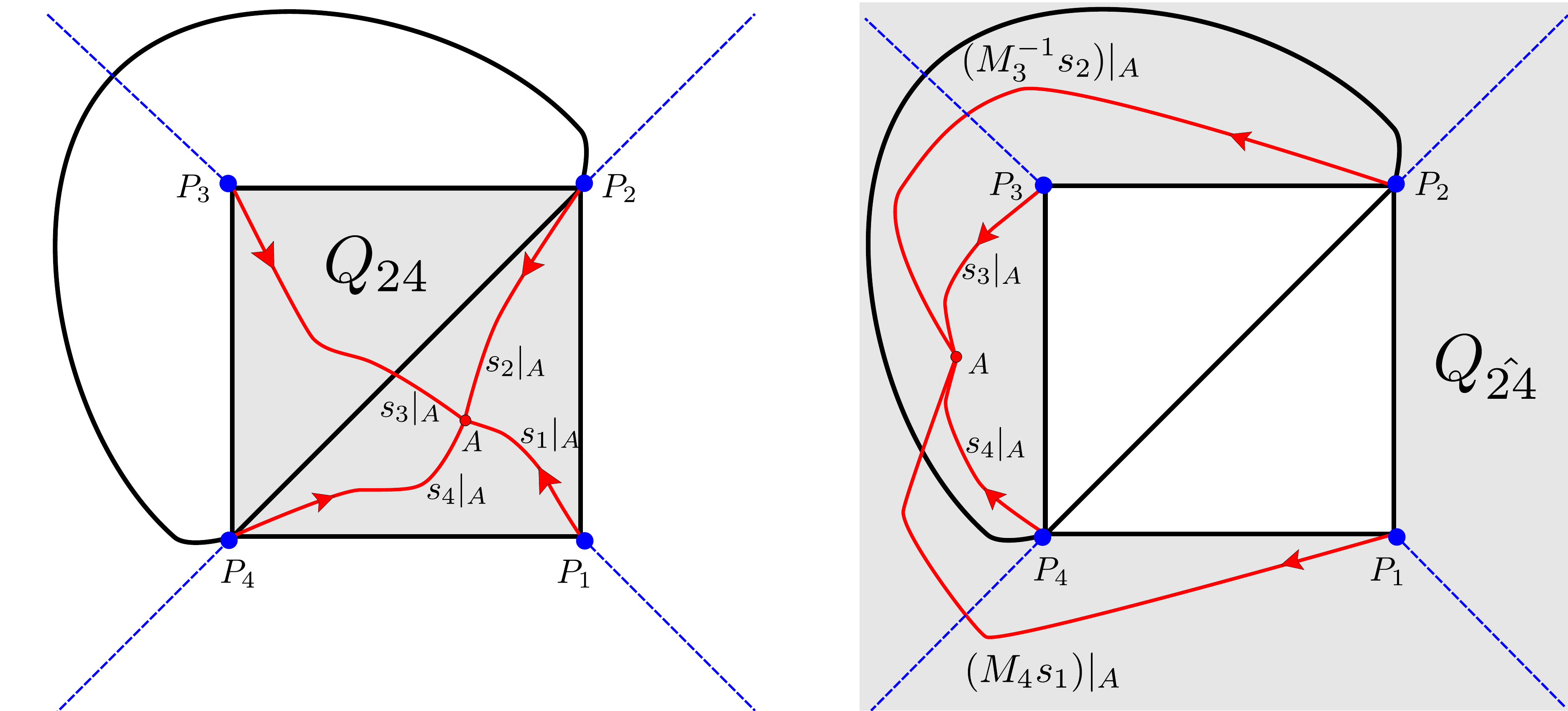}
\caption{Here we show how to construct the coordinates $\chi_{24}$ (left panel) and $\chi_{\hat{24}}$ (right panel) of the triangulation of figure \ref{ExampleTriangulation}.  The gray shaded regions represent $Q_{24}$ and $Q_{\hat{24}}$ respectively.  These figures should be pictured on the sphere.  The dashed blue lines emanating from the punctures indicate our conventions for defining the sheets of the small solutions as explained in section \ref{defining globally}.  The red lines indicate how we globally define the solutions $s_a$ by transporting away from $P_a$ using the connection.  We use paths that never leave the quadrilateral such that the solutions used to form the coordinates are guaranteed to be single-valued and smooth throughout the quadrilateral, as required.}  \label{ChiConstruction1}
\end{center}
\end{figure}
\begin{figure}[h!]
\begin{center}
\includegraphics[width=0.50\linewidth]{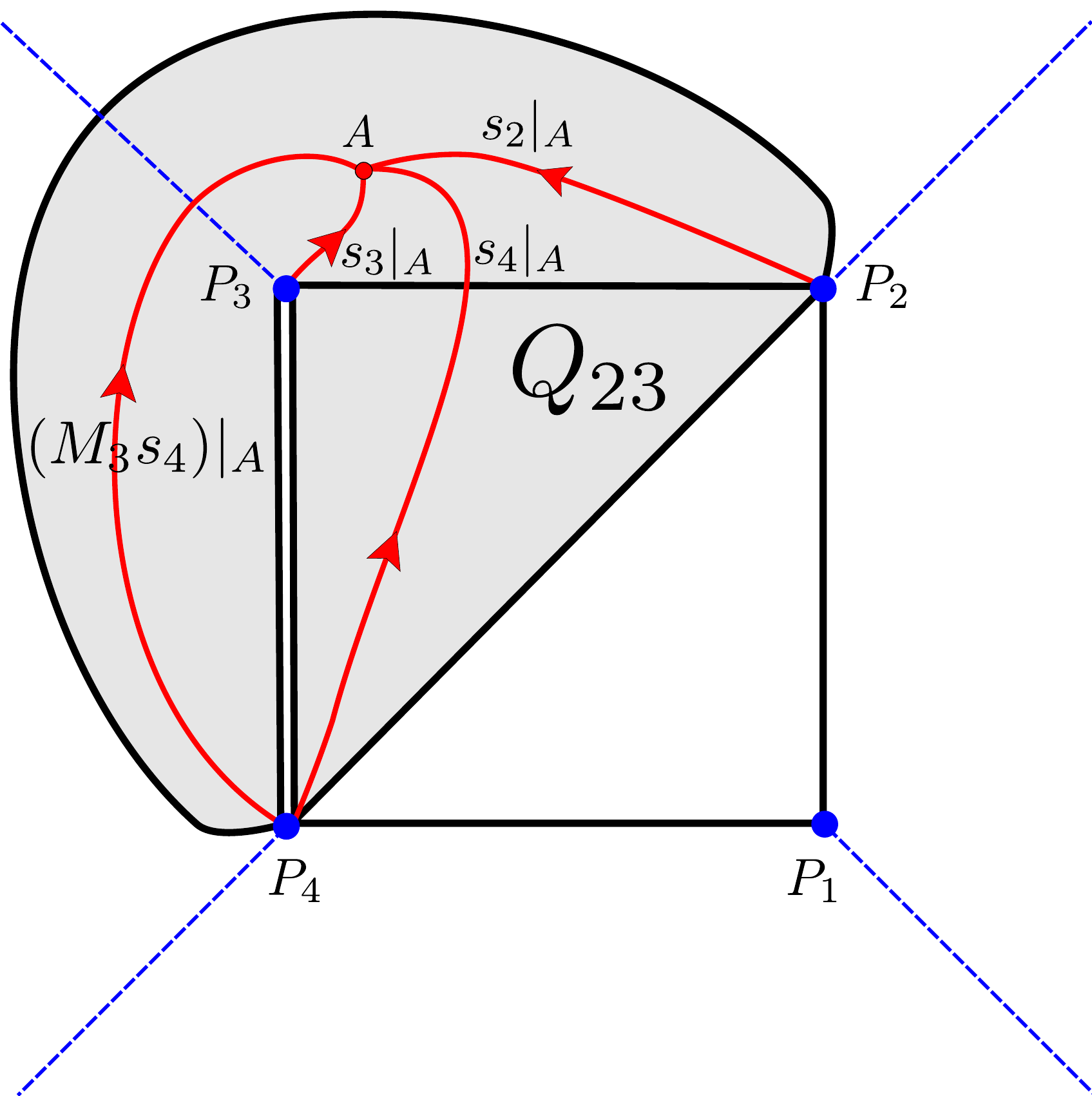}
\end{center}
\caption{Here we describe the construction of the coordinate for the slightly degenerate case where the coordinate corresponds to an edge ending at a vertex that has only two incident edges (e.g. $P_2$ has only 2 incident edges: $E_{12}$ and $E_{23}$).  We construct the coordinate for edge $E_{23}$ of the triangulation shown in figure $\ref{ExampleTriangulation}$.  The quadrilateral prescription described above still applies, but one must take care to correctly define $Q_E$  and the solutions inside $Q_{E}$.  First of all,  in order to have single-valued and smooth solutions throughout $Q_{23}$ we must exclude a region between $P_3$ and $P_4$.  Otherwise $Q_{23}$ would contain $P_3$ and thus the solutions could not be single valued in $Q_{23}$ (there would be a monodromy around $P_3$).  Since the boundaries of the quadrilateral must be edges of the triangulation, the only choice is to remove a thin region running along edge $E_{34}$ and then to treat the two `sides' of $E_{34}$ as different edges.  In the figure we have represented this process by showing $E_{34}$ as doubled and with the region between the new edges excluded from $Q_{23}$. We then define the solutions throughout $Q_{23}$ in the same way as described in figure $\ref{ChiConstruction1}$, by analytically continuing the solutions along paths from $P_a$ to $A$ that stay within $Q_{23}$ which is represented as the shaded region.  Once we have defined the solutions at a common point we form the coordinate $\chi_{23}$ given in equation \eqref{chis2}.}\label{ChiConstruction2}
\end{figure}
\indent As a concrete example consider the triangulation of the 4-punctured sphere shown in figure $\ref{ExampleTriangulation}$.   
In figure $\ref{ChiConstruction1}$ we show how to apply the procedure just described to construct the coordinates corresponding to edges $E_{24}$ and $E_{\hat{24}}$.  Consider first the left panel of $\ref{ChiConstruction1}$.  We define each solution $s_a$ throughout $Q_{24}$ by parallel transporting from each $P_a$ where the explicit form of the solutions is known -- see $(\ref{psi near p})$.  The red lines indicate the parallel transport of each $s_a$ from $P_a$ to a common point $A$; clearly we can define the small solutions at any point $A \in Q_{24}$ in this way.  Further, if the paths never leave the quadrilateral $($or at least is always homotopically equivalent to a paths that never leave the quadrilateral$)$ then the solution defined in this way is guaranteed to be single-valued and smooth throughout the quadrilateral, as required. With the solutions defined at a common point in the quadrilateral we can construct the coordinate $\chi_{24}$, which is independent of the choice of $A \in Q_{24}$.  Now consider the right panel of figure $\ref{ChiConstruction1}$ where the grey shading indicates the quadrilateral associated to edge $Q_{\hat{24}}$.  These figures should be imagined on the sphere. Now to transport the small solutions to a common point one cannot avoid passing under a cut onto new sheets of some of the small solutions.  For example $s_2$ must pass onto a new sheet in order to be smoothly continued to the point $A$.  This is because if we were to compare the $s_2$ of the left panel and the $s_2$ of the right panel $($by moving each respective $A$ to a common point $A'$ along the edge $E_{34}$, for example$)$ the two paths of continuation would differ by a holonomy around $P_3$, and thus the values at the point $A'$  would not coincide but would differ by the action of $M_3^{\pm1}$.  Of course which solution we call $s_2$ and $M_3^{\pm1}s_2 $ is purely a matter of convention.  Similarly, which solutions acquire factors of $M_a$ depends on the choice of the point $A$.  We stress that the coordinates are independent of all such ambiguities, as one can easily check using identities such as $ \(M_c s_a \wedge s_b\)=\(s_a \wedge M_c^{-1} s_b\)$, etc.  Then from figure $\ref{ChiConstruction1}$ we read off \beq
\chi_{24} = (-1)\frac{(s_2 \wedge s_3)(s_4 \wedge s_1)}{(s_3\wedge s_4)(s_1 \wedge s_2)}, \;\;\;\;\;
\chi_{\hat{24}} = (-1)\frac{(M_{3}^{-1}s_2\wedge M_{4}s_1)(s_4 \wedge s_3)}{(M_{4}s_1 \wedge s_4)(s_3 \wedge M_{3}^{-1}s_2 )} \, .\label{chis1}
\eeq
\indent We will also need to construct coordinates in the slightly degenerate case where the coordinate corresponds to an edge ending at a vertex that has only two incident edges (including the edge under consideration) for example all edges in figure $\ref{ExampleTriangulation}$ except $E_{24}$ and $E_{\hat{24}}$.  We show how to construct this coordinate in figure $\ref{ChiConstruction2}$.  Using the procedure described there we find
\beq
\chi_{23} = (-1)\frac{(s_2 \wedge M_3 s_4)(s_3 \wedge s_4)}{(M_3 s_4 \wedge s_3)(s_4 \wedge s_2)}, \;\;\;\;\;
\chi_{12} = (-1)\frac{(s_1 \wedge M_1^{-1} s_4)(s_2 \wedge s_4)}{(M_1^{-1} s_4 \wedge s_2)(s_4 \wedge s_1)}\label{chis2}
\eeq
The other two coordinates $\chi_{34}$ and $\chi_{14}$ are computed in a similar way. \\
\indent We have now completed our discussion of how to construct the coordinates.  Before we continue, let us comment on a useful property of these objects.   Consider multiplying all of the coordinates associated with edges meeting a given puncture $P$.  For example, the edges ending at $P_2$ in the triangulation of figure \ref{ExampleTriangulation} are $E_{12}$, $E_{\hat{24}}$, $E_{23}$ and $E_{24}$.  Using \eqref{chis1}-\eqref{chis2} we have
\beq
\chi_{12}\chi_{\hat{24}}\chi_{23}\chi_{24}=\mu^2_{2} \, .
\eeq
This property is true in general since the inner-products in the coordinates telescopically cancel in the product and the only thing that remains is the effect of the monodromy around the puncture which produces a $\mu^2_P$ factor.  Thus we have the general rule \cite{GMN}
\beq\label{mu rule}
\prod_{E \; \small{\mbox{meeting}}\; P}\chi_E = \mu_P^2\, .
\eeq
\subsection{WKB asymptotics of the coordinates}\label{coord asymp}  
\begin{figure}[t!]
\begin{center}
\includegraphics[width=0.85\linewidth]{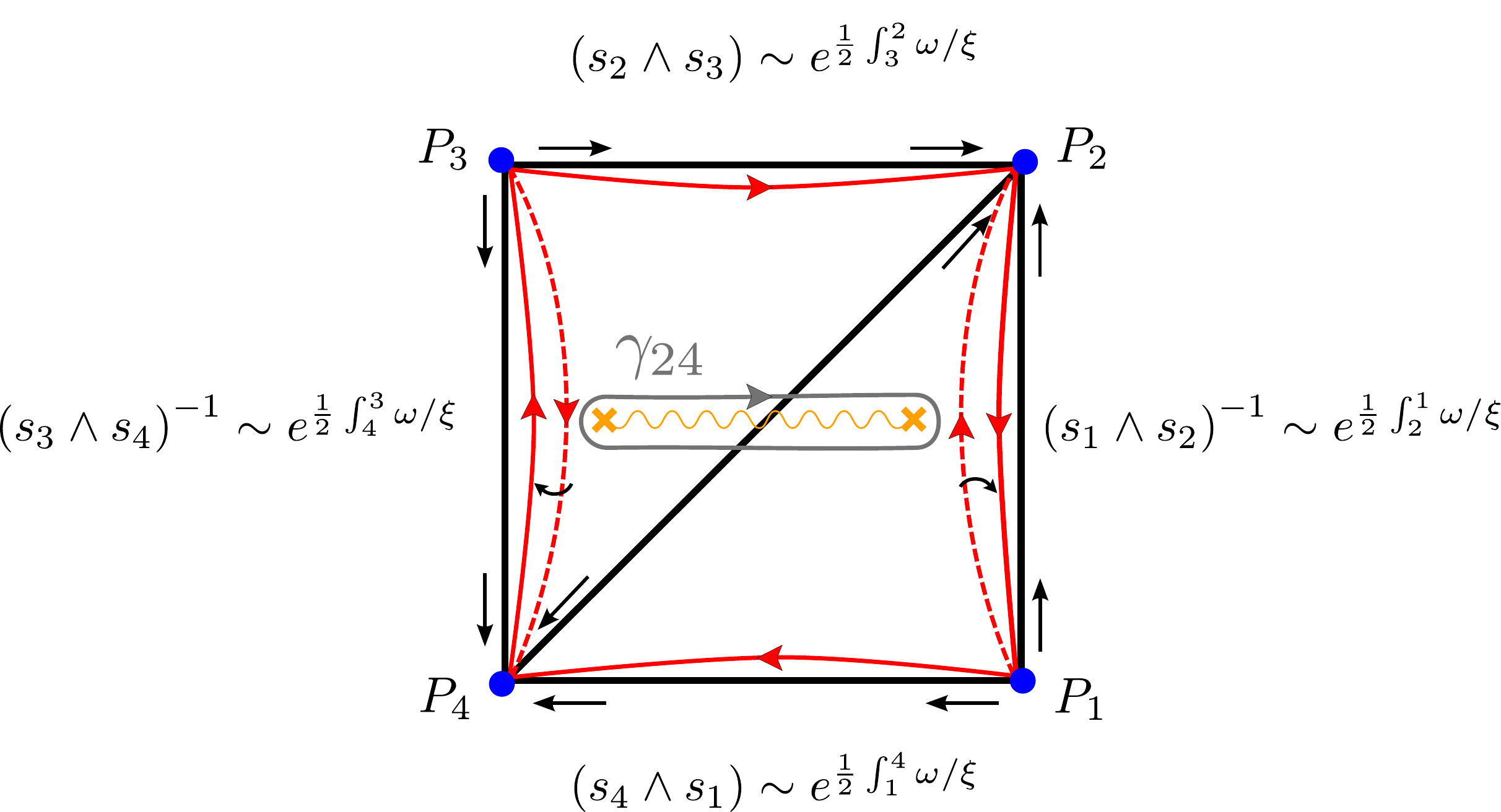}
\end{center}
\caption{Computing the $\xi \ra 0$ asymptotic of the coordinate $\chi_{24}$ for a typical WKB triangulation.  The blue disks represent the punctures and the black lines represent edges of the triangulation.  A yellow $\times$ represents a zero of $\o$ and the wavy yellow line shows our convention for defining the sheets of $\widetilde{\Sigma}$. The black arrows running along the edges indicate the choice of the direction for the edges.  Each red curve indicates the transport of a small solutions in the limit $\xi \ra 0$.  The dashed red lines correspond to the transport of a solution appearing in the \emph{denominator} of the coordinate.  The transports used to form the coordinate combine into the continuous integral of $\o$ near the boundary of $Q_{24}$, which can then be deformed into the cycle integral $\g_{24}$ shown in gray.  }\label{ExWKBCycle}
\end{figure}
The advantage of using the WKB triangulation is that the $\xi \ra 0, \infty$ asymptotics of the coordinates of the triangulation are easily extracted given the discussion of section $\ref{WKB curves}$.  That is, because we have maximum control over the large/small $\xi$ asymptotics of the small solutions when we transport along WKB curves. We give only the basic idea of the derivation of these asymptotics here and refer the reader to appendix $\ref{WKB}$ and \cite{GMN} for details. \\
\indent To obtain the asymptotic of some $\chi_E$ one simply needs to use expression $(\ref{WKB singularity})$ for each inner-product of the coordinate, taking care to account for the direction of the WKB lines.  Consider the coordinate associated with edge $E_{24}$ in figure \ref{ExWKBCycle}.  The expression for this coordinate in terms of the small solutions is given in \eqref{chis1}.   We will now use formula \eqref{WKB singularity} to compute the asymptotic of this coordinate in the $\xi \ra 0$ limit.  Let us take the directions of the WKB lines to be as given in figure \ref{ExWKBCycle}.  To evaluate the inner product $\(s_2 \wedge s_3\)$  we must transport the solutions to a common point.  Since there is a WKB line flowing from $P_3$ to $P_2$ we can safely use (to leading order) expression $(\ref{WKB singularity})$ to transport $s_3$ to the neighborhood of $P_2$, giving a contribution of the form $\(s_2 \wedge s_3 \) \sim e^{\frac{1}{2}\int_{3}^{2} \o/\xi}$.  To evaluate $\(s_3 \wedge s_4\)$ we must transport $s_3$ to $P_4$ since that is the direction of the WKB line and thus we get the contribution $\(s_3 \wedge s_4 \)\sim e^{\frac{1}{2}\int_{3}^{4} \o/\xi}$.  We may then reverse the order of integration in $\(s_3 \wedge s_4\)$ and also move it to the numerator of the coordinate.  Then the integrations from $\(s_2 \wedge s_3\)$ and $\(s_3 \wedge s_4\)$ combine nicely into a continuous integral running just inside the boundary of $Q_{24}$ from the neighborhood of $P_2$ to $P_3$ to $P_4$.  Repeating this analysis for the remaining two brackets one obtains a closed cycle integral passing along the boundary of $Q_{24}$.  Recall from the discussion of section \ref{WKB Triangulation} that each triangle in the WKB triangulation encloses one zero of $\o$.  The integral of $\o$ thus encloses two zeros and so it can be deformed to the cycle integral shown in figure \ref{ExWKBCycle}.  Thus the non-vanishing contribution in the limit $\xi \ra 0$ is given by
\beq
\chi_E \sim (-1)\exp\(\frac{1}{2} \xi^{-1}\int_{\g_E}\o + C_E^{\(0\)} \) \label{small xi}
\eeq
The contour $\g_E$ is the cycle encircling the two zeros contained in $Q_E$ and its direction is the same as that of the WKB lines corresponding to the brackets in the numerator of the coordinate.  The term $C_E^{\(0\)}$ is the $\mathcal{O}\!\(\xi^0\)$ contribution to the WKB expansion, which we will discuss momentarily.  The overall $(-1)$ prefactor in ($\ref{small xi}$) is the same $(-1)$ appearing in the definition of the coordinate ($\ref{FG coord}$). \\
\indent To derive the subleading WKB corrections (in the $\xi \ra 0$ limit, for example) is essentially a matter of perturbation theory once the singular contribution has been extracted.  We give a detailed discussion of this in appendix \ref{WKB}.  Here we will simply focus on the result and its implications.  We find the first subleading contribution is given by
\beq\label{C0}
C_E^{\(0\)} = \log (-1)^{u_E} \pm i \pi 
\eeq
where $u_E$ is the number of $u$-spikes enclosed by $\g_E$. \\
\indent Finally the $\xi \ra \infty$ asymptotic follows in the same way as the $\xi\ra 0$ and leads to a cycle integral around $Q_E$ of $\xi \bar{w}$.    \\
\indent To summarize, the $\xi \ra 0,\infty$ asymptotics for $\chi_E$ are given by 
\beq
\chi_E \sim (-1)^{u_E}\exp\[\frac{1}{2}\int_{\g_E} \(\xi^{-1}\o+\xi \bar{\o} \) \]\label{WKB asymp}
\eeq  
where $\g_E$ is the cycle encircling the two zeros contained in $Q_E$ and its direction is the same as that of the WKB lines corresponding to the brackets in the numerator of the coordinate.  Now it is clear how the choice of spikes (i.e. the choice of signs in \eqref{g at zeros}) is encoded into the coordinates -- via the constant term in the WKB expansion which contributes the $(-1)^{u_E}$ factor in \eqref{WKB asymp}. Recall that $u_E$ is the number of $u$-spikes encircled by $\g_E$.

\subsection{Shift relation.}  In section $\ref{basic properties}$ we explained that there are two special solutions $s_P$, $\tilde{s}_P$ associated with each puncture $P$ and that they are related to each other by a shift in the spectral parameter: $\tilde{s}_P(\xi)=\sigma^3 s_P(e^{-i \pi}\xi )$.  Here we give an alternative relation between the small and big solutions that does not involve shifting the spectral parameter.  The solutions $s_P$ and $\tilde{s}_P$ are linearly independent and thus we can expand any solution $s_Q$ in terms of them.  In particular we have
\beqy
s_Q &=& \(\frac{\tilde{s}_P \wedge s_Q}{\tilde{s}_P \wedge s_P}\) s_P +\(\frac{s_P \wedge s_Q}{s_P \wedge \tilde{s}_P}\) \tilde{s}_P \\
M_P s_Q &=& \(\frac{\tilde{s}_P \wedge s_Q}{\tilde{s}_P \wedge s_P}\)\mu_P s_P +\(\frac{s_P \wedge s_Q}{s_P \wedge \tilde{s}_P}\) \mu^{-1}_P \tilde{s}_P
\eeqy
For the second equality we have used \eqref{mu}-\eqref{mut}.  Combining these two equations it follows that
\beq
\(\frac{M_P s_Q\wedge s_Q}{M_P s_Q \wedge s_P}\)=(1-\mu_{P}^{2})\(\frac{\tilde{s}_P\wedge s_Q}{\tilde{s}_P \wedge s_P}\)\label{shift relation}
\eeq
The utility of this equation is that it allows us to replace certain wronskians involving big solutions (as on the RHS of  ($\ref{shift relation}$)) in terms of small solutions with monodromies.  This will play a key role in the derivation of the functional equations that we present in the following section. 
\subsection{$\chi$-system.}\label{chi system}  
We will now derive a set of functional equations for the coordinates which, together with certain analytic properties, allows us to determine the coordinates completely.  Our inspiration comes from the solution of the bosonic Wilson-loop problem at strong coupling \cite{AMSV} where the solution involves a set of functional equations of the schematic form\footnote{The linear problem associated with that problem is very similar to the one considered here and the $Y_a$ are (up to shifts in the spectral parameter) the coordinates associated with that problem.  We are referring here to the special case where the Wilson loop lives in an $\mathbb{R}_{1,1}$ subspace.}
\beq
Y_a^{+} Y_a^{-} = F_a(Y)\label{Y system}
\eeq   
where $f^{n \times \pm} \equiv f(\th \pm n i \pi/2)$.   On the RHS of $(\ref{Y system})$ the function $F_a$ can depend on all of the $Y_a$, but with their arguments un-shifted.  The only shifts in the spectral parameter occur on the LHS of $(\ref{Y system})$.  For the Wilson-loop problem the $F_a$ are such that $(\ref{Y system})$ takes the form of a so-called Y-system which commonly appear in the context of $1+1$ dimensional integrable QFT's.  Here, using the general formalism of \cite{GMN}, we will arrive at a set of functional equations with the same schematic form as $(\ref{Y system})$ but with the $F_a$ of a different form than that occurring in the Wilson-loop  problem.  We will call this type of functional equation a $\chi$-system. \\
\indent To derive a relation of the form $(\ref{Y system})$ we begin with the LHS.  Using $(\ref{big solutions})$ we have
\beqy
\chi_E \widetilde{\chi}_E =\chi_E \chi_E^{++}\label{LHS chi system}
\eeqy 
where $\widetilde{\chi}_E$ is defined by taking $\chi_E$ and replacing each small solutions $s_a \ra \tilde{s}_a$.   To obtain the schematic form $(\ref{Y system})$ we need to rewrite ($\ref{LHS chi system}$) in terms of only un-shifted small solutions.  That is, we need to get rid of all the tildes without introducing any shifts in the spectral parameter.  For this we can use $(\ref{shift relation})$ after applying the Schouten identity\footnote{\label{schouten} $(s_a \wedge s_b)(s_c \wedge s_d)+(s_a\wedge s_c)(s_d \wedge s_b)+ (s_a \wedge s_d)(s_b \wedge s_c)=0.$ } to $(\ref{LHS chi system})$ to obtain
\beq
\chi_E \chi_E^{++}=\chi_E \widetilde{\chi}_E = \frac{(1+A_{ab})(1+A_{cd})}{(1+A_{bc})(1+A_{da})}\label{chi to A}
\eeq 
where we have defined the useful auxiliary quantity 
\beqy
A_{PQ} &=& (-1)\frac{\(s_Q \wedge \tilde{s}_P\)\(s_P \wedge \tilde{s}_Q\)}{\(s_P \wedge \tilde{s}_P\)\(s_Q \wedge \tilde{s}_Q\)}\label{defn A}\\
 &=& 
(-1)\(1+\mu_P^2\)^{-1}\(1+\mu_Q^2\)^{-1}\(\frac{M_P s_Q\wedge s_Q}{M_P s_Q \wedge s_P}\)\(\frac{M_Q s_P\wedge s_P}{M_Q s_P \wedge s_Q}\label{A using shift}\)
\eeqy
Here, the edge $E$ is the edge $ac$ in $Q_E$ where the vertices are labeled $abcd$ in counter-clockwise order.  To go from ($\ref{defn A}$) to ($\ref{A using shift}$) we used the shift relation $(\ref{shift relation})$.  The last step is to rewrite the wronskians appearing in $(\ref{A using shift})$ in terms of the coordinates.  Once this is done, combining $(\ref{LHS chi system})-(\ref{A using shift})$, we can assemble a functional equation of the form $(\ref{Y system})$.  To do this (following \cite{GMN}) we introduce the quantity
\beq
\Sigma\(P;Q\ra Q\) = 1+\chi_{P,a}\(1+\chi_{P,a-1}\(1+...\chi_{P,2}\(1+\chi_{P,1}\)\)\)\label{def Sigma}
\eeq
\begin{figure}[t!]
\begin{center}
\includegraphics[width=0.350\linewidth]{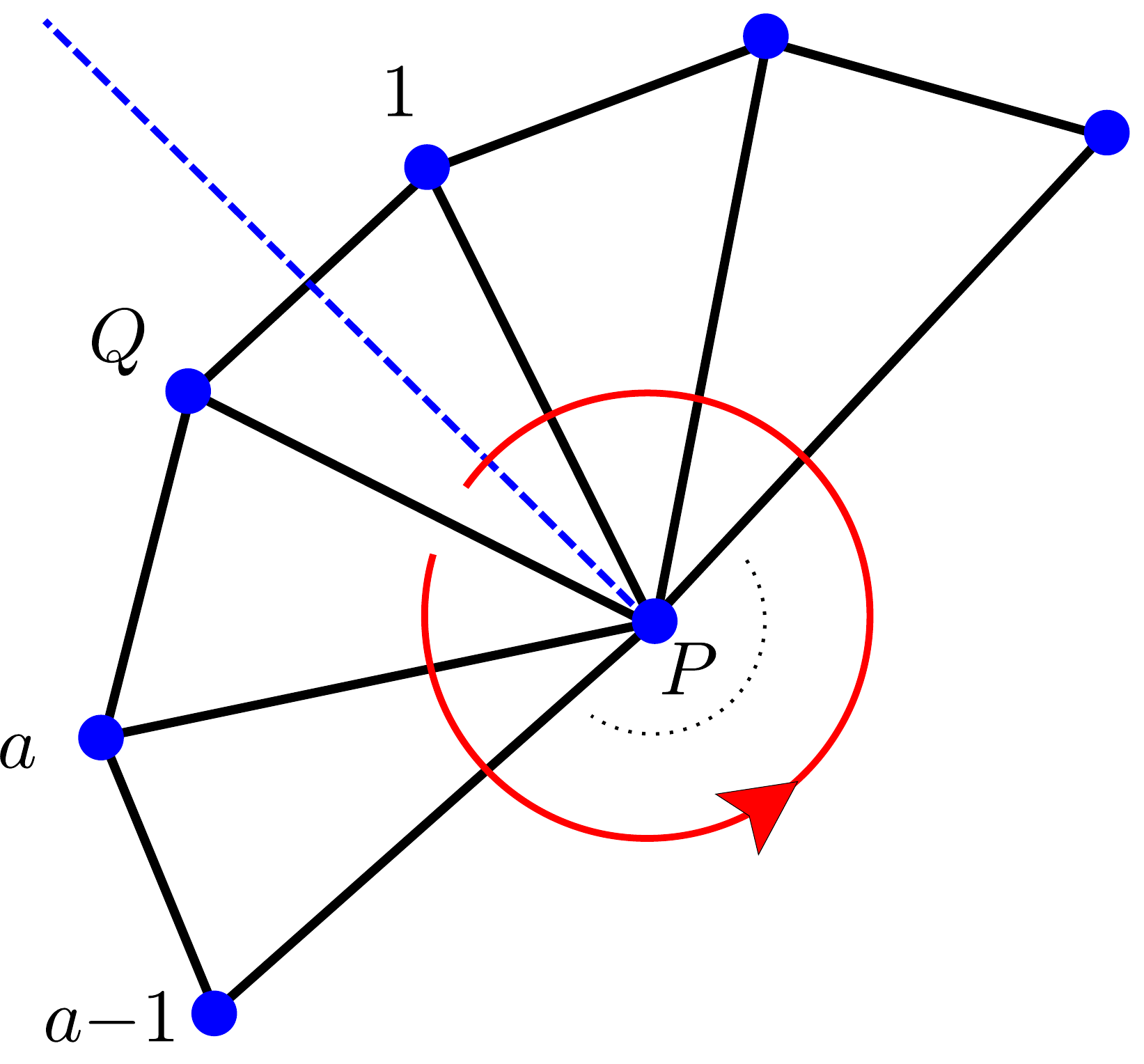}
\end{center}
\caption{Graphical rules for constructing $\Sigma\(P;Q\ra Q\)$.  Start at edge $E_{PQ}$ and continue in a \emph{counterclockwise} fashion about $P$ forming the nested product $(\ref{def Sigma})$ by multiplying the coordinates for each edge encountered along the way (i.e. the coordinates associated with each edge intersected by the red line in the order indicated by the arrow).  The dashed blue line indicates our convention for cutting the solutions to account for the monodromy around $P$.  The small solutions used to form the coordinates are defined in the vicinity of $P$ by analytically continuing them throughout the triangles along the direction indicated by the red arrow and thus if we use $s_Q$ in $\chi_{P,a}$ then we must include a monodromy matrix when the solution is continued around $P$ to form $\chi_{P,1}$.}\label{ComputingSigma}
\end{figure}
The coordinates appearing in this object are shown in figure $\ref{ComputingSigma}$.  By repeatedly applying the Schouten identity (see footnote \ref{schouten}) starting with $\(1+\chi_{P,1}\)$ one can see that the Wronskians in $(\ref{def Sigma})$ telescopically cancel so that\footnote{An  easy way to see this in general is to use induction \cite{GMN}.  The case $a=1$ is simple to prove using Schouten identity.  Then one can show (again using Schouten) that $\Sigma\(P;Q_{a+2}\ra Q_0\) = 1+\chi_{P,a+1}\Sigma\(P;Q_{a+1}\ra Q_0\) $.}
\beq
\Sigma\(P;Q\ra Q\) = \frac{(s_0 \wedge s_{a+1})(s_P \wedge s_a)}{(s_{a+1}\wedge s_a)(s_0 \wedge s_P)} = 
\frac{(s_P \wedge s_{a})(M_P s_Q \wedge s_Q)}{(s_{Q}\wedge s_a)(M_P s_Q \wedge s_P)}\label{Sigma with mono}
\eeq
In going from the first equality to the second in ($\ref{Sigma with mono}$) we have accounted for the monodromy acquired by the small solutions when they are analytically continued around $P$ (see figure $\ref{ComputingSigma}$).  Then, from $(\ref{Sigma with mono})$ and $(\ref{A using shift})$ we have
\beq
(1+\mu_P)^2(1+\mu_Q)^2 A_{PQ} = \chi_{PQ} \Sigma\(P; Q\ra Q\)\Sigma\(Q; P\ra P\)\label{A to Sigma}
\eeq
Finally, using $(\ref{A to Sigma})$ in $(\ref{chi to A})$ and noting $(\ref{def Sigma})$ we obtain a closed functional equation for $\chi_E$ of the form $(\ref{Y system})$.  Repeating this procedure for the coordinate associated to each edge in a given triangulation gives the desired set of functional equations.  Note that this procedure can be applied to derive the $\chi$-system for an arbitrary number of punctures.  In section \ref{AdS action} we will apply this procedure to the triangulation $(\ref{ExampleTriangulation})$, which is one of the triangulations of interest for the four-point function computation.  
\subsection{Inverting $\chi$-systems}\label{inversion}
In the previous section we showed how to derive the $\chi$-system associated with a given triangulation of the $N$-punctured sphere.  In this section we will discuss how to use the $\chi$-system along with certain analytic properties of the coordinates to obtain integral equations that determine the $\chi_E$ uniquely.  \\
\indent The basic idea behind the inversion of a $\chi$-system is to Fourier transform (the $\log$ of) each equation since in Fourier space these nonlocal relations become local as the shifts in the parameter $\th$ can be undone in the usual way.   For such a procedure to be successful one must have a certain amount of control of the analytic properties of the coordinates.  Let us discuss this carefully.  The equations that we want to Fourier transform have the form
\beq
\log \chi^-_E\(\th+ i \phi \) + \log\chi^+_E\(\th+ i \phi\) = \log F_E\(\chi^{\pm}\(\th+ i \phi\)\) \label{log chi system}
\eeq
where $F_E\(\c\)$ is an explicit function of the coordinates which follows from the discussion of section \ref{chi system}.  We have introduced the arbitrary shift $\phi$ for reasons that will be explained momentarily.  Note that $\chi_E \chi_E^{++}=\chi_E \chi_E^{--}$ since the small solutions are $2\pi i$-periodic, which is why we can have either shift $F_E\(\chi^{\pm}\)$ on the RHS. The choice of this shift is arbitrary since the objects we will eventually compute (the $\eta$-cycles) are functionals of the coordinates only through $A_{PQ}$ which is $i \pi$-periodic and thus does not care about the choice of shift.  As a convention we choose the shift $- i\pi/2$.  \\
\indent To Fourier transform the relationship $(\ref{log chi system})$ one must be sure that the transform converges.  Moreover, to undo the shifts on the LHS, one must account for the singularities (if any) of $\log \chi_E$ in the strip of width $\pi$ centered along the line where the transform has been performed. We will now discuss each of these issues in turn. \\
\indent The information from the WKB analysis will allow us to ensure the convergence of the Fourier transform, provided certain conditions are satisfied.  First consider the LHS of $(\ref{log chi system})$.  We need to ensure that the transform of each \emph{individual} term converges. We can ensure this if we know the asymptotics of the coordinates in the full strip $\mbox{Im}\(\th\)\in \(\phi-\pi/2,\phi+\pi/2\)$.   The coordinates should be derived from the triangulation that one has at $\mbox{Im} \(\th\)= \phi$.  Then the WKB analysis guarantees that the asymptotics are given by $(\ref{WKB asymp})$ in a strip that includes the region $\mbox{Im}\in \(\phi-\pi/2,\phi+\pi/2\)$.   Each term on the LHS can be made safe to transform by making (on the LHS only) the replacement $\c_E \ra \c_E/\c_E^{(0)}$ where $\c_E^{(0)}$ is the asymptotic $(\ref{WKB asymp})$.  This replacement does not modify the equation since $\(\c_E^{(0)}\)\(\c_E^{(0)}\)^{++}=1$. \\ 
\indent Now consider the RHS of $(\ref{log chi system})$, which has the form (see equation \eqref{chi to A})
\beq
\log F_E\(\chi^{\pm}\(\th+i\phi\)\)=\log \[\frac{(1+A_{ab})(1+A_{cd})}{(1+A_{bc})(1+A_{da})}\(\th\pm i\pi/2+i\phi\)\]\label{RHS of log chi system}
\eeq 
Each $A_{PQ}$ is computed by $(\ref{A to Sigma})$ and $(\ref{def Sigma})$. For the RHS of \eqref{RHS of log chi system} to be decaying it is sufficient for all of the $A_{PQ}$ in $(\ref{RHS of log chi system})$ to be decaying.  If all the $\chi$-functions are decaying then from \eqref{A to Sigma} and \eqref{def Sigma} it is clear that all of the $A_{PQ}$ will decay; the $\mu$-factors will decay by virtue of the rule \eqref{mu rule}.   On the other hand, if all the $\chi$-functions are growing the $\mu$-factors in $(\ref{A to Sigma})$ will dominate the RHS of $(\ref{A to Sigma})$ so that $A_{PQ}$ is still decaying; to see this one should re-express the $\mu$-factors in terms of the coordinates using \eqref{mu rule}.  Thus the RHS of \eqref{RHS of log chi system} will decay if all of the $\chi$-functions are growing, or alternatively if they are all decaying.  For generic $\phi$ it will generally not be true that the RHS of $(\ref{RHS of log chi system})$ is well behaved, and one must try to find a range of $\phi$-values for which the $\chi_{E}$ are all decaying or are all growing.  If a suitable $\phi$ can be found, then $(\ref{log chi system})$ can be directly solved by Fourier-transform.  In all of the examples we have considered (in particular, those relevant for the 4-point function) it has been possible to find such a $\phi$.\\  
\indent Concerning the issue of singularities within the strip of inversion, it follows from $(\ref{LP})$ that the Wronskians $\(s_a \wedge s_b\)\(\th\)$ are (in an appropriate normalization) analytic away from $\th =\pm \infty$.  It is, however, possible for these objects to have \emph{zeros} and in the following it is an assumption that there are no zeros in the strip where we do the inversion.\footnote{ In the limit where the WKB approximation holds, i.e. when $\th \ra \pm \infty$ or in the limit of large zero modes $|Z_{E}| \ra \infty$ \cite{GMN}, it is clear that (in an appropriate normalization)  the Wronskians will not have any zeros since (suppose we compute the Wronskian near $P_b$) then $s_a$ will be the \emph{big} solution near $P_b$ and is thus linearly independent of $s_b$ which is small at $P_b$.  For finite values of $\th$ (or alternately of $|Z_E|$) we have no concrete way of arguing that these zeros are not present.}  In section \ref{num tests} we perform numerical tests that support this assumption.    \\ 
\indent Finally, we use the Fourier analysis to obtain
\beq
\log X_E\(\th\) = \log X_E^{(0)}\(\th\) - \int_{\mathbb{R}} \frac{d\th'}{2 \pi i} \frac{\log F_{E}\(X\(\th'\)\)}{\mbox{sinh}\(\th'-\th +i0\)}\label{chi TBA}
\eeq
 where $X_E\(\th\) = \c_E\(\th+ i \phi-i\pi/2\)$ and $X_E^{(0)}$ is the (shifted) asymptotic $(\ref{WKB asymp})$ and $F_E\(X\)$ is an explicit function of the coordinates which follows from the discussion of the previous section.  \\
\indent The equations \eqref{chi TBA} can easily be solved for the $X_E$ by iterating them in a computer.  In the next subsection we will show how to extract the $\eta$-cycles of formula $(\ref{action as cycles})$ from the $X_E$ which are computed using $(\ref{chi TBA})$.  We will then perform some numerical tests in section \ref{num tests}.
\subsection{Extracting $\eta$-cycles}\label{extract}
Once the coordinates are computed according to the prescription of the preceding section we extract the $\eta$-cycles as follows.  What we need to compute are the individual Wronskians $\(s_a \wedge s_b\)$.  For this, note that from \eqref{defn A} and footnote \ref{schouten} we have
\beqy
\(1+A_{ab}\) &=& \frac{\(s_a\wedge s_b \)\(\tilde{s}_a \wedge \tilde{s}_b\)}{\(s_a \wedge \tilde{s}_a\)\(s_b \wedge \tilde{s}_b\)} \label{inner prods from A}
\eeqy
We can choose a guage where $\(s_P \wedge \tilde{s}_P\) =1$.  The final result will be gauge independent.  With this gauge choice we have
\beq\label{wedge to A}
\log \(s_a\wedge s_b\)^{-}+\log\(s_a\wedge s_b\)^{+} = \log\(1+A^{-}_{ab}\)
\eeq
Here we will use the notation $\th \ra \th+i\phi$ where $\th$ and $\phi$  are real.  We then insert the zero-modes on the LHS in the same way as for the $\chi$-system (see section $\ref{inversion}$).  We are only interested in $P_a$ and $P_b$ that are connected by a WKB line when $\mbox{Arg}\(\xi\)=\phi$, and thus we have good control over the asymptotics in the required strip.  Performing the Fourier transforms we obtain
\beq\label{extracting eta 1}
\log\(s_a\wedge s_b\)\(\th+i\phi\) 
= \(\frac{1}{2}e^{-\th-i\phi}\varpi_{ab}+ \frac{1}{2} e^{\th+i\phi}\bar{\varpi}_{ab}\) + \int_{\mathbb{R}}\frac{d\th'}{2\pi}\frac{\log\(1+A^{-}_{ab}\(\th'+i\phi\)\)}{\cosh\(\th-\th'\)} 
\eeq
where we have defined 
\beq\label{omega reg 2}
\varpi_{ab} \equiv 
\lim_{w'_a \ra w_a}\lim_{w'_b \ra w_b}\left[\int_{E_{ab}} \sqrt{T} dw + \frac{\D_a}{2} \log(w_a-w'_a)+\frac{\D_b}{2} \log(w_b -w'_b) \right]
\eeq
The integration in \eqref{omega reg 2} is performed along edge $E_{ab}$.  The direction of integration is the same as the direction of the edge $E_{ab}$ (see appendix \ref{WKB}).  Note that the logarithmic terms precisely cancel the divergence from the endpoints of integration in \eqref{omega reg 2} so that the $\varpi_{ab}$ are finite. In going from \eqref{wedge to A} to \eqref{extracting eta 1} we have used the asymptotics for $\(s_a \wedge s_b\)$ derived in appendix \ref{WKB}.  \\
\indent Expanding $(\ref{extracting eta 1})$ around $\th \ra -\infty$, and comparing with $\(\ref{prod WKB}\)$  with $\xi = e^{\th+i \phi}$ we read off
\beq\label{extracting eta 2}
\int_{E_{ab}}\eta = \int_{\mathbb{R}}\frac{d\th}{\pi}e^{-\th-i\phi}\log\(1+A^{-}_{ab}\(\th+i\phi\)\)
\eeq
The contour of integration in $\int_{E_{ab}} \eta$ is along the WKB line connecting $P_a$ and $P_b$ and the direction of integration is the same as the direction of the edge $E_{ab}$. This formula allows us to compute the $\eta$-cycles from the $\chi$-functions since the $A_{PQ}$ are explicit functions of the coordinates.  
\section{The AdS action}\label{AdS action}
\subsection{Regularized $AdS$ action}\label{sec:Afin}
Now that we have introduced the needed tools we are ready to calculate the action $(\ref{action as cycles})$.  We will demonstrate for the case of the $4$-point function, but the method is general and could be performed for any number of operators inserted along a line.  The computation will be as follows.  First we will introduce the relevant WKB triangulation which will be topologically equivalent to the triangulation shown in figure $\ref{ExampleTriangulation}$.  Second, using the procedure of section $\ref{chi system}$ we will derive the $\chi$-system satisfied by the coordinates of this triangulation.  Supplementing these functional relations by the WKB asymptotics we will invert these functional relations using the technique of section $\ref{inversion}$ to obtain a set of integral equations that uniquely determine the coordinates.  Finally, from coordinates we extract the $\eta$-cycles using the method of section $\ref{extract}$.  Once we have the $\eta$-cycles, we compute the action using $(\ref{action as cycles})$. 
\subsubsection{Stress-energy tensor and WKB triangulation}
\begin{figure}[t!]
\begin{center}
\includegraphics[width=1.0\linewidth]{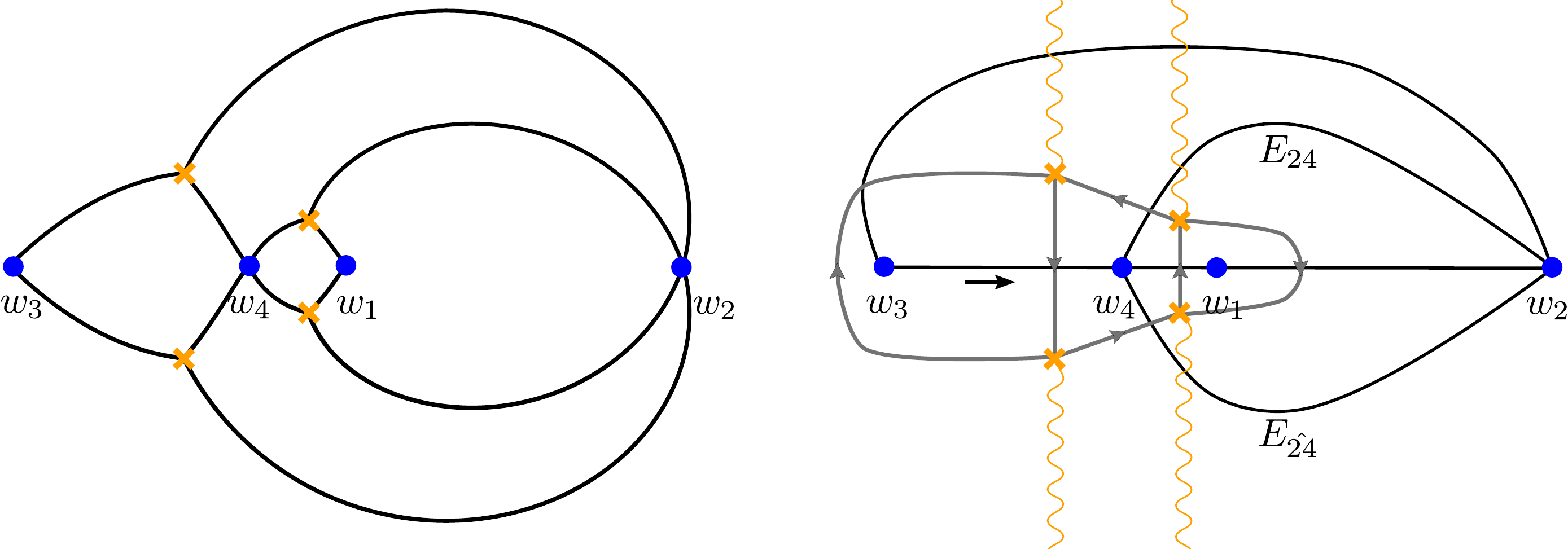}
\end{center}
\caption{Constructing the triangulation for the 4-point function.  In the left panel we show the WKB cells for $\mbox{Arg}\(\xi\)=0$.  The cell walls are formed by the separating WKB curves as described in section \ref{WKB Triangulation}; as described there,  inside each cell there is a 1-parameter family of generic WKB curves and by taking a representative curve from each family we obtain the triangulation shown in the right panel. In the right panel the black lines are the edges of the WKB triangulation and the wavy yellow lines show our convention for defining the branches of $\o$. Notice that this triangulation is topologically equivalent to the one shown in figure $\ref{ExampleTriangulation}$.  This means that we can borrow the results derived for that example.  In particular, the coordinates can be carried over from that example by making the proper identifications.  The cycles corresponding to each coordinate are represented by the gray curves -- we show only the portion of each cycle on the sheet of $\o$ where the edge $E_{34}$ has orientation \emph{towards} $P_4$ as indicated by the black arrow along edge $E_{34}$.}\label{WKBCells1}
\end{figure}

For the purpose of the following computation, a useful parameterization of the stress energy tensor is
\beq\label{4ptT}
T(w)=\frac{1}{(w-w_4)^2}\(c_{\infty}+\frac{c_0+c_1 w+c_2 w^2 + U w^3}{(1+w)^2(1-w)^2}\)
\eeq
Here we have fixed three of the insertion points at $w_1 =+1$, $w_2=\infty$, $w_3=-1$ using the world-sheet conformal symmetry.  The fourth insertion point is left at the position $w_4$ which should be fixed at the saddle point $w_4^*$ once the full action is assembled.  For the purpose of demonstration we will take $w_4$ to be between $w_3=-1$ and $w_1 = +1$.  When the dominant saddle point is located in one of the other intervals one can proceed by a similar procedure.  The constants $c_a = c_a\(w_4,\D\)$ are functions of $w_4$ and dimensions of the operators and are fixed by the condition $(\ref{T near P})$.  Their explicit expressions are given in appendix $\ref{detailed area}$. The parameter $U$ is unfixed by the condition  $(\ref{T near P})$ and implicitly parameterizes the cross ratio of the four operators (recall that they are inserted along a line in the boundary theory so that there is only one cross ratio).  The analytic structure of $T$, the resulting WKB-structure and the WKB triangulation are shown in figure $\ref{WKBCells1}$. \\
\subsubsection{$\chi$-system for the 4-point function}
From equation $(\ref{chi to A})$ and figure $\ref{WKBCells1}$ we have
\beqy
\chi_{24} \chi_{24}^{++} &=& \(\chi_{\hat{24}} \chi_{\hat{24}}^{++}\)^{-1}
=\frac{\(1+A_{23}\)\(1+A_{14}\)}{\(1+A_{34}\)\(1+A_{12}\)} \label{chi system 1}\\
\chi_{12} \chi_{12}^{++} &=& \(\chi_{14} \chi_{14}^{++}\)^{-1}=\chi_{34} \chi_{34}^{++} = \(\chi_{23} \chi_{23}^{++}\)^{-1} 
=\frac{\(1+A_{24}\)}{\(1+A_{\hat{24}}\)}\label{chi system 2}
\eeqy
To compute each $A_{PQ}$ we use formulas $(\ref{A to Sigma})$ and $(\ref{def Sigma})$ along with the rules given in figure $\ref{ComputingSigma}$.  In that way we find
\beqy
A_{24}&=& \frac{\c_{24}\(1+\c_{12}\(1+\c_{\hat{24}}\(1+\c_{23}\)\)\)\(1+\c_{43}\(1+\c_{\hat{42}}\(1+\c_{41}\)\)\)}{\(1-\mu_{2}^2\)\(1-\mu_4^2\)}\label{A24} \\
A_{23}&=& \frac{\c_{23}\(1+\c_{34}\)\(1+\c_{24}\(1+\c_{12}\(1+\c_{\hat{24}}\)\)\)}{\(1-\mu_{2}^2\)\(1-\mu_3^2\)}\label{A23}
\eeqy
with the rest of the $A_{PQ}$ being related by relabeling (see appendix $\ref{detailed area}$ for the explicit formulas).  These expressions and equations $(\ref{chi system 1})-(\ref{chi system 2})$ provide a closed system of functional equations for the 6 coordinates associated with the triangulation shown in figure $\ref{WKBCells1}$. \\
\indent These functional equations can be converted into integral equations of the form $(\ref{chi TBA})$ using the technique described in section $\ref{inversion}$.  To apply the procedure of section $\ref{inversion}$ one must find a $\phi$ such that the RHS of  (\ref{log chi system}) is decaying, and for this one should appeal to the WKB analysis.  The WKB cycles which determine the asymptotics of the coordinates are shown in figure $\ref{WKBCells1}$.  When $\D_1 \sim \D_3$ and $U\sim0, w_4\sim 0$ the cycles shown in figure $\ref{WKBCells1}$ all have $\mbox{Arg} (\oint_{\g_E}\o ) \sim \pi/2$.\footnote{Interestingly, when  $\D_1=\D_3$ and $U=w_4=0$ there is a symmetry which causes the RHS of the $\chi$-system to trivialize (i.e. to reduce to $1$ for all $\chi_E$) and the $\chi$-functions can be computed explicitly (they are just equal to their zero-mode part).  This is reminiscent of the case for the three-point function and, in fact, there is also a change of coordinates that maps the specific case $\D_1=\D_3$ and $U=w_4=0$ to two copies of a three-point function. \label{footsym}}   In this case $\phi=0$ is a suitable choice since then all $\chi_E^{-}$ will be growing and $(\ref{RHS of log chi system})$ will decay rapidly.\footnote{This will continue to be the case as long as the $\mbox{Arg} (\oint_{\g_E}\o )$ remain in the upper-half plane.  In other-words, the inversion procedure will be valid for all $U$ and $w_4$ such that the triangulation is unchanged since the triangulation will jump precisely when one of the $\oint_{\g_E}\o$ crosses the real-axis \cite{GMN}.}  In summary, the integral equations in the region of present interest are given by equations $(\ref{chi TBA})$ with $F_E$ given by $(\ref{chi system 1})-(\ref{A23})$.  These equations will remain valid for all values of the parameters $\D_a$, $U$, and $w_4$ such that the triangulation is unchanged.  If one deforms these parameters too much the triangulation will jump.  One can then easily write the $\chi$-system for the new triangulation and apply the same procedure to obtain the integral equations for that region of parameters.\footnote{Another (more elegant) approach would be to find a systematic way of analytically continuing the integral equations from one region of parameters to another as was done for the TBA equations of \cite{AMSV}. }  \\
\indent By numerically iterating these equations (using $\chi_E^{(0)}$ as the initial iterate for each $\chi_E$) we obtain the $\chi$-functions.  The $\eta$-cycles are then extracted from the $\chi$-functions using the procedure of section $(\ref{extract})$.  In the following section we will write the regularized $AdS$ action in terms of these $\eta$-cycles.   

\subsubsection{Finite part of $AdS$ action}\label{sec:finite}
Now that we are able to compute the $\eta$-cycles (see previous subsection) we can use the formula
\beq\label{area rbi}
 A_{fin}
=\int_{\Sigma} \sqrt{T \bar{T}}\(\cosh \g-1\) 
=\frac{\pi}{3}-\frac{i}{2} \(\oint_{\g_a} \!\! \o \) I^{-1}_{ab} \(\oint_{\g_b} \!\! \eta \)\,.
\eeq
(see section \ref{action formula} and equation ($\ref{action as cycles}$)) to compute the regularized part of the $AdS$ action.  To use $(\ref{area rbi})$ there are few steps.  These steps are simple but tedious and we will only list them here (see appendix $\ref{detailed area}$ for a detailed implementation).  As described in section $\ref{action formula}$ one should first modify $T$ by spreading the double poles slightly such that $\o = \sqrt{T}dw $ has an additional square-root cut at each of these points.   Then one should choose a complete basis of $a$- and $b$-cycles (five of each is needed for the 4-point function).  One can then apply formula $(\ref{action as cycles})$ and then take the limit in which the small cuts close to form simple poles in $\o$.  Once this is done the area will generically be expressed in terms of three different types of $\eta$-cycles:  cycles connecting two punctures, cycles connecting a puncture with a zero and cycles connecting two zeros.  The latter two can be expressed as linear combinations of the puncture-puncture cycles as described in appendix $\ref{detailed area}$.  Once this is done, the final result takes the elegant form
\beq
A_{fin}=\frac{\pi}{3}-i \sum_{E\in \mathcal{T}} \o_E \, \eta_{E} \label{action final}
\eeq
where the sum runs over the edges in the triangulation (see figure \ref{WKBCells1}), $\eta_{E_{ab}}$ is defined in \eqref{extracting eta 2} while $\o_{E_{ab}}$ is the $\o$-cycles that intersects edge $E_{ab}$ (i.e. the integral of $\o$ that is associated with the coordinate $\chi_{ab}$; these integrals are shown as the gray contours in figure $\ref{WKBCells1}$).\footnote{Note that in formula \eqref{action final} both integrals $\o_E$ and $\eta_E$ are the \emph{segment} integrals between the appropriate limits.  For example, the $\o_E= \frac{1}{2} \oint_{\g_E} \o$.  In this sense we are abusive with the term `cycle'.} \\
\indent Formula $(\ref{action final})$ and the procedure of section $\ref{Linear Problem}$ for computing the $\eta$-cycles solve the problem of computing the regularized $AdS$ contribution to the 4-point function.  In the next section we present some numerical tests of the procedure.  Let us note that the procedure of section $\ref{Linear Problem}$ is general and can be implemented for any number of punctures.  Further, while we have only proved equation ($\ref{action final}$) for the case of the 4-point function, given its simplicity one might suspect that the formula holds in general (with $\pi/3 \ra \pi/12 \times \(\# \mbox{number of zeros of T}\)$, of course).\footnote{It would be a simple matter to check this, but we have not pursued this issue.  We did check that the formula holds for the 3-point function (see appendix \ref{app 3 point}). }  Even if the general result does not take the simple form $(\ref{action final})$, for a given $T$ (i.e. for any number of punctures) the procedure described in section $\ref{action formula}$ is still valid and one can still write $A_{fin}$ in terms of the $\eta_{E}$ for the corresponding triangulation).  In principle this solves the problem of computing the regularized $AdS$ contribution to the $N$-point function.  We have performed numerical tests only for the case of the 4-point function.  We present these numerical results in the following section.
\subsubsection{Numerical tests}\label{num tests}
We now present numerical tests of the method described above. We solved numerically the modified sinh-Gordon equation \eqref{SGE} for the function $\gamma$ and then using this numerical solution to directly compute $A_{fin}$ via
\beq\label{Afin2}
\int_{\Sigma} \sqrt{T \bar{T}}\(\cosh \g-1\)
\eeq 
The general set-up of the numerical problem essentially follows that of \cite{JW2} with some modifications.  However, the numerical method that we use to solve the PDE \eqref{SGE} is quite different from that of \cite{JW2}.\footnote{We are very grateful to Romuald Janik for providing us with a copy of the code used in \cite{JW2} which was very useful in helping us to develop and test our own numerics.}  
\begin{table}[t!]
\begin{center}
  \begin{tabular}{| c | c | c | c | c | c | c |}
    \hline
     U             & $\Delta_3$ & $\Delta_4$ & $\Delta_1$ & $\Delta_2$ &  \text{Numerics} & \text{$\chi$-system} \\ \hline \hline
    $1/5$ & 1           & 2           & 1           & 2            & 0.84807  & 0.84812   \\  
    $1/2$ & 1           & 2           & 1           & 2            & 0.82421  &  0.82423 \\ \hline 
  \end{tabular}
 \caption{Comparison of the $A_{fin}$ obtained by numerically integrating \eqref{SGE} and the area computed from the $\chi$-system.  The results are for the spike configuration of figure \ref{SpikeConfigsMainText}B.}
\label{table1}
\end{center}
\end{table}
\begin{figure}[t!]
\begin{center}
\includegraphics[width=0.6\linewidth]{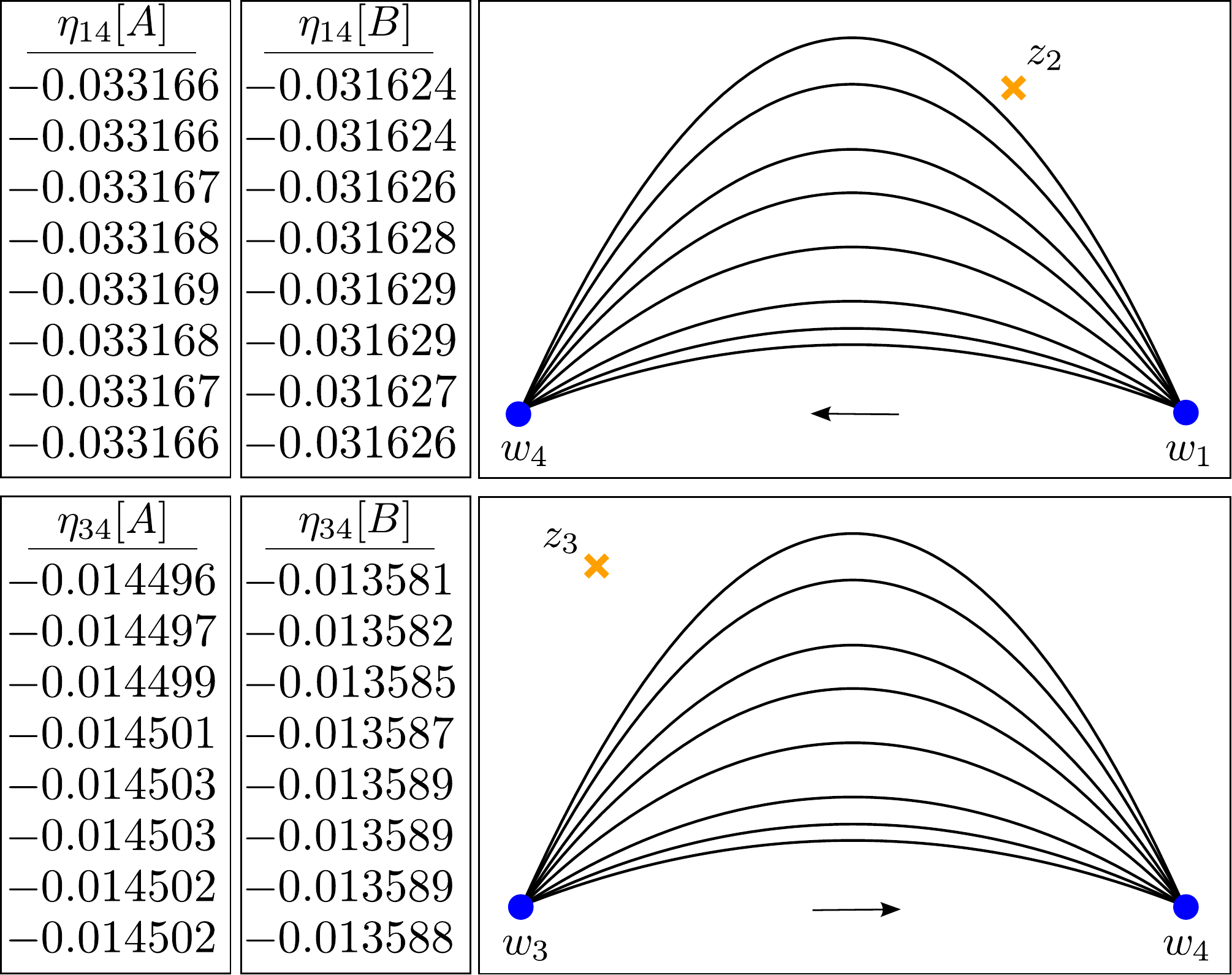}
\caption{Here we show the values of $\eta_{14}$ and $\eta_{34}$ evaluated along several different contours.  For example, the column labeled $\eta_{14}[A]([B])$ shows the values of $\eta_{14}$ for the spike configuration of figure \ref{SpikeConfigsMainText}$A(B)$ for each of the contours shown to the right of the column. We use the parameter values $\Delta_3=\Delta_1=1$, $\Delta_2=\Delta_4=2$ and $U=1/5$ for both spike configurations. There are five digits that we trust since they are unchanged for the different contours and they should be compared with our result from the functional equations that is $\eta^{\chi-\mbox{system}}_{14}[A] \approx -0.033169$, $\eta^{\chi-\mbox{system}}_{34}[A] \approx -0.014503$ and $\eta^{\chi-\mbox{system}}_{14}[B] \approx -0.031628$, $\eta^{\chi-\mbox{system}}_{34}[B] \approx -0.013588$. In the digits where the forms are closed there is perfect agreement with the analytic results.  }\label{holonum}
\end{center}
\end{figure}
We place the punctures at $w_3=-1$, $w_4=0$, $w_1=1$, and $w_2=\infty$. We then map the sphere to a square domain with the point at infinity mapping to the boundary of the square and the real axis mapping onto itself.  Since $\g$ must vanish at the punctures, we should impose $\g=0$ along the boundary of the square domain since $w_2$ maps to the boundary of the square in the new coordinates.   Further, since for either configuration of spikes (see section \ref{folds and spikes} and figure \ref{SpikeConfigsMainText}) there is a fold-line along the real axis, we know $\g\(x,0\)=0$ where we are using the coordinates $w=x+iy$ and writing $\g=\g(x,y)$.  Thus we can solve the problem in half of the square with the Dirichlet boundary conditions $\g=0$ on the boundaries.  Lastly, we must remove the logarithmic singularities \eqref{g at zeros} in order to have a nice smooth function to solve for.  A suitable function is
\beq\label{logreg}
2\g_{reg}=\g +\frac{1}{2} \sum_{a} \s_a  \log\[\frac{\(w-z_a\)\(\bar{w}-z_a\)}{\(1+w \bar{w}\)}\]
\eeq
where we $\s_a=\pm1$ is determined by $\g \sim -\s_a \frac{1}{2} \log T\bar{T}$ at $z_a$.  The numerator of \eqref{logreg} removes the log divergences \eqref{g at zeros} in $\g$ while the denominator is included to kill off these additional log terms at infinity.  In the numerical implementation we fix the spike configuration we want to describe by choosing the set of $\{\s_a\}$.  Finally, to numerically integrate the equation \eqref{SGE} (re-written in terms of $\g_{reg}$, of course) we use a standard relaxation method with an uniform grid. \\
\indent In table \ref{table1} we compare the numerical results with the analytic results.  The numerical results are obtained by the area computed using \eqref{Afin2} with the numerical solution for $\g$.  The analytic result is obtained from \eqref{action final} with the $\eta$-cycles computed using the $\chi$-system procedure.  These results show a good agreement of our formula with the numerics.  \\
\indent A sharper measure of the agreement between the analytics and numerics is to compare directly the $\eta$-cycles.  In figure \ref{holonum} we show the numerical results for $\eta_{14}$ and $\eta_{34}$ computed along several different contours.  This allows us to test the closure of the numerical $\eta$ which we obtain from the numerical $\g$ via \eqref{eta}.  Note that closure of $\eta$ implies that $\g$ must obey \eqref{SGE} and thus this is a good measure of the numerical error.  Indeed, one can see in figure \ref{holonum} that the numerical cycles agree with the analytical predictions in all digits in which they are closed.  That is, the numerics is in agreement with the analytics in all of the digits for which the numerics can be trusted.  \\
\indent Finally, it would be interesting to perform numerical tests for a larger portion of the parameter space (i.e. more values of the $\D_a$, $U$ and $w_4$).  To perform a systematic study will probably require an improvement of our numerical method as our current method, while extremely simple, has very slow convergence.  
\subsection{Divergent part}
In section $\ref{sec:Afin}$ we completed the task of computing the first term in formula (\ref{action}). In this section we will discuss the second term 
\beq
- \int_{\Sigma \backslash \{\e_a\}} d^2 w \sqrt{T \bar{T}} 
= -\frac{\pi}{2}\sum_{a}\Delta^{2}_{a} \log \epsilon_a - A_{reg}\label{Adiv}
\eeq
where $A_{reg}$ is finite at $\e_a \ra 0$.  The contribution $A_{reg}$ can be computed by simple but tedious application of the Riemann bilinear identity and there are many ways to write the result.  For example
\beq\label{Areg}
A_{reg} = i\sum_{E \in \mathcal{T}} \varpi_{E}\o_E - i\frac{1}{2} (\varpi_{24}-\varpi_{\hat{24}})(\o_{24}-\o_{\hat{24}})
\eeq \\
where the sum is over the triangulation shown in figure \ref{WKBCells1} and $\varpi_{E_{ab}}\equiv \varpi_{ab}$ is defined in \eqref{omega reg 2}. The $\o_E$ are defined in the same way as in \eqref{action final}.  One can check this formula by comparing with the direct 2D numerical integration of $\sqrt{T \bar{T}}$ with small circular disks cut out around the puncture (in Mathematica one can use NIntegrate along with the Boole command, for example). \\
\indent We recall that $(\ref{Adiv})$ came from the regularization of the string action where we have added and subtracted $\sqrt{T\bar{T}}$ from the integrand of the $AdS$ action.  This integral depends explicitly on the cut-off $\epsilon_a$ around the punctures. It will be important to understand the connection with the physical cut-off $\mathcal{E}$ at the boundary of $AdS$.  Fortunately we can extract the needed information from the linear problem since we have good analytic control over the solutions near the insertion points.  To proceed by this route (which parallels the discussion of \cite{JW2} for the 3-point function) we must first describe how the string embedding coordinates are recovered from the linear problem formalism, which is via the aptly-named \emph{reconstruction formulas}.  We will discuss this in the next subsection, \ref{recform}. After that, in section \ref{ST dep} we will use the reconstruction formulas to eliminate the $\e_a$ in favor of $\mathcal{E}$.  From this procedure we will recover the standard spacetime dependence in \eqref{can4pt} along with a contribution to the function $f(u,v)$.  This will complete the computation of the semiclassical $AdS$ contribution to \eqref{can4pt}.
\subsubsection{Reconstruction formulas}\label{recform}
The reconstruction formulas allow us to express the string embedding coordinates in terms of solutions of the linear problem. This point is crucial in our construction for the following reasons. First, we have introduced some regulators in the world-sheet, $\epsilon_a$, that must be related to the physical cut-off in the boundary of $AdS$, $z=\mathcal{E}$. Second, by using them we will be able make the spatial dependence explicit in the final result, namely the insertion points $x_a$ of the operators in the gauge theory. \\
\indent Consider two solutions of the linear problem, $\psi_A$ and $\psi_B$ normalized as $ \(\psi_A \wedge \psi_B\) =1$, and construct a matrix $\Psi$ as
\beq\label{matrix notation}
\Psi=(\psi_A\,\, \psi_B)\, .
\eeq
The matrix $\Psi$ obeys the same equations of motion as $\psi_{A,B}$ (\ref{LP}), namely 
\beq
(\partial + J_w)\Psi=0,\,\,\,\,\,(\bar{\partial} + J_{\bar{w}}) \Psi=0\, .
\eeq
where $J_w$ and $J_{\bar{w}}$ are defined in \eqref{LP}-\eqref{connection1}. One can verify using (\ref{L}) that the quantity
\beq\label{yI}
y^{I}\equiv-\frac{1}{2}\,\tr \left(\tilde{\sigma}^I \sigma^2 \Psi^{T} \sigma^1 \Psi \right)\!\big|_{\th=0}\,
\eeq
with $\tilde{\sigma}^1=\sigma^1,\, \tilde{\sigma}^2=-i \sigma^2,\, \tilde{\sigma}^3= \sigma^3 $, satisifes the same equations of motion as $Y^{I}$ and also the constraint $y \cdot y=-1$ (with the $AdS$ metric). In this way we establish a correspondence between target space coordinates and solutions of the linear problem,
\beq\label{rec}
\frac{1}{z}=Y^2-Y^1=2i \, \Psi_{11}\Psi_{21},\,\,\,\,\,\,\,\, \frac{x}{z}=Y^3=i\, (\Psi_{11}\Psi_{22}+\Psi_{12}\Psi_{21})
\eeq
In order to relate the operator insertion points $x_a$ and physical cut-off $\mathcal{E}$ with the linear problem data, it is convenient to express $\psi_A$ and $\psi_B$ in terms of the elementary solutions $s_a$ and $\tilde{s}_{a}$ whose behavior close to the punctures is given by (\ref{psi near p}),
\beq
\psi_A = \(\psi_A \wedge \tilde{s}_a \) s_a + \( s_a \wedge \psi_A \) \tilde{s}_{a}, \,\,\,\,\,\,\,\,  
\psi_B = \(\psi_B \wedge \tilde{s}_a \) s_a + \( s_a \wedge \psi_B \) \tilde{s}_{a}
\eeq
Close to the punctures the solution $\tilde{s}_{a}$ becomes dominant. Then, using (\ref{rec}) and the explicit form of $\tilde{s}_{a}$ close to the puncture $P_a$ we get that
\beq\label{cuttoff}
z=\frac{1}{i \( s_a \wedge \psi_A \)_0^2}|w-w_a|^{\Delta_a}
\eeq
where the subscript $0$ indicates that the solutions are evaluated at $\th=0$ (recall that this is the value where the physical problem is recovered -- see equation \eqref{yI}). Equation \eqref{cuttoff} is the relation needed to make the connection between the world-sheet and physical cut-off's
\beq\label{reccut}
\D_a \log \epsilon_a = \log \mathcal{E} +\log |\( s_a \wedge \psi_A \)|_0^2
\eeq
Finally, using once again (\ref{rec}) we express the insertion points $x_a$ of the operators in the gauge theory as
\beq\label{recx}
x_{a}= \frac{\( s_a \wedge \psi_B \)_0 }{\( s_a \wedge \psi_A \)_0}
\eeq


\subsubsection{Physical regulator and spacetime dependence}\label{ST dep}
We can now use $(\ref{reccut})$ to eliminate the $\e_a$ in $(\ref{Adiv})$ in favor of the physical cut-off at the boundary of AdS
$z= \mathcal{E}$. We have 
\beq\label{eps terms}
\sum_{a}\Delta^{2}_{a} \log \epsilon_a = \left(\sum_{a}\Delta_{a} \log \mathcal{E} + \sum_{a} \Delta_{a} \log|\( s_a \wedge \psi_A \) |_0^{2} \right)
\eeq
where $a$ and $A$ refer respectively to the small solution $s_{a}$ and one generic solution $\psi_{A}$ appearing in the reconstruction formulas. Now we will eliminate the factors $|\( s_a \wedge \psi_A \) |_0$ in terms of objects that we can compute.  \\
\indent The terms $|\( s_a \wedge \psi_A \) |_0^{2}$  can be related to the insertion points $x_{a}$ in target space and overlaps of the elementary solutions evaluated at $\theta = 0$  through expression (\ref{recx}). Using Schouten's identity one can verify that
\beq \label{sol}
|\( s_a \wedge \psi_A \)|_0^2 = \frac{x_{b c}}{x_{b a} x_{c a}} \frac{|\( s_b \wedge s_a \)|_{0} |\( s_c \wedge s_a \)|_{0}}{|\( s_c \wedge s_b \)|_{0}} 
\eeq
for $a,b,c$ distinct. This solution is unique up to different ways of rewriting the spatial dependence using the cross-ratio
\beq \label{cr}
u
=\frac{x_{14} x_{23}}{x_{12} x_{34}}=\frac{\( s_1 \wedge s_4 \)_{0}\, \( s_2 \wedge s_3 \)_{0}}{\( s_1 \wedge s_2 \)_{0}\, \( s_3 \wedge s_4 \)_{0}}
\eeq
where we have used \eqref{recx}.  Note that we can compute the brackets appearing in \eqref{reccut}-\eqref{recx} using \eqref{extracting eta 1}.  In particular we have
\beq\label{ab_0}
\log\(s_a\wedge s_b\)_0 = \(\frac{1}{2} \varpi_{ab}+\frac{1}{2} \bar{\varpi}_{ab}\) + \int_{\mathbb{R}}\frac{d\th}{2\pi}\frac{\log\(1+A^{-}_{ab}\)}{\cosh\th} 
\eeq 
This formula is valid when there is a WKB line connecting $P_a$ and $P_b$. If a bracket appears for which we do not have a WKB line, we can simply use the cross ratio \eqref{cr} to eliminate it in terms of brackets that can be computed using \eqref{ab_0}. \\
\indent Finally, using \eqref{sol} in \eqref{eps terms} and massaging the resulting spacetime dependence by extracting multiples of $u$ and $\(1+u\)$ we find 
\beq
e^{2\times\frac{\sqrt{\l}}{2} \D_a^2 \log \e_a} =
\prod^4_{a>b} \(|s_a \wedge s_b|_0\)^{-\sqrt{\l}\D_{ab}}\(\frac{x_{ab}}{\mathcal{E}}\)^{\sqrt{\l}\D_{ab}}\label{ST final}
\eeq
where $\D_{ab}=\(\sum_c \D_c\)/3-\D_a-\D_b$.  The extra factor of $2$ in the exponent on the left hand side of \eqref{ST final} anticipates the sphere regularization which turns out to be similar to the $AdS$ part and will be treated in section \ref{sphere}. \\
\indent We recognize in \eqref{ST final} the canonical spacetime dependence in the 4-point function of a conformal field theory (compare with equation \eqref{can4pt}). The appearance of the cut-off in \eqref{ST final} is related to the renormalization of the operators. In fact, if we define $\tilde {\mathcal{O}}_{\Delta_a}\equiv \mathcal{E}^{\Delta_a} \mathcal{O}_{\Delta_a}$ this will cancel the $\mathcal{E}$ factors in \eqref{ST final}.   To be more precise, we should define a 4-point function that is independent of the operator renormalization.  The standard procedure is to divide by the appropriate product of 2-point functions such that normalization factors cancel.  The same factors of $\mathcal{E}$ will appear in these 2-point functions and will cancel with those in \eqref{ST final}.  We will thus drop the factors of $\mathcal{E}$ in the formulas below. \\


\subsection{Summary of the $AdS$ and divergent contributions}
We have now computed all the parts of $\eqref{action}$.  In this section we summarize the full result.  The semiclassical limit of the 4-point function \eqref{can4pt} is given by
\beq\label{4ptsum}
\left(f^{AdS}_{fin}f^{AdS\times S}_{div}f^{S}_{fin}\right)^* \prod_{a<b}^4 (x_{ab})^{\D_{ab}}\,,
\eeq
where the $*$ denotes evaluation at $w_4=w^{*}_4$ and we define
\beqy
f^{AdS}_{fin}\(w_4\) &=& e^{-\frac{\sqrt{\l}}{\pi}A_{fin}}\label{Aregs1} \\
f^{AdS\times S}_{div}\(w_4\)&=& e^{-2\frac{\sqrt{\l}}{\pi}A_{reg}}\prod^4_{a>b} \(|s_a \wedge s_b|_0\)^{-\sqrt{\l}\D_{ab}}\label{Aregs2}
\eeqy
and $f^S_{fin}$ will be defined momentarily.   The contribution $A_{fin}$ is given by \eqref{action final}, $A_{reg}$ is given in \eqref{Areg}, the brackets in $f^{AdS\times S}_{div}$ are given by \eqref{ab_0}.   \\
\indent The sphere part of the correlation function contains divergences of the same type as $AdS$. We therefore regularize it also by subtracting $\sqrt{T \bar{T}}$. Such finite contribution is what we denote by $f^{S}_{fin}$
\beq\label{sfin}
f^{S}_{fin}\equiv e^{-\frac{\sqrt{\lambda}}{\pi}\int_{\Sigma}\left(S^{5}\,\text{contribution}-\sqrt{T\bar{T}}\right)}\,.
\eeq
where $S^{5}\,\text{contribution}$ stands for the $S^5$ Lagrangian and wavefunctions \cite{JW2}. To compensate this subtraction, we include the factor of 2 in front of $A_{reg}$ in expression (\ref{Aregs2}). In general we cannot complete the construction of the 4-point function because we are unable to compute the contribution $f^{S}_{fin}$.  Fortunately, for correlators involving only BPS operators of the same type (e.g. only $Z$ and $\bar{Z}$) the sphere part is known and we can assemble the full result.  This is the subject of the next section.
\section{Full correlation function for BMN operators}
\indent In this section we compute the full correlation function for operators of the type $\tr Z^{\Delta}$ when $\Delta$ scales as $\sqrt{\lambda}$. For these type of operators, the sphere part $f^{S}_{fin}$ was already known \cite{TSEYTLINSPHERE} and therefore we can complete our computation. We stress that, unlike the three point function, this four point correlator is not protected. In section \ref{saddle point}, we fix the location of the puncture $w_4$ by the saddle point method and discuss some issues on the multiple string embedding configurations. In section \ref{extremal} we perform an analytical check of our procedure by studying the extremal limit where $\Delta_2=\Delta_1+\Delta_3+\Delta_4$, which is known to be protected from quantum corrections.
\subsection{Sphere part}\label{sphere}
The sphere part of the correlation function involves the classical wave-functions associated to the external states. We consider specifically the correlation function of four BMN operators\footnote{We are using the following notation for the dimensions of the operators $\hat\Delta_{a}=\sqrt{\lambda}\Delta_{a}$.}
\beq\label{scorr}
\langle  \tr Z^{\hat\Delta_1}(x_1)\; \tr Z^{\hat\Delta_{2}} (x_2)\; \tr  \bar{Z}^{\hat\Delta_{3}}(x_3)\; \tr \bar{Z}^{\hat\Delta_4} (x_4)\; \rangle\, ,
\eeq
for which the wave-functions are known \cite{POLYAKOV,TSEYTLINVERTEX}. The string dual of these operators corresponds geometrically to a string that is point-like in the sphere and rotates around an equator\,\cite{GUBSER2002}. The surface developed by the worldsheet is not extended in the sphere. \\ 
\indent Let $X_i$ ($i=1,..., 6$) be the coordinates in $S^5$. This particular string state can be expressed as
\beqa
X_1 + i X_2 = e^{i \varphi} \;\;\;\;\;\;\;\;\;\;\;\;\;\;\;\;\;\;\;\; X_i=0,\;\;\;\  i=3,\dots, 6
\eeqa
where $\varphi$ is an azimuthal angle of the sphere. The wave-functions for $\tr Z^{\hat \Delta_a}$ and $\tr \bar{Z}^{\hat \Delta_a}$ are given respectively by
\beq
\Psi_{\hat\Delta_a}=e^{i \hat\Delta_a \varphi(w_a,\bar{w}_a)},\,\,\,\,\,\,\,\,\,\,\, \bar{\Psi}_{\hat\Delta_a}=e^{-i \hat\Delta_a \varphi(w_a,\bar{w}_a)}
\eeq
where the field $\varphi$ is evaluated at the puncture corresponding to the respective operator insertion. \\
\indent As the wave-functions scale exponentially with $\sqrt{\lambda}$, they will act as sources for the equations of motion for $\varphi$. The total sphere contribution is then given by
\beq \label{saction}
\exp \left[-\frac{\sqrt{\lambda}}{\pi} \left(\int d^2 w\,\partial \varphi \bar{\partial} \varphi +i \pi \left( \Delta_{3} \varphi_{w=-1} +  \Delta_{4} \varphi_{w=w_4}- \Delta_{1} \varphi_{w=1}-  \Delta_{2} \varphi_{w=\infty}\right)\right)\right].
\eeq
Considering both the contributions from the $S^5$ action and wave-functions as an effective action, we obtain the equations of motion for $\varphi$ which are solved by
\beq \label{sollaplace}
 \varphi(w,\bar{w})= i\left( \Delta_{3} \log |w+1| + \Delta_4 \log |w-w_4| - \Delta_1 \log |w-1| \right)\, .
\eeq
This solution has an additional singularity at infinity with charge $-\Delta_{3} - \Delta_4+ \Delta_1  (\equiv-\Delta_{2})$, corresponding to the wave-function inserted at infinity. This is consistent with $R$-charge conservation.
We may now plug (\ref{sollaplace}) into (\ref{saction}), introducing cut-off's around the punctures to regulate this contribution. This amounts to evaluate the solution at a distance $\epsilon$ away from the punctures. As in the case of the $AdS$ action, the logarithmic divergences 
\beq
\exp \left[ \frac{\sqrt{\lambda}}{2}\sum_{i}\Delta^{2}_{i} \log \epsilon_i \right]
\eeq
need to be regularized. We do this by subtracting $\sqrt{T\bar{T}}$ from the integrand. To compensate, we add a similar contribution to the divergent part, that was already treated in the previous section (indeed, this regularization procedure is responsible for the factor of $2$ appearing in front of $A_{reg}$ in expression for $f^{AdS\times S}_{div}$, see (\ref{Aregs2})). The dependence on the cut-off's then disappears yielding the following expression for the regularized sphere action and wave-functions
\beq
f^{S}_{fin} =\exp\Bigl[\sqrt{\lambda }\,\Bigl(A_{reg}-  \log 2^{\Delta _{3} \Delta _{1}}-\log\frac{|w_4-1|^{\Delta _{1}\Delta _{4} }}{|w_4+1|^{\Delta _{3}\Delta _{4} }}\Bigr) \Bigr]\,,
\eeq
where $f^{S}_{fin}$ was defined in (\ref{sfin})


\subsection{Saddle point determination}\label{saddle point}
We have shown how to compute the quantities \eqref{Aregs1}-\eqref{Aregs2} as a general function of $w_4$.  However, to compute \eqref{4ptsum} we must evaluate at the saddle point $w_4=w_4^*$.  Before discussing how to locate the saddle point for generic values of the parameters,  let us discuss a very symmetrical situation in which we can guess its position.  \\
\begin{figure}[ht!]
\begin{center}
\label{saddle}
\includegraphics[width=1\linewidth]{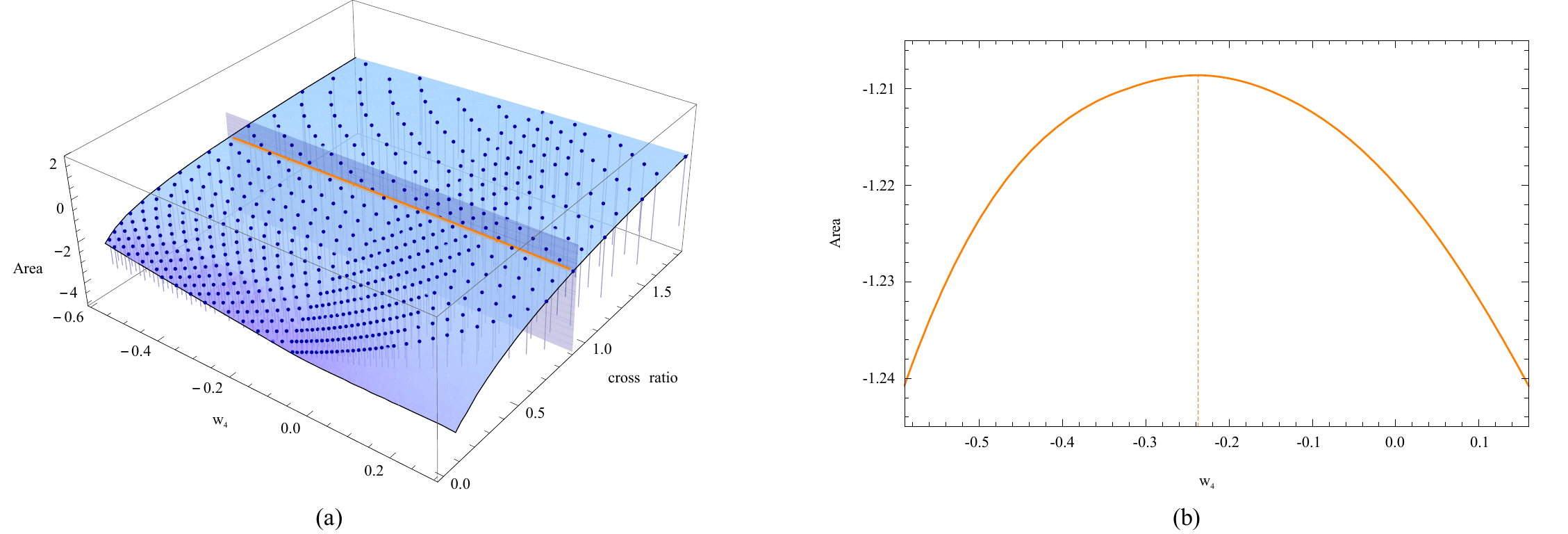}
\caption{In (a) we computed several points for a region around $u \sim 0.9$ and interpolated the surface in blue. We then took a constant cross ratio slice. The intersection is given by the orange line. In (b) we have plotted the area as function of $w_4$ for cross ratio $u=0.9$, where we have a saddle point at $w^*_4 \approx-0.24$.}
\end{center}
\end{figure}\indent Using conformal symmetry, we can fix three of the points in the target space at $x_3=-1,x_1=1$ and $x_2=\infty$ and also in the world-sheet at $w_3=-1$, $w_1=1$ and $w_2=\infty$ . The fourth point $x_4$ will then be related to the cross ratio. For the particular choice of the cross ratio (\ref{cr}) equal to $1$, the fourth point will be located at zero. Moreover if we choose the points at $x_3=-1$ and $x_1=+1$ to have the same conformal dimension and the same type of fields (say $Z$'s) then this is a very symmetrical configuration. Going back to the world-sheet coordinates, by symmetry we expect to find the saddle point at $w^*_4=0$.\\
\indent Let us now describe how we determine the saddle point in practice. All the physical information that we input is contained on the stress-energy tensor. Besides the conformal dimensions of the operators and the position of the punctures, there is the extra parameter $U$ that translates the additional degree of freedom of the cross ratio. The two are implicitly connected by the expression
\beq
\chi_{24}(\theta=0;U)=\frac{x_{14}\, x_{23}}{x_{12}\, x_{34}} \equiv u
\eeq 
by formulas (\ref{cr}) and (\ref{chis1}). Therefore, our strategy will be to compute the area and cross ratio for many points in a region in the $(w_4,U)$ plane and then translate that information to the $(w_4,u)$ plane. Finally, we take slices of constant cross ratio, and determine for which $w_4$ the area is stationary.\footnote{In this way one can confirm that indeed 
$w^{*}_4=0$ for the very symmetrical configuration described above.}  We can very easily perturb away from this very symmetrical configuration and track the location of the saddle point. For example, in figure \ref{saddle}, we have found a saddle point for $u=0.9$ at $w^*_4 \approx-0.24$.  One can compute the location of $w^*_4$ to arbitrary numerical accuracy by iterating the $\chi$-system; here we present only $2$ digits since we just want to demonstrate the procedure for locating the saddle point.  In the following section we will discuss the issue of the multiple saddle points and connect with the discussion of section \ref{folds and spikes} about the different configurations for the string embeddings in $AdS_2$.
\subsection{Saddle points and multiple string configurations}\label{saddle and configs}
In section \ref{4ptgen} and appendix \ref{gamma props}, we discussed the different $AdS_2$ string embedding geometries and its connection with the different boundary conditions \eqref{g at zeros} that one can impose on $\g$. At the level of the functional equations, we have seen that the different boundary conditions manifest in the different $\xi\rightarrow0,\infty$ asymptotics of the coordinates. More precisely they will affect the constant $C_E^{(0)}$ in the expression (\ref{WKB asymp}). One may ask which of the configurations in figure \ref{stringconfigs2} we should find given a cross ratio and a set of conformal dimensions. We have already introduced this question but let us make it more precise now that we have all the tools in hand. 
\\
\indent Consider the example of the WKB triangulation we have been studying. There exist \emph{a priori} two choices for the orientations of the spikes, as discussed in section \ref{folds and spikes} and appendix \ref{gamma props}.  Consider the spike configuration of figure \ref{SpikeConfigsMainText}A, which is the one used for the saddle-point analysis of section \ref{saddle}.  Recall that for this case the cross ratio $u=\chi_{24}(\theta=0)\approx 0.9$  is \emph{positive}.  This means that the point $x_4$ is located between $x_3$ and $x_1$.   Furthermore, we have found a saddle point located in $-1<w^*_4<+1$.  Looking at figure \ref{insertions}  we see that this situation corresponds to a case in which the insertions do not cross since $x_3<x_4<x_2$ and $w_3<w^*_4<w_1$.\footnote{Be aware that the ordering of the $x_a$ in this discussion is different from that used in figure \ref{insertions}.}  Since the insertions do not cross, we expect the string embedding to be that of figure \ref{stringconfigs2}A.  
\\
\indent Now, one might expect that the configuration in figure \ref{stringconfigs2}B can be described simply by considering the spike configuration of figure \ref{SpikeConfigsMainText}B \emph{but keeping the same saddle point} $w_3<w^*_4<w_1$ and the same value of the cross ratio as in the previous spike configuration.  However, this is not possible since in this new spike configuration the cycle for $\chi_{24}$ connects two spikes of the same type and thus from \eqref{WKB asymp} we see that it will acquire an overall factor of $(-1)$ so that $u<0$.\footnote{Of course the corrections to $\chi_{24}$ from iterating the integral equations will differ for the two different spike configurations but they should not change the overall sign of $\chi^{\(0\)}_{24}$.  We are taking this as a physically motivated assumption in this discussion.  We have checked this assumption in a few examples and found that it holds. }  
\\
\indent The point is that if we \emph{fix} a cross ratio $u$ and then consider a specific saddle-point $w_4^*$ then the orientation of the spikes \emph{is fixed} and thus the configuration of the string embedding is also fixed. This is in perfect agreement with the mapping between figure \ref{insertions} and \ref{stringconfigs2} and it is non-trivial that the integral equations encode this  mapping.\\
\indent Given the above discussion, we are confronted with a very interesting possibility.  Generically we do not expect the saddle point $w_4^*$ to be unique and it's likely that there are actually several saddle points $w^{\(i\)}_{4^*}$ on the 4-punctured sphere.  As per the above discussion, for fixed $u$, $\D_a$ and a given $w^{\(i\)}_{4^*}$ the corresponding string embedding is fixed.\footnote{Note that for fixed $u$, $\D_a$ the triangulation will depend on which $w^{\(i\)}_{4^*}$ one is considering.  This is not a problem as one can apply the method of section \ref{Linear Problem} to each of these triangulations individually.}  In particular, if there is a saddle point in each of the three intervals of the real axis we should examine each of these in turn.\footnote{Here we are not considering the possibility of complex-conjugate pairs of saddle points located off of the real axis (see footnote \ref{foot: saddle caveat}).}  For a fixed $u$, two of these will be double-folded and one will be single-folded.  One should find all of these saddle points and determine which is the dominant one,  which is equivalent to ask which string embedding configuration is dominant. 
One may even find a dependence of the dominant string configuration on the dimensions of the operators thus giving rise to phase transitions between configurations. The issue of finding the different saddle points and their dependence on the parameters of the theory certainly deserves a deeper study.  
\\ 

\subsection{Extremal Limit}\label{extremal}
In this section, we study the correlation function
\beq\langle  \tr \bar{Z}^{\hat\Delta}(x_1)\; \tr Z^{\hat\Delta_{2}} (x_2)\;\tr \bar{Z}^{\hat\Delta}(x_3)\; \tr \bar{Z}^{\hat\Delta_4} (x_4) \rangle \label{extrz}
\eeq
in the extremal limit when
\beq\label{extcond}
 \Delta_{2}=2\Delta+\Delta_{4}.
\eeq
Such correlator is protected from quantum corrections as conjectured in \cite{RASTELLI} and later proved in \cite{SOKATCHEVPROOF}. Thus, we expect to obtain the tree level gauge theory result which in the planar limit is simply given by Wick contractions
\beq\label{extr}
\frac{1}{x_{12}^{2\hat\Delta}\,x_{23}^{2\hat\Delta}\,x_{24}^{2\hat\Delta_{4}}}.
\eeq
The $AdS$ part of our formula is universal in the sense that it only depends on the dimensions of the operators. On the other hand, the sphere part of the correlation function involves the precise details of the operators inserted. Compared to the previous sphere calculation (\ref{scorr}), computing (\ref{extrz}) just amounts to take the complex conjugate of the wave function located at $x_1$, due to the replacement of $Z \rightarrow \bar{Z}$. 
\begin{figure}
\centering
\def\svgwidth{12cm}
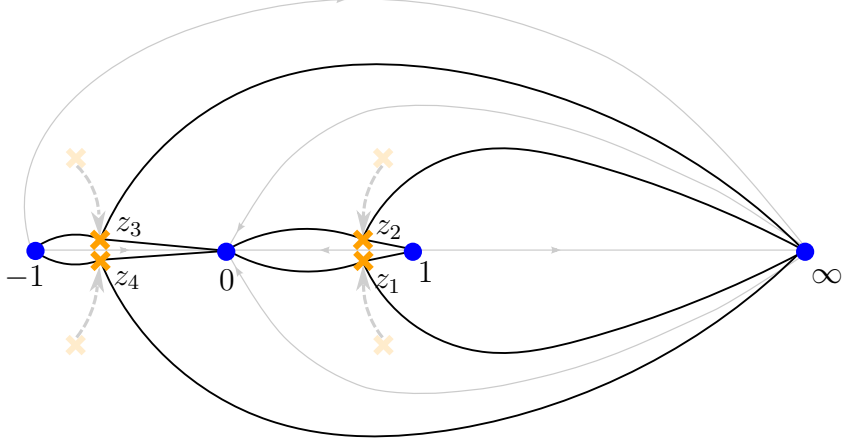
\caption{In the extremal limit, the main feature is that the zeros collide on the real axis. The black lines represent the WKB cells whereas the gray lines represent the WKB triangulation. At the exact extremal configuration, there are no WKB lines connecting $1$ to $0$ nor $-1$ to $0$. We interpret this as a manifestation of the field theory fact that at tree level all operators are Wick contracted only to the fourth operator.}
\label{fig:extremal}
\end{figure}\\
\indent Let us start by studying the case when the cross ratio is $u=1$, where we know the saddle point is $w^*_4=0$. From this we will be able to see the general mechanism that gives the expected simplification of our result. 
The first important observation is that in this limit the zeros of $T(w)$ collide on the real axis as depicted in figure \ref{fig:extremal}.  Let us start by analyzing what this implies at the level of the $\chi$-system. As the integrals $\omega_{14}$ and $\omega_{34}$ vanish, the $\chi$'s associated to these cycles, namely $\chi_{34}$ and $\chi_{14}$, tend to $-1$. 
This observation has the remarkable consequence that the right hand side of all equations in the $\chi$-system becomes trivially equal to 1 as one can easily verify \footnote{This trivialization of the $\chi$-system is general and follows \emph{just} from the fact that the two cycles $\omega_{14}$ and $\omega_{34}$ vanish which implies that the $\chi$-functions $\chi_{34}$ and $\chi_{14}$ become -1. In the specific case of $U=w_4=0$ and $\Delta_1=\Delta_3$, which turns out to correspond to cross ratio 1, the $\chi$-system is already trivial because of the symmetry of the stress energy tensor in this particular point of the parameter space, see footnote \ref{footsym}). Nevertheless, we emphasize that the trivialization of the $\chi$-system in general does not rely on this specific symmetry of the stress energy tensor.}. As a result, all $\chi$-functions are $\it{exactly}$ given by leading term of the WKB expansion (\ref{WKB asymp})\footnote{Indeed, when the right hand side of the $\chi$-system is $1$, the kernel term in equation (\ref{chi TBA}) vanishes and we are left with leading WKB contribution.}. 
For convenience, let us introduce an infinitesimal $\delta$ defined by the condition $\delta=2\Delta+\Delta_{4}-\Delta_{2}$. At the end of the day, we will take $\delta \rightarrow 0$.
In this limit, the solutions of the $\chi$-system are then given by
\beq
\chi_{23}=\chi_{12}=-e^{-\frac{ \pi (4\Delta-\delta)}{2}\cosh\theta}, \,\,\,\,\,\, \chi_{34}=\chi_{14}=-e^{-\frac{ \pi \delta}{2}\cosh\theta},\,\,\,\,\,\, \chi_{24}=\chi_{\hat{24}}= e^{-\frac{ \pi (2\Delta_{z}-\delta)}{2}\cosh\theta}
\eeq
We may now plug this solution in the expression (\ref{A to Sigma}) and extract the cycles using as described in section \ref{extract}.  We find that all $A$'s vanish in the limit $\delta\rightarrow 0$ except for $A_{14}$ and $A_{34}$, which tend to $-1$ as $\delta$ goes to zero. This implies that all $\eta_{E_{ab}}$ vanish except for $\eta_{14}$ and $\eta_{34}$, which diverge since the integrand of these cycles becomes singular in this limit. However, one must go back to the area formula (\ref{action final}) and realize that such cycles are multiplied by a vanishing quantity. Indeed, (\ref{action final}) simplifies to
\beqa\label{simplearea}
\frac{1}{4} \pi  \delta  \eta _{14}+\frac{\pi}{3}\, .
\eeqa
In the limit $\delta\rightarrow 0$, the first term of this expression is explicitly given by
\beq
\delta\,\int_{0}^{\infty}d\theta\,  \cosh \theta  \log\left(1-e^{-\frac{1}{2} \pi  \delta  \cosh\theta}\right) +\mathcal{O}(\delta)=-\frac{\pi}{3}+\mathcal{O}(\delta).
\eeq
Hence, it turns out that the finite $AdS$ contribution vanishes in the extremal limit. We believe this is the general mechanism for any value of the cross ratio. \\
\indent The computation of the sphere contribution follows the same steps as before, with a slight change on one vertex operator (recall that to get the extremal case, we replaced the operator located at $x_1$ in (\ref{scorr}) by $\tr \bar{Z}^{\hat\Delta}$). The new solution for the equations of motion is
\beq
 \varphi(w,\bar{w})= i\left( \Delta \log |w+1| + \Delta_4 \log |w| + \Delta \log |w-1| \right)\, .
\eeq
Now when we compute the contribution of the sphere action and wavefunctions on this solution, we find that it $\it{exactly}$ cancels the term $\sqrt{T\bar{T}}$ for $\Delta$'s satisfying  (\ref{extcond}).  Consequently, the sphere part of the correlation function also vanishes in the extremal limit.

The divergent piece in the extremal becomes simply
\beq
e^{-\frac{\sqrt{\l}}{\pi}A_{reg}}
\prod^4_{a>b} \(|s_a \wedge s_b|_0\)^{-\sqrt{\l}\D_{ab}}\rightarrow\delta\, \int_{-\infty}^{\infty}\frac{d\theta}{2 \pi \cosh\theta}\log (1-e^{-\frac{1}{2} \pi  \delta  \cosh\theta }) +\mathcal{O}(\delta)= \frac{\delta}{\pi}\log \frac{\pi \delta}{2}+\mathcal{O}(\delta)
\eeq
which goes to zero as $\delta\rightarrow 0$. We are left with the spatial dependent part which, using that the cross-ratio is 1, can be written as
\beq
\frac{1}{\left(\frac{x_{12}}{\mathcal{E} }\right)^{2\hat\Delta} \left(\frac{x_{23}}{\mathcal{E}}\right)^{2\hat\Delta} \left(\frac{x_{24}}{\mathcal{E} }\right)^{2\hat\Delta_4  }}.
\eeq
This is nothing but the tree level result (\ref{extr}) of the gauge theory.


\section{Discussion and future directions}\label{conclusion}
In this paper, we have computed the $AdS$ part of the four point function for heavy scalar operators in $\mathcal{N}=4$ SYM in the classical limit. For the particular case of BPS operators on a line with a single scalar field, the sphere part is known and thus we can construct the full strong coupling four point function. \\
\indent The main ingredient of our method is the integrability of the string equations of motion in $AdS_2$. Specifically,  we use the method of Pohlmeyer reduction to map the problem to that of solving a certain modified Sinh-Gordon equation which is known to be integrable.  We construct the linear problem associated with this equation, which has the form of an $SU(2)$ Hitchin system. This approach was used in the solution of the Null Polygonal Wilson-Loop problem at strong coupling \cite{AMSV} as well as in the study of three-point functions of heavy operators at strong coupling \cite{JW2,KK1,KK2}.  \\
\indent Let us mention that while our approach was inspired by these previous works, to solve the $N>3$ point function problem required significant generalization of \cite{AMSV,JW2,KK1} as well as nontrivial new ingredients.  For the case of the Null-Polygonal Wilson loop the world-sheet has the topology of a disk, whereas in our problem it is that of an $N$-punctured sphere and this changes the boundary conditions that one imposes.  This issue was addressed in \cite{JW2,KK1} for the case of the 3-punctured sphere, however in those works the total monodromy condition was enough to derive the functional equations that determine the necessary objects.  These functional equations are linear and can be easily inverted using standard techniques.  For the case of 4 or more punctures the situation becomes significantly more complex.  First of all, the total monodromy condition no longer provides enough information to fix the necessary objects. We have made heavy use of the formalism developed in \cite{GMN} to derive the functional equations. Second, the inversion of these functional equations is more subtle due to their complexity. The result turns out to be some integral equations resembling the usual TBA equations.\\
\indent Perhaps the most exciting aspect of this work is the multitude of interesting applications and extensions of the results.  Let us consider each of these in turn.
\begin{itemize}
\item \emph{Multiple configurations and phase transitions.} An important physical outcome of this paper is the emergence of multiple string configurations in $AdS_2$. Each of these configurations is associated to the existence of several saddle points. A natural question is to figure out whether the dominant saddle point depends on the parameters of the theory. If so it would be interesting to study the phase diagram and the possible transitions. We have already made some preliminary progress in this direction and we hope to make a more extensive study in a future publication.
\item \emph{OPE.} A natural question to ask given any 4-point function in a conformal field theory is what can be learned from its OPE decomposition.  In particular, important information about the spectrum and structure constants of the theory can be extracted.  
\item \emph{GKP string.} An interesting aspect of \cite{JW2,KK1} is the similarity between the mathematical formalism employed despite the differences in the physical problem: \cite{JW2} describes strings without spin in $AdS_2$ whereas \cite{KK1} describes spinning strings in $AdS_3$. In the formalism of \cite{KK1,KK2}, one expresses the $N$-point function of GKP string in terms of a universal AdS contribution and a contribution from vertex operators, both of which can be computed for the case of the three-point function.  It is possible that one could use the formalism developed in this work to calculate the AdS contribution to the $N$-point function of GKP strings.     
\item \emph{$N$-point functions.}  The formalism developed here does not depend in any special way on having only 4 punctures and in principle one could use the same methods to study the $N$-punctured sphere for any $N$.  It would be interesting to understand how the functional equations generalize to higher $N$.  Furthermore, since for the $N$-point function there will be $N-3$ unfixed insertion points, the moduli space of possible configurations should be quite interesting.   
\item \emph{TBA equations.}  We should note that the techniques developed in \cite{GMN}, in principle, allow one to write the functional equations derived in this paper in the usual form of a Y-system.  Typically this Y-system will involve an infinite number of Y-functions.  This form of the equations could be useful for various applications including analytic continuation of parameters and generalization to $N$-point functions.
\item \emph{Generalizing out of the line and WL/CF duality} A natural step would be to generalize this work for operators not inserted on a line. In this case the string is embedded in a higher dimensional $AdS$ space, which involves a more complicated Pohlmeyer reduction scheme. It would be interesting to study the question of whether the multiple string configurations/ saddle points we have found is special to $AdS_2$ case. Another promising application of such generalization would be the possibility of studying the OPE for the Null Polygonal Wilson Loop \cite{nullope1}-\cite{nullope3}. One could also investigate the duality between Null Polygonal Wilson Loops and Correlation functions of null separated local operators at strong coupling \cite{WLDUALITY1}-\cite{WLDUALITY6}.
\end{itemize}
Many of these points present interesting opportunities to try to learn about finite coupling and weak/strong coupling interpolation and this is probably the most stimulating reason for pursuing them.

\section*{Acknowledgments}
We thank Pedro Vieira for numerous invaluable discussions, motivation and inspiration throughout this work.  We also thank Jorge Escobedo, Davide Gaiotto, Kolya Gromov, Romuald Janik, Amit Sever, Kostya Zarembo and Miguel Zilh\~{a}o for many useful discussions.  Finally, we thank Davide Gaiotto for valuable comments on the first version of this paper.  JC is funded by the FCT fellowship SFRH/BD/69084/2010. This work has been supported in part by the Province of Ontario through ERA grant ER 06-02-293. Research at the Perimeter Institute is supported in part by the Government of Canada through NSERC and by the Province of Ontario through MRI. This work was partially funded by the research grants PTDC/FIS/099293/2008 and CERN/FP/116358/2010 and by Fund. Calouste Gulbenkian.
\emph{Centro de F\'{i}sica do Porto} is partially funded by FCT under grant PEst-OE/FIS/UI0044/2011.   


\appendix
\section{The linear problem}\label{App LP}
\subsection{Summary of the linear problem}
\indent  The linear problem associated with \eqref{SGE} is given by
\beqy
\(\pd+J_w\)\psi=0, \;\;\;\;\; \(\pdb+J_{\bar{w}}\)\psi=0\label{LP1}
\eeqy
where the connection has the form
\beqy
J_w = \frac{1}{\xi}\Phi_w+A_w, \;\;\;\;\; J_{\bar{w}} = \xi \Phi_{\bar{w}}+A_{\bar{w}}\label{LP2} 
\eeqy
\beqy
\Phi_w &=& 
\left(
\begin{smallmatrix}
0  			                                                     &  -\frac{1}{2}e^{\widetilde{\g}}   \\
 -\frac{1}{2}T e^{-\widetilde{\g}} 	&  0   
\end{smallmatrix}
\right)\label{LP app 1} \\
\Phi_{\bar{w}} &=& 
\left(
\begin{smallmatrix}
0  			                                                     &-\frac{1}{2}\bar{T} e^{-\widetilde{\g} }    \\
 -\frac{1}{2}e^{\widetilde{\g} }        	&  0   
\end{smallmatrix}
\right) 
\eeqy
\beqy
A_{w} &=&
\pd_{w}\left(
\begin{smallmatrix}
\frac{1}{2}\widetilde{\g} 	 & 0   \\
0                                                                                    & -\frac{1}{2}\widetilde{\g}   
\end{smallmatrix}
\right) \\
A_{\bar{w}} &=&
\pd_{\bar{w}}\left(
\begin{smallmatrix}
-\frac{1}{2}\widetilde{\g} 	 & 0   \\
0                                                                                    & \frac{1}{2}\widetilde{\g}    
\end{smallmatrix}
\right)\label{LP app 2}
\eeqy
For compactness we have introduced the combination $\widetilde{\g}=1/2(\g+\log\sqrt{T \overline{T}}\,)$. The function $\g$ is defined as the solution of the following problem
\beqy
\pd \bar{\pd} \g &=& \sqrt{T \bar{T}}\sinh \g\nn   \\
\g &\ra& \pm \frac{1}{2} \log T\bar{T}\;\;\;\;\; (w \ra z_a ) \label{SG}\\
\g &\ra& 0 \;\;\;\;\;\;\;\;\;\;\;\;\;\;\;\;\;\;\;\; (w \ra w_a) \nn
\eeqy
where $z_a$ and $w_a$ are the zeros and the poles of $T$, respectively.  \\
\indent For the near-puncture analysis as well as the WKB analysis it is useful to make the field redefinition $\psi \ra \hat{\psi} = \hat{G} \psi$ where 
\beqy
\hat{G}=
\frac{1}{2}\left(\begin{smallmatrix}
+e^{-\g/2}T^{1/4}\bar{T}^{-1/4} 	 & 1   \\
-e^{-\g/2}T^{1/4}\bar{T}^{-1/4}           & 1    
\end{smallmatrix}
\right)
\eeqy
This is usually refered to as `diagonal gauge' in the literature.  In diagonal gauge we have
\beqy
\hat{\Phi}_w &=& 
\frac{1}{2} \sqrt{T}\left(
\begin{smallmatrix}
-1  & 0   \\
0   & 1   
\end{smallmatrix}
\right) \\
\hat{\Phi}_{\bar{w}} &=& 
\frac{1}{2}\sqrt{\bar{T}}\left(
\begin{smallmatrix}
-\cosh \g                  &  \sinh \g    \\
-\sinh \g        	&  \cosh \g   
\end{smallmatrix}
\right) 
\eeqy
\beqy
\hat{A}_{w} &=&
\pd_{w}\left(
\begin{smallmatrix}
\frac{1}{4}\g -\frac{1}{8}\log(T\bar{T}) 	 & -\frac{1}{2} \g   \\
-\frac{1}{2} \g                                                       & \frac{1}{4}\g -\frac{1}{8}\log(T\bar{T})    
\end{smallmatrix}
\right) \\
\hat{A}_{\bar{w}} &=&
\pd_{\bar{w}}\left(
\begin{smallmatrix}
\frac{1}{4}\g +\frac{1}{8}\log(T\bar{T}) 	 & 0   \\
0                                                                          &\frac{1}{4}\g +\frac{1}{8}\log(T\bar{T})    
\end{smallmatrix}
\right) \label{Awb diag}
\eeqy
We are now ready to consider the behavior of the solutions near the points $w_a$ and $z_a$.


\subsection{Solutions near $w_a$}\label{App near punc}
Let us first consider the solutions of the linear problem in the neighborhood of one of the punctures.  From ($\ref{SG}$) and the explicit expressions for $\hat{\Phi}$ and $\hat{A}$ for $w \ra w_a$ we have
\beqy
\hat{\Phi}_w \ra 
\frac{1}{2} \sqrt{T}\left(
\begin{smallmatrix}
-1  & 0   \\
0   & +1   
\end{smallmatrix}
\right), \;\;\;\;\;
\hat{\Phi}_{\bar{w}} \ra 
\frac{1}{2}\sqrt{\bar{T}}\left(
\begin{smallmatrix}
-1        &  0    \\
0        	& + 1   
\end{smallmatrix}
\right) \label{DG near P 1}
\eeqy
\beqy
\hat{A}_{w} \ra
\pd_{w}\(-\frac{1}{8}\log T\bar{T}\)\left(
\begin{smallmatrix}
 +1 	 & 0   \\
0          & +1    
\end{smallmatrix}
\right), \;\;\;\;\;
\hat{A}_{\bar{w}} \ra
\pd_{\bar{w}}\(\frac{1}{8}\log T\bar{T}\)\left(
\begin{smallmatrix}
+1 	 & 0   \\
0          &+1    
\end{smallmatrix}
\right)\label{DG near P 2}
\eeqy
Then the solution in the vicinity of puncture $P_a$ is given by:
\beqy\label{near punc in app}
\hat{\psi}^{\pm}\(w\) \equiv \(T/\bar{T}\)^{1/8} e^{\pm \frac{1}{2} \int^w \xi^{-1}\o +\xi \bar{\o} } |\pm\rangle 
                           \sim \(w-w_a\)^{\pm \frac{1}{4}\D_a \xi^{-1}-\frac{1}{4}}\(\bar{w}-\bar{w}_a\)^{\pm \frac{1}{4}\bar{\D}_a \xi +\frac{1}{4}}|\pm \rangle
\eeqy
where $|\pm\rangle$ are the eigenvectors of the Pauli matrix $\s^3$.    Note the characteristic monodromy of the solutions about $w_a$. 


\subsection{Solutions near $z_a$}\label{App near zero}

Now we will consider the behavior of the solutions near the zeros $z_a$ of $T$.  Notice from \eqref{LP app 1} - \eqref{LP app 2} and \eqref{SG} that the connection is regular or singular at $z_a$ depending on the direction of the spike in $\g$ at $z_a$.  More specifically,  the connection is regular if $\g \sim - \log |w-z_a|$ and thus the solution will be regular in the vicinity of a $d$-spike. However, the connection has a singularity if $\g \sim + \log |w-z_a|$ and at the $u$-spikes one can check that in gauge  \eqref{LP app 1} - \eqref{LP app 2} there are two linearly independent solutions behaving as
\beqy\label{near za}
\Psi \sim \Psi_{z_a}\equiv \left(
\begin{smallmatrix}
\(w-z_a\)^{-1/4}\(\bar{w}-\bar{z}_a\)^{+1/4}     & 0 \\
 0                                                                              &  \(w-z_a\)^{+1/4}\(\bar{w}-\bar{z}_a\)^{-1/4}
\end{smallmatrix}
\right) 
\eeqy 
where we have written the two solutions in matrix form as in \eqref{matrix notation}.  Notice that $\Psi$ has square-root type singularity at $z_a$ since it has a monodromy of $\Psi \ra \(-1\)\Psi $ about $z_a$.  The solutions associated with the punctures $\{s_P\}$ and $\{\tilde{s}_P\}$ inherit this square-root singularity as one can see by expanding them in the basis \eqref{near za} near $z_a$. \\
\indent In our analysis it is crucial to account for the additional monodromies originating from $u$-spikes.  Let us explain our conventions for doing this.  If there is a $u$-spike at $z_a$, one can always make the gauge-transformation $\Psi \ra \Psi^{-1}_{z_a} \Psi$ that removes the square root singularity ($\Psi_{z_a}$ is given in \eqref{near za}).  Of course this gauge transformation contains the same multivaluedness and one must still account for it at the end of the day.   In the main text we use the point of view that this gauge transformation has been performed for each $u$-spike.  The connection in this gauge will only have singularities at the punctures and the solutions in this gauge will only have non-trivial monodromies around the punctures.  In this way we can define small solutions that are single valued throughout some $Q_E$, as is the prescription of \cite{GMN}.  We must then be sure to account for the multivaluedness of these gauge transformations whenever we have a holonomy that encloses an odd number of $u$-spikes.  Such holonomies arise in the WKB expansion of the coordinates and we will return to this issue below.


\section{WKB analysis}\label{WKB}
\subsection{Statement of the WKB approximation}\label{WKB1}  As we have discussed above, it is essential to have control over the $\xi \ra 0,\infty$ asymptotics of the inner products.  It is clear from $(\ref{LP1}-\ref{LP2})$ that these are both singular limits, and the basic idea of extracting this singularity is as follows.  As discussed above, we have good control over the solutions in the neighborhood of the punctures.   Thus what we would like to study is the transport
\beq
\mbox{Pexp}\[ - \int_{\mathcal{C}\(w'_a \ra w\)} \frac{1}{\xi} \Phi + A + \xi \bar{\Phi} \]\psi\(w'_a\)
\eeq
where $\mathcal{C}\(w'_a \ra w\)$ is a curve starting at $w'_a$, a point in the neighborhood of $w_a$, and terminating at a generic point $w$.  Note that at any point on the punctured sphere $C$ the Higgs field $\Phi$ has the two eigenvalues $\mp \o/2 = \mp \sqrt{T}/2 \; dw$ (which are single valued on the double cover $\widetilde{\Sigma}$), and thus  we can choose a gauge along $\mc{C}$ where $\Phi$ is diagonal and given by 
\beqy
\Phi = 
\frac{1}{2} \left(
\begin{matrix}
-\o  & 0   \\
0    & \o   
\end{matrix}
\right)\label{Diag Phi}
\eeqy
Now consider the $\xi \ra 0$ limit.  First consider an infinitesimal segment of $\mc{C}$ in the neighborhood of $P_a$. In the neighborhood of $P_a$ the connection (in diagonal gauge) becomes diagonal (see \eqref{DG near P 1}-\eqref{DG near P 2}) and thus one can break apart the path-ordered exponential.  In particular, one can isolate the singular part $e^{-\int \Phi/\xi} |\pm\>$ which will have one component growing exponentially and one component decaying.  Let us choose the branch of $\Phi$ such that the $|+\>$ component is the one that is growing as we transport along $\mc{C}$ \emph{away} from $P_a$ (although for the moment we are still working in a neighborhood of $P_a$).  This will correspond to the small solution at $P_a$ since it is exponentially decaying as it is transported \emph{toward} from $P_a$.    The \emph{WKB approximation} is the statement that the exponentially growing part of the solution as $\xi \ra 0$ will continue to be given by  $e^{-\int \Phi/\xi} |+\>$  as we transport away from the neighborhood of $P_a$ (now leaving the neighborhood of $P_a$) as  long as we follow a curve such that at every point we have
\beq
\mbox{Re}\(\o/\xi\) >0\label{WKB cond}
\eeq
This condition is satisfied most strongly along a curve such that
\beq
\mbox{Im}\(\o/\xi\) = 0 \label{WKB curve}
\eeq    
Condition $(\ref{WKB cond})$ is called the \emph{WKB condition} and curves satisfying ($\ref{WKB curve}$) are called \emph{WKB curves} \cite{GMN}.  Along a WKB curve defined for $\mbox{Arg}\(\xi\)=\th$ the WKB condition is satisfied for $\mbox{Arg}\(\xi\) \in \(\th-\pi/2,\th+\pi/2\)$ and the WKB approximation is guaranteed to hold in this range.  For example, suppose there is a WKB line connecting $P_a$ to $P_b$ for $\th \in \(\th_-,\th_+\)$ but not outside that range.  Then the WKB approximation will reliably give the $\xi \ra 0, \infty$ asymptotic for $\th \in \(\th_- - \pi/2, \th_+ +\pi/2\)$.  These statements are proven in \cite{GMN} and we refer the reader there for a more detailed discussion.
\subsection{Subleading WKB}
\indent We will now consider the $\xi \ra 0$ limit of the inner products (or Wronskians) $\(s_b \wedge s_a\)\(\xi\)$.  We consider the case when $P_a$ and $P_b$ are connected by a WKB line which will be an edge $E_{ab}$ in the WKB triangulation.   From the analysis of \ref{App near punc} we know $s_a$ and $s_b$ in the neighborhood of $P_a$ and $P_b$ respectively.  In order to evaluate the Wronskian we need to know the solutions  at a common point.  The approach  here is to use the connection to transport the solution $s_a$  along $E_{ab}$ to a point $w'_b$ in the neighborhood of $P_b$ and then to evaluate the Wronskian at $w'_b$.  That is, we want to study the $\xi \ra 0$ behavior of 
\beqy
\<s_b|\mbox{Pexp}\[-\int_{0}^{1} \!\! dt \; \frac{1}{\xi}H_0+V\]\! \!|s_a\>\label{transport}
\eeqy   
where we defined
\beqy\label{H0 and V}
H_0 = \dot{w} \hat{\Phi}_w, \;\;\;\;\; V = \dot{w}\hat{A}_w+\dot{\bar{w}}\hat{A}_{\bar{w}}+\xi \dot{\bar{w}} \hat{\Phi}_{\bar{w}} 
\eeqy
The contour of integration in \eqref{transport} is the edge $E_{ab}$ and the components of \eqref{H0 and V} are defined in appendix \ref{App LP}.  The basic idea of the computation is to expand in a perturbative series where $\xi^{-1}H_0$ acts as the free Hamiltonian.  Such a procedure will be valid so long as the free part of the Hamiltonian is sufficiently larger than $V$ for all points along the curve, which will be true along the edges of the WKB triangulation.  Then we can expand ($\ref{transport}$) in the Born series
\beqy
(-1)\hat{\psi}_b^- \hat{\psi}_a^+\Bigg(\!
\<+|e^{-\int_0^1 H_0/\xi}|+\> 
\!\!\! &-& \!\!\! \int_{0}^1 \!\!\! dt_1\<+|e^{-\int^1_{t_1}H_0/\xi}V(t_1)e^{-\int_0^{t_1}H_0/\xi}|+\>\label{Born series} \\
\!\!\!&+&\!\!\! \int^{1}_{0}\!\!\! dt_2 \! \int^{t_2}_{0}\!\!\! dt_1 \<+|e^{-\int_1^{t_2}H_0/\xi}V(t_2)e^{-\int^{t_2}_{t_1} H_0/\xi}V(t_1)e^{-\int^{t_1}_0 H_0/\xi}|+\>\Bigg) \nn
\eeqy
Let us explain a subtle point regarding the `external states' in the above expression.  We start with the small solution at $P_a$ which we take to be $\psi_a^+$.  We then transport it to $P_b$ and then extract the coefficient of the exponentially growing part -- that is, we take the inner product with the small part of this transported solution.  Since $\psi_a^+ \sim |+\>$ grows as we transport it along a WKB curve (i.e. it decays as one follows the curve into $P_a$ and thus grows as we transport it away from $P_a$) and $H_0$ is diagonal, we infer that the small part of the solution at $P_b$ is the solution proportional to $|-\>$.  Thus we take the out-state to be $\<-|\psi^-_b$.  Finally, since the inner product is the antisymmetric the $\<-|$ gets flipped to a $\<+|$.  \\
\indent Using the fact that $|\pm\>$ are eigenstates of the free Hamiltonian we can easily evaluate the order $\mathcal{O}\!\(V^0\)$ and $\mathcal{O}\!\(V^1\)$ terms in \eqref{Born series}.  For the $\mathcal{O}\!\(V^2\)$ term, we insert the identity $|+\>\<+|+|-\>\<-|$ between the two insertions of $V$.  We find  
\beqy
 (-1)\hat{\psi}_b^- \hat{\psi}_a^+ \; e^{+\frac{1}{2}\int_0^1 \o/\xi}\Bigg(1-\int^{1}_{0}dt_1\<+|V(t_1)|+\>
 +\frac{1}{2} \[\int^{1}_{0}dt_1\<+|V(t_1)|+\>\]^2 + 
\\ \int^{1}_{0}dt_2 \int^{t_2}_{0}dt_1 e^{-\int^{t_2}_{t_1} \o/\xi}\<+|V(t_2)|-\>\<-|V(t_1)|+\> \Bigg)\nn
\eeqy
%
%
Now concentrate on the second term on the $\mathcal{O}\(V^2\)$ contribution.  As $\xi \ra 0$ the factor $\exp\(-\int^{t_2}_{t_1}\o/\xi\)$  will suppress the integrand except for the small range $t_2 = t_1 + \mathcal{O}(\xi)$ and thus the result of the first integration will already be $\mathcal{O}\!\(\xi\)$.  So to order $\xi$ we can take $\o$ to be constant and $V(t_1) \ra V(t_2)$. We then find for the second term in the $\mathcal{O}\!\(V^2\)$ contribution
\beqy
 e^{\frac{1}{2}\int^1_{0}\o/\xi}\int^{1}_{0}  dt_2 \xi \frac{|\<+|V(t_2)|-\>|^2}{\dot{w}\sqrt{T}}
\eeqy
%
%
Putting everything together, we see that the result re-exponentiates and we find
\beqy
  (-1)\hat{\psi}_b^- \hat{\psi}_a^+ \exp\[
+\frac{1}{\xi}\int^1_0 dt \frac{1}{2}\sqrt{T} - \int^{1}_{0}dt\<+|V(t)|+\>+ \xi \int^1_0 dt \frac{|\<+|V(t)|-\>|^2}{\dot{w}\sqrt{T}} \] \label{WKB to O1}
\eeqy
Grouping each term based on its order in $\xi$ (including the prefactors $\hat{\psi}_b^- \hat{\psi}_a^+$ whose explicit expression are given in \eqref{near punc in app}) we find
\beqy
 \(s_b \wedge s_a\)\(\xi\) \sim \exp\[
+\frac{1}{2}\xi^{-1}\varpi_{ab} + \a_{ab} + \frac{1}{2}\xi \overline{\varpi}_{ab}+\xi \eta_{ab}\] \label{WKB to O1 pretty}
\eeqy
where 
\beqy
\varpi_{ab} &=&
\lim_{w'_a \ra w_a}\lim_{w'_b \ra w_b}\left[\int_{E_{ab}} \sqrt{T} dw + \frac{\D_a}{2} \log(w_a-w'_a)+\frac{\D_b}{2} \log(w_b -w'_b) \right]\label{form 1} \\
\a_{ab}
&=&-\int_{E_{ab}} \( \frac{1}{4} \pd_w \(\g-\log \sqrt{T \bar{T}}\)dw+\frac{1}{4}  \pd_{\bar{w}} \(\g+\log \sqrt{T \bar{T}}\)d\bar{w} \)\label{form 2} \\
\eta_{ab} &=& \int_{E_{ab}} \( \frac{1}{2}\sqrt{\bar{T}} \(\cosh \g-1\) \; d\bar{w}+\frac{1}{4 \sqrt{T}}\(\pd \g\)^2 \; dw \)\label{form 3}
\eeqy
This completes the derivation of formula \eqref{prod WKB} used in the main text.  The integral $\varpi_{ab}$ is defined as in \eqref{omega reg 2}.  Note that the logarithmic terms in $\varpi_{ab}$ in \eqref{WKB to O1 pretty} are due to the prefactor $\hat{\psi}_b^- \hat{\psi}_a^+$.  These terms precisely cancel the singularities at the endpoints of the integral $\int_{a}^{b} \o$ so that $\varpi_{ab}$ is finite as we continue the limits of integration all the way up the to punctures at $w_a$ and $w_b$ \cite{JW2}.   \\


\subsection{WKB expansion of the coordinates}
In the previous section we derived the $\xi \ra 0$ WKB expansion of $\(s_a \wedge s_b\)$ up to order $\mathcal{O}\! \(\xi\)$.  To compute the WKB expansion of the coordinate $\chi_E$ we simply combine the expansions for each edge of the quadrilateral $Q_E$, taking care to account for the directions of the WKB lines as discussed in section \ref{coord asymp}.  When this is done each of the integrals \eqref{form 1} - \eqref{form 3} become closed integrals along the cycle $\g_E$.  The asymptotics of the $\chi$-functions are needed for the inversion of the $\chi$-system described in section \ref{inversion}.  For that purpose only the non-vanishing contributions are needed in the $\xi\ra 0,\infty$ limits.   \\ 
\indent There is one very important subtlety that must be addressed here, which is that of the monodromy around $u$-spikes discussed in appendix \ref{App near zero}.  We take the point of view that we have made the (multi-valued) gauge-transformation \eqref{near za} that removes the monodromy about each $u$-spike.   The small solutions in this gauge are single valued throughout $Q_E$, but we must account for the monodromy of the gauge transformation about $Q_E$.  This monodromy is simply $(-1)^{u_E}$ where $u_E$ is the number of $u$-spikes in $Q_E$. \\
\indent Combining the above discussion with \eqref{form 2}, the constant term in the WKB expansion of $\chi_E$ is given by
\beq\label{C final}
C_E^{\(0\)}=\log (-1)^{u_E} - \frac{1}{4}\int_{\g_E} \( d\g + *d \log \sqrt{T \bar{T}} \) = \log (-1)^{u_E} \pm i \pi 
\eeq
To arrive at the last equality \eqref{C final} we have used the fact that $\g$ is single-valued on the 4-punctured sphere so that the integral of $d \g$ on any closed contour is zero.  The integral of $*d \log \sqrt{T \bar{T}}$ is simple to do explicitly and gives the $\pm i \pi$ factor.\footnote{The $\pm$ depends on the orientation of $\g_E$ but both signs have the same overall effect so that the $\pm$ is irrelevant.}  \\
\indent The discussion of the $\xi \ra \infty$ limit follows along the same lines as the $\xi \ra 0$ limit.  The singular term is given by $e^{\xi \int_{\g_E}\bar{\o}/2}$.  The constant term is the same.  Thus the full non-vanishing WKB asymptotic is given by
\beq
\chi_E \sim  (-1)^{u_E}\exp\[\frac{1}{2} \int_{\g_E}\( \xi^{-1}\o+\xi \bar{\o}\) \]
\eeq
where we recall that $u_E$ is the number of $u$-spikes enclosed in $\g_E$.  This is the expression \eqref{WKB asymp} used in the main text.


\section{Fold lines and Properties of $\g$}\label{gamma props}
In this appendix we discuss some properties of the function $\g$ and how they are related to geometric features of the string embedding.  In appendix \ref{fold lines app} we show that the world-sheet contours where $\g=0$ map to the fold-lines of the target space solution; in appendix \ref{app g near wa} we discuss how the geometry of the string embedding near the boundary is deduced from the structure of these $\g=0$ contours near the points $w_a$; finally, in section \ref{app g=0 structure} we show how the global structure of the $\g=0$ contours is deduced from the choice of spikes in $\g$.  The point of this appendix is to give the background details that were omitted in the discussion of section \ref{folds and spikes}.
\subsection{Fold lines}\label{fold lines app}
In this section we show that the contours on the worldsheet where $\g=0$ map to the fold lines of the string embedding.  This was pointed out in \cite{JW2}.  Recall the relation between $\g$ and the world-sheet metric
\beqy
\sqrt{T \bar{T}} \cosh \g = \frac{\pd x \bar{\pd} x+\pd z \bar{\pd} z}{z^2}
\eeqy
Furthermore, we have
\beqy
T(w) = \frac{\(\pd x\)^2+\(\pd z\)^2}{z^2}, \;\;\;\;\; \bar{T}(\bar{w}) = \frac{\(\bar{\pd} x\)^2+\(\bar{\pd} z\)^2}{z^2}
\eeqy
Now, suppose that $\mathcal{C}$ is a curve on the worldsheet that maps to a fold-line of the string and consider a point $\mathcal{P}$ in that curve.  We can choose local coordinates at $\mathcal{P}$  so that the derivative takes the form $\pd_w \ra e^{i\phi}\(\pd_t+i \pd_n\)$ where the direction $\pd_n$ is chosen such that $\pd_n z =0$.  The prefactor $e^{i\phi}$  is the Jacobian of the coordinate transformation (just a translation and rotation). The defining property of the fold-line is then that the $x$-coordinate reaches an local extrema and thus we also have $\pd_n x =0$ as we cross the fold.  Therefore along the fold-line we have (with $\pd_t x = \dot{x}$)
\beqy
\sqrt{T \bar{T}} \cosh \g \ra \frac{\dot{x}^2+\dot{z}^2}{z^2}, \;\;\;\;\;
T(w) \ra  e^{2 i \phi}\frac{\dot{x}^2+\dot{z}^2}{z^2}, \;\;\;\;\;
\bar{T}(\bar{w}) \ra  e^{-2 i \phi}\frac{\dot{x}^2+\dot{z}^2}{z^2} \label{FoldConds}
\eeqy
Using the last two equations to solve for $\sqrt{T \bar{T}}$ we see that they are consistent with the first equation only if $\g =0$. Therefore, the worldsheet contours where $\g=0$ map to the fold-lines of the string-embedding.  For this reason, we frequently refer to the contours where $\g=0$ as fold-lines. 
\subsection{Structure of $\g$ near $w_a$}\label{app g near wa}
To gain some intuition about the structure of the contours where $\g=0$ it is useful to study the behavior of $\g$ near the points $w_a$. Recall that $\g$ is defined as the solution to the PDE:
\beq
\pd \pdb \g = \sqrt{T \bar{T}}\sinh \g \label{SGE 2}
\eeq
subject to the boundary conditions
\beqy
\g &\ra& \pm \frac{1}{2} \log T\bar{T}\;\;\;\;\; (w \ra z_a )\label{g at zeros 2} \\
\g &\ra& 0 \;\;\;\;\;\;\;\;\;\;\;\;\;\;\;\;\;\;\;\; (w \ra P_a)\label{g at punctures 2} 
\eeqy
The boundary condition $(\ref{g at punctures 2})$ simply imposes that $\g$ is non-singular at the singularities of $T$ and this condition is automatically imposed if we demand the solution be regular away from the zeros of $T$. \\
\indent  Since we know that $\g$ must vanish at singularities of $T$, it's natural to study the function in the neighborhood of these points. Let us consider some $P_a$ and use polar coordinates $\(r,\phi\)$ in which the origin is at $w_a$.  Since $\g$ is vanishing, we can linearize the RHS of $(\ref{SGE 2})$.  Further, we can take $\sqrt{T \bar{T}} \sim |\D|^2/(4 r^2)$.   The PDE becomes linear and separable and using standard techniques one finds the series solution
\beq
\g \sim g_0 \; r^{\frac{1}{2}\D} +\sum_{m=1}^{\infty} g_m \sin \(m \phi+\d_m\) r^{\frac{1}{2}\sqrt{\D^2+4m^2}} \label{g series}
\eeq
Now consider a small circle centered at $r=0$.  As $r\ra 0$ the series $(\ref{g series})$ is dominated by the lowest mode in the expansion.  Thus  along an infinitesimal circle centered at $r=0$ the series $(\ref{g series})$ will vanish $2m^{*}$ times, where $g_{m^*}$ is the smallest non-zero coefficient $g_m$, $m=0,1,2,...$ in the series.   Thus, if $g_0$ is the smallest non-zero coefficient then the series will vanish only at the point $w_a$ which will be a local extrema.  If $m^*=1$ then the series will vanish along a single curve passing through $P_a$; if $m^{*}=2$ then $\g$ will vanish along two curves that intersect at $P_a$, and so on.  \\
\begin{figure}[t!]
\begin{center}
\includegraphics[width=0.65\linewidth]{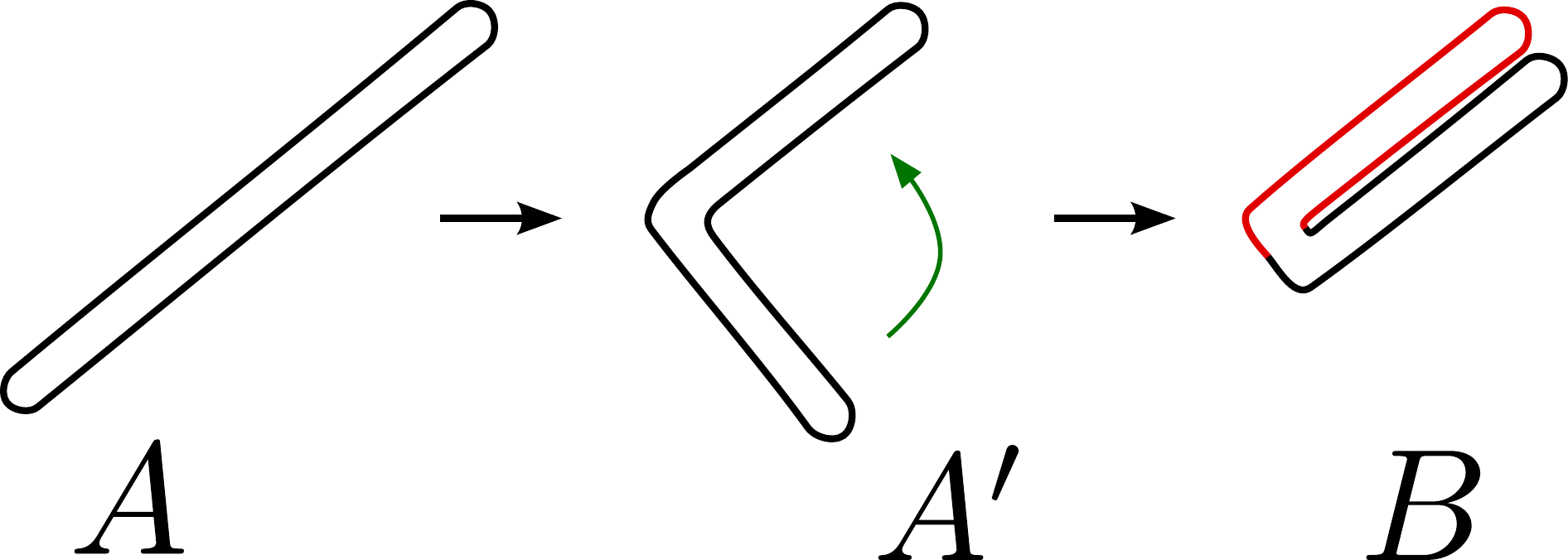}
\caption{Single-folded and double-folded string in panels $A$ and $B$ respectively.}\label{DoubleFold}
\end{center}
\end{figure}
\indent The fact that the contours where $\g=0$ map to fold-lines of the target-space solution gives a clear geometric meaning to each possible behavior $m^*=1,2,...$ near an insertion point.  For $m^*=1$ we will cross two fold-lines as the world-sheet coordinate traverses a small loop around the point $w_a$.  This means that near the insertion point the string is single-folded as shown in figure \ref{DoubleFold}$A$.  For $m^*=2$ we will cross two fold-lines as the world-sheet coordinate traverses a small loop around the point $w_a$.  This means that near the insertion point the string is double-folded as shown in figure \ref{DoubleFold}$B$, for example.  In general for $n>0$ the case $m^*=n$ should correspond to an $n$-folded string.   The only subtle case seems to be $m^*=0$. Apparently if $m^*=0$, as we traverse a closed loop around $w_a$ the contour swept out in the target space does not close since there is no point at which the coordinates $\(x,z\)$ can `turn around'.  In this paper we are only interested in solutions that are closed (i.e. the embedding coordinates have trivial monodromies around operator the insertion points $x_a$) and thus we will only study cases for which $m^*>0$ at all $w_a$.  This is further discussed in appendix \ref{app g=0 structure}.   \\
\indent It is important to keep in mind that (as we mentioned above) the behavior of $\g$ at $P_a$ is not our choice, and  is determined by regularity and the conditions $(\ref{g at zeros 2})$. In other words, for fixed $T$ the only remaining conditions one can specify are the choice of signs in $(\ref{g at zeros 2})$.  For each choice of signs there will be a unique $m^{*}$ for each $P_a$.  In the next section we demonstrate how this works using the $T$ of the 4-point function discussed in the main text.   \\
\subsection{Structure of contours where $\g=0$}\label{app g=0 structure}
\begin{figure}[t!]
\begin{center}
\includegraphics[width=0.55\linewidth]{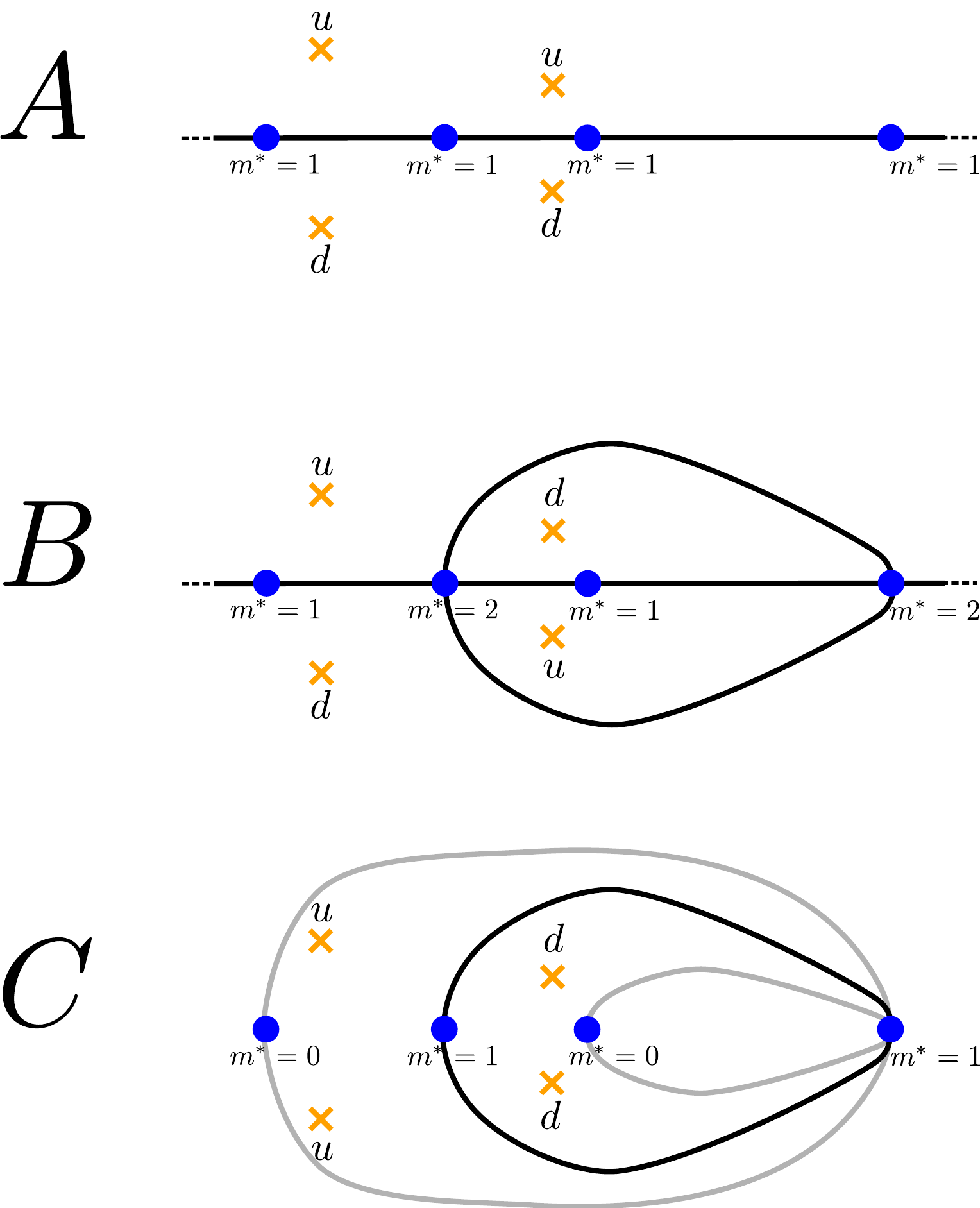}
\caption{Three different spike configurations and the corresponding structure of the $\g=0$ contours.   The black lines schematically represent the contours where $\g=0$ and one can read off the $m^{*}$ associated with each puncture.  Panels $A$ and $B$ show the physically relevant configurations studied in the main text.  Panel $C$ shows a third spike configuration which is not physical due to the presence of $m^*=0$ behavior at two of the insertion points. The gray contours in panel $C$ indicate contours that cannot cannot correspond to fold-lines due to the restriction that $\g=0$ contours must encircle at least one zero of $T$ (see footnote \ref{encircle zeros}).  In this figure we are not indicating the location saddle-point $w_4^*$ because it is not relevant for the present discussion (so long as it is located somewhere on the real axis).  }\label{SpikeConfigs}
\end{center}
\end{figure}

\indent In this section we describe why the spike configurations of figure \ref{SpikeConfigsMainText} are the only two physically relevant configurations. Furthermore, we deduce the structure of the contours where $\g=0$ for each of these spike configurations.\\
\indent Consider $T$ fixed to be that of the 4-point function discussed in the main text (see equation \eqref{4ptT} and figure \ref{WKBCells1}).  There are 4 zeros and therefore $2^4$  ways to choose the signs in $(\ref{g at zeros 2})$.  Because of the symmetry of $\(\ref{SGE 2}\)$ under $\g \ra -\g$, without loss of generality we can fix one of the spikes to be up which leaves $2^3$ choices.  Now, because the string is embedded in $AdS_2$ we know that it must be folded.  Moreover we know that the operator insertions $x_a$ will sit along the fold-lines of the target-space solution.  In the world-sheet coordinates this translates to the statement that we should require $m^*>0$ at each $w_a$.  That is, there should be at least one contour where $\g=0$ running through each insertion point $w_a$.  For the 4-point function $T$ (see equation \eqref{4ptT} and figure \ref{WKBCells1}) the only obvious way to accomplish this in general is to choose the spikes such that $\g \ra -\g$ under reflection about the real axis.  This leaves only the spike configurations of figure \ref{SpikeConfigs}$A$,$B$, which are those of figure \ref{SpikeConfigsMainText} used in the main text.  We will now discuss the global structure of the $\g=0$ contours for these two choices of spikes.\\
\indent In figure \ref{SpikeConfigs} we show the fold-structure for three different spike configurations.  The black lines schematically represent the contours where $\g=0$ and one can read off the $m^{*}$ associated with each puncture.  The structure of these contours is determined purely by the choice of the directions of the spikes of $\g$.  We refer to these contours as `fold lines' since they map onto the fold-lines of the target-space embedding (see appendix \ref{fold lines app}).  We guess the structure of the fold lines for each choice of the spikes as follows:   $u$ spikes must be separated from $d$ spikes by at least one fold line; we use the minimum number of fold lines needed to accomplish this for all spikes.  Note that fold lines must encircle at least one zero of $T$.\footnote{\label{encircle zeros} Consider a closed contour along which $\g=0$ and suppose (for a contradiction) that it does not enclose any zeros of $T$.  Let $\mathcal{D}$ be the region enclosed by the contour.  This contour must separate positive values from negative values (i.e. it cannot sit at the bottom of a `valley' since this locally violates the equation \eqref{SGE 2}).  Suppose for simplicity that $\g<0$ in $\mathcal{D}$.  Since $\g$ is regular away from the zeros of $T$, there must be at least one local minimum inside $\mathcal{D}$, and therefore at least one point where $\(\pd^2_x+\pd^2_y\) \g \ge 0$.  Thus at such a point the LHS of $(\ref{SGE 2})$ is positive or zero, but the RHS is strictly less than zero by assumption, which is the desired contradiction.} This restriction is useful because, for example, it allows one to rule-out the possibility of fold-lines corresponding to the gray contours in figure \ref{SpikeConfigs}C.  This is important because if it was possible for the gray contours to be fold-lines then it might be possible to have a solutions with all $m^*>0$ for configuration $C$.  Configurations $A$ and $B$ are the physical configurations that we study in this paper and we have checked the fold structures of figure \ref{SpikeConfigs}$A$,$B$ numerically.  Configuration $C$ is an example of a spike-configuration that does not correspond to a target-space solution with the desired properties; the corresponding fold structure is only our best guess but we have not checked it numerically.   \\
\indent To summarize this appendix, in appendix \ref{fold lines app} we showed that the world-sheet contours where $\g=0$ map to the fold-lines of the target space solution; in appendix \ref{app g near wa} we discussed how the geometry of the string embedding near the boundary is deduced from the structure of these $\g=0$ contours near the points $w_a$; finally, in section \ref{app g=0 structure} we discussed how the global structure of the $\g=0$ contours is deduced from the choice of spikes in $\g$.  From all of this one can deduce some qualitative global features of the string embedding, which is discussed in detail in section \ref{folds and spikes}.

\section{Details of the 4-point function computation}\label{detailed area}
\subsection{Explicit expression for stress-energy tensor coefficients}
For completeness we present the coefficients $c_a$ of the stress-energy tensor in formula (\ref{4ptT}),
\beqa
c_{\infty}&=&\frac{\Delta _{\infty }^2}{4} \nonumber\\
c_0&=&\frac{1}{4} \left[4 \,U w_4+2 w_4 \left(1+w_4\right) \Delta _3^2+\left(-1+w_4\right) \left(2 w_4 \Delta _1^2+\left(1+w_4\right) \left(\Delta _2^2-\Delta _4^2\right)\right)\right]\nonumber\\
c_1&=&\frac{1}{2} \left[-2\, U+\left(-1+w_4\right){}^2 \Delta _1^2-\left(1+w_4\right){}^2 \Delta _3^2\right]\nonumber\\
c_2&=&\frac{1}{4} \left[-4\, U w_4+2 \left(1+w_4\right) \Delta _3^2+\left(-1+w_4\right) \left(-2 \Delta _1^2+\left(1+w_4\right) \left(-\Delta _2^2+\Delta _4^2\right)\right)\right]
\eeqa
\subsection{Explicit expressions for $\chi$-functions and $A_{PQ}$}
For reference, we include here the explicit expressions for the $\chi$-functions for the triangulation of figure \ref{WKBCells1}.  They are given by
\beqa
\chi_{12} &=& (-1)\frac{(s_1 \wedge M_1^{-1} s_4)(s_2 \wedge s_4)}{(M_1^{-1} s_4 \wedge s_2)(s_4 \wedge s_1)}\label{app coords 1}\\
\chi_{23} &=& (-1)\frac{(s_2 \wedge M_3 s_4)(s_3 \wedge s_4)}{(M_3 s_4 \wedge s_3)(s_4 \wedge s_2)}\\
\chi _{34}&=&(-1)\frac{\left(s_4\wedge s_2\right)\left(s_3\wedge M_3^{-1}s_2\right)}{\left(s_2\wedge s_3\right)\left(M_3^{-1}s_2\wedge s_4\right)}\\
\chi _{14}&=&(-1)\frac{\left(s_4\wedge M_1s_2\right)\left(s_1\wedge s_2\right)}{\left(M_1s_2\wedge s_1\right)\left(s_2\wedge s_4\right)}
\eeqa
\beqa
\chi_{24} &=& (-1)\frac{(s_2 \wedge s_3)(s_4 \wedge s_1)}{(s_3\wedge s_4)(s_1 \wedge s_2)}\\
\chi_{\hat{24}} &=& (-1)\frac{(M_{3}^{-1}s_2\wedge M_{4}s_1)(s_4 \wedge s_3)}{(M_{4}s_1 \wedge s_4)(s_3 \wedge M_{3}^{-1}s_2 )} \label{app coords 2}
\eeqa
One can check that these coordinates satisfy the rule \eqref{mu rule} at each puncture. The $\chi$-system obeyed by these coordinates is given by
\beqy
\chi_{24} \chi_{24}^{++} &=& \(\chi_{\hat{24}} \chi_{\hat{24}}^{++}\)^{-1}
=\frac{\(1+A_{23}\)\(1+A_{14}\)}{\(1+A_{34}\)\(1+A_{12}\)} \label{app chi system 1}\\
\chi_{12} \chi_{12}^{++} &=& \(\chi_{14} \chi_{14}^{++}\)^{-1}=\chi_{34} \chi_{34}^{++} = \(\chi_{23} \chi_{23}^{++}\)^{-1} 
=\frac{\(1+A_{24}\)}{\(1+A_{\hat{24}}\)}\label{app chi system 2}
\eeqy
where the $A_{PQ}$ are given by 
\beqa
A_{12}&=&\frac{\chi _{12}\left(1+\chi _{14}\right)\left(1+\hat{\chi }_{24}\left(1+\chi _{23}\left(1+\chi _{24}\right)\right)\right)}{\left(1-\mu _1^2\right)\left(1-\mu _2^2\right)}\\
A_{23}&=& \frac{\c_{23}\(1+\c_{34}\)\(1+\c_{24}\(1+\c_{12}\(1+\c_{\hat{24}}\)\)\)}{\(1-\mu_{2}^2\)\(1-\mu_3^2\)}\\
A_{34}&=&\frac{\chi _{34}\left(1+\chi _{23}\right)\left(1+\hat{\chi }_{24}\left(1+\chi _{14}\left(1+\chi _{24}\right)\right)\right)}{\left(1-\mu _3^2\right)\left(1-\mu _4^2\right)}\\
A_{14}&=&\frac{\chi _{14}\left(1+\chi _{12}\right)\left(1+\chi _{24}\left(1+\chi _{34}\left(1+\hat{\chi }_{24}\right)\right)\right)}{\left(1-\mu _1^2\right)\left(1-\mu _4^2\right)}\\
A_{24}&=& \frac{\c_{24}\(1+\c_{12}\(1+\c_{\hat{24}}\(1+\c_{23}\)\)\)\(1+\c_{43}\(1+\c_{\hat{42}}\(1+\c_{41}\)\)\)}{\(1-\mu_{2}^2\)\(1-\mu_4^2\)}\label{A24} \\
A_{\hat{24}}&=&\frac{\hat{\chi }_{24}\left(1+\chi _{23}\left(1+\chi _{24}\left(1+\chi _{12}\right)\right)\right)\left(1+\chi _{41}\left(1+\chi _{42}\left(1+\chi _{43}\right)\right)\right)}{\left( 1-\mu _2^2\right)\left(1-\mu _4^2\right)}
\eeqa
Using the explicit expressions for the coordinates \eqref{app coords 1}-\eqref{app coords 2}, schouten identity and the shift relation \eqref{shift relation} one can directly verify the functional equations \eqref{app chi system 1}-\eqref{app chi system 2}.
\subsection{Finite part of $AdS$}
\begin{figure}
\centering
\def\svgwidth{13cm}
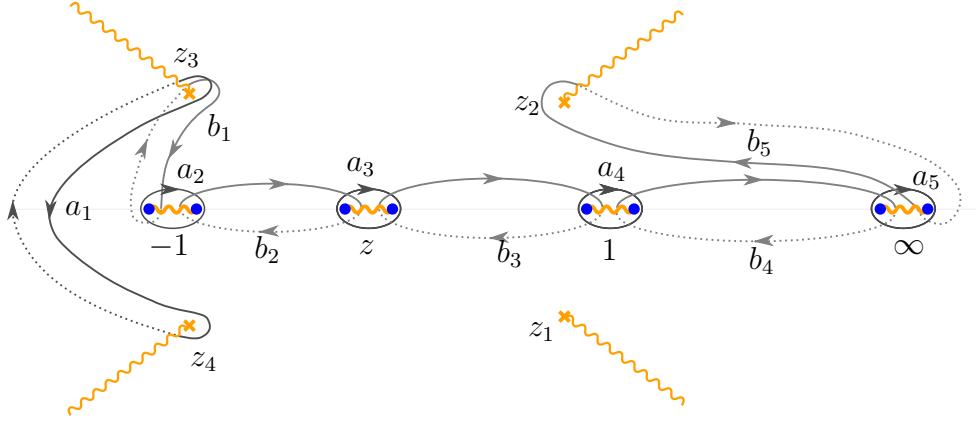
\caption{The cycles for Riemann bilinear identity. The dashed line represents a contour in a different Riemann sheet. The wavy lines represent a choice of branch cuts. From this picture we also read the intersection matrix $I_{ab}$ of the cycles. For each pair of cycles, say $\gamma_a$ and $\gamma_b$, intersecting at a point with tangent vectors $\partial_a$ and $\partial_b$ respectively, we assign $I_{ab}=+1$ $(-1)$ if $\det \left[ \{ \partial_a, \partial_b \} \right]>0\, (<0)$.}
\label{fig:cycles}
\end{figure}
\begin{figure}[t!]
\centering
\def\svgwidth{17cm}
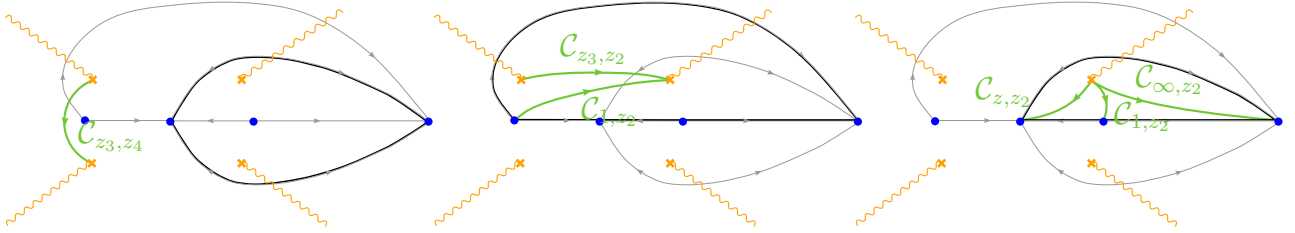
\caption{To extract line integrals connecting a zero to puncture or connecting two zeros we combine products of elementary solutions that have WKB expansions involving integrals over the paths indicated by the black lines. The resulting closed contours can be continuously deformed into the contour that we want, indicated by the green lines. The precise way of combining these products is dictated by the direction of the WKB lines indicated by the gray arrows.}
\label{fig:lines}
\end{figure}

In this section, we present some intermediate steps in the derivation of our formula (\ref{action final}) for the finite part of the $AdS$ contribution.  We want to compute 
\beq\label{area rbi 2}
 A_{fin}=\frac{\pi}{3}-\frac{i}{2} \(\oint_{\g_a} \!\! \o \) I^{-1}_{ab} \(\oint_{\g_b} \!\! \eta \)\,.
\eeq
according to the steps outline in section $\ref{sec:finite}$.  The complete basis of five a-cycles and five b-cycles that we chose is depicted in figure \ref{fig:cycles}. From this figure we also read-off the intersection matrix $I_{ab}=\(\delta_{a+1,b}-\delta_{a-1,b}\)$ using the conventions described in the caption. The only other ingredient we need is
\beq
\int_{a_i} \eta = 0,\,\,\,\,\,\,\, i=2,\dots,5
\eeq
which follows from the regularity of $\eta$ at the poles of $T$.  Plugging into $(\ref{area rbi 2})$ and computing we find
\beq\label{action1}
A_{fin}=\frac{\pi}{3}+ i \left(\o_{a_1}\eta_{z_3,z_2}+\o_{a_2} \eta_{-1,z_2}+ \o_{a_3} \eta_{z,z_2}+ \o_{a_4} \eta_{1,z_2}+\o_{a_5} \eta_{\infty ,z_2}\right) - i\(\sum_{i=1}^{5}\o_{b_i}\)\eta_{z_3,z_4} 
\eeq
where we are using the notation $\eta_{ab}=\int_a^b \eta$ and $\o_{\mathcal{C}} = \int_{\mathcal{C}}\o$ and the contours are defined in figures \ref{fig:cycles} and \ref{fig:lines}.  \\
\indent Each of these $\eta_{ab}$ can be written as a linear combination of the $\eta_{E_{ab}}=\int_{E_{ab}}\eta$ where the integral is taken along the WKB-line from $P_a$ to $P_b$ and the direction of the contour is the same as that of the WKB line.  The idea is to combine the $\eta_{E_{ab}}$ to form the contour that we want. 
Let us exemplify with $\eta_{1,z_2}$. From the WKB configuration, see figure \ref{fig:lines}, we see that the large $\theta$ expansion of the ratio 
$\frac{\(s_1 \wedge s_2\) \(s_1 \wedge s_4\)}{\(s_2 \wedge s_4\)}$ involves a cycle that can be continuously deformed into $\it{twice}$ the line integral connecting the puncture at $w=1$ and the zero at $w=z_2$. Therefore we have
\beq\label{etacycles2}
\eta_{1,z_2}=\frac{1}{2}\int_{-\infty}^{\infty}\frac{d\theta}{\pi}e^{-\theta}\log\left[\frac{\left(1+A^{-}_{12}\right)\left(1+A^{-}_{14}\right)}{\left(1+A^{-}_{24}\right)}\right]
=\frac{1}{2}\(\eta_{E_{12}}+\eta_{E_{14}}-\eta_{E_{24}}\)
\eeq
In the same way we obtain 
\beqa\label{etacycles3}
\eta_{z_3,z_4}
&=&\frac{1}{2}\(\eta_{24}-\eta_{\hat{24}}\)\\
\eta_{-1,z_2}
&=&\frac{1}{2}\(2\eta_{34}+\eta_{12}-\eta_{14}-\eta_{24}\)\\
\eta_{z,z_2}
&=&\frac{1}{2}\(\eta_{12}-\eta_{14}-\eta_{24}\)\\
\eta_{\infty,z_2}
&=&\frac{1}{2}\(\eta_{14}-\eta_{12}-\eta_{24}\)\\
\eta_{z_3,z_2}
&=&\frac{1}{2}\(\eta_{12}+\eta_{34}-\eta_{14}-\eta_{23}\)
\eeqa
where the notation is the natural simplification of that used in (\ref{etacycles2}).  Plugging these expressions into $(\ref{action1})$ and re-collecting each $\eta_{E}$, one finds that the coefficient of $\eta_E$ is simply $\o_E$ where $\o_E$ is the $\o$-cycle the intersects edge $E$, \emph{not} the integral of $\o$ along edge $E$ (which would be divergent).  That is, it's (1/2 of) the $\o$-cycle associated with the coordinate $\chi_E$ which are shown in figure \ref{WKBCells1}.  Thus we have
\beq\label{action final 2}
A_{fin}=\frac{\pi}{3}-i\sum_{E\in \mathcal{T}}\o_E \, \eta_E 
\eeq  
which is formula $(\ref{action final})$ as desired. \\
\indent Equation ($\ref{action final 2}$) is perhaps the simplest possible result one could write from the triangulation data.  Given this simplicity, it is probably possible to derive the result in a much more elegant way and perhaps even for any number of punctures.  We have not pursued this issue but feel that it merits further exploration.  


\section{Three-point function in GMN language}\label{app 3 point}
\begin{figure}[t!]
\begin{center}
\includegraphics[width=0.5\linewidth]{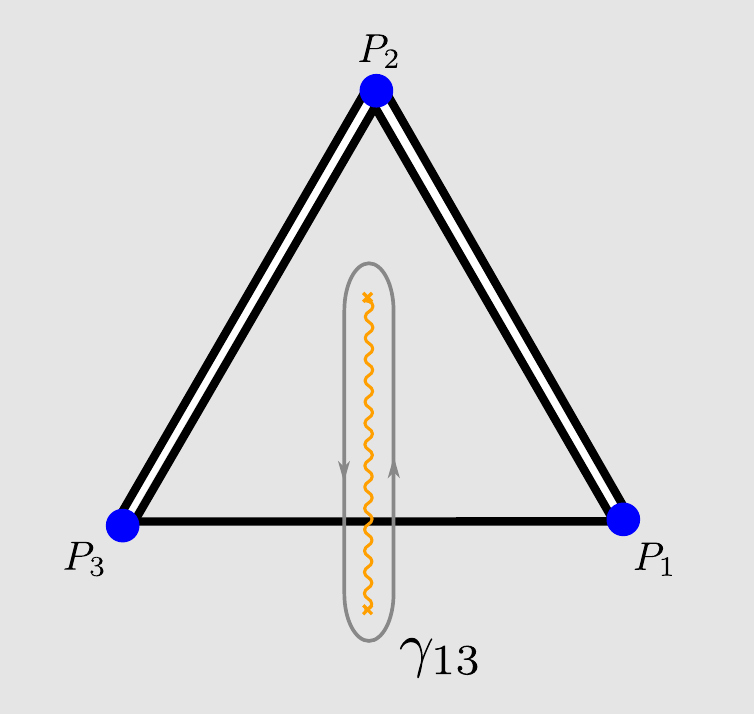}
\caption{The WKB triangulation for the 3-point function is composed of 3 edges forming two triangles on the sphere.  Here we show the construction of the coordinate $\chi_{13}$.  We are using the edge-splitting procedure discussed in section \ref{coordinates} (in particular, see figure \ref{ChiConstruction2}).  The gray contour shows the cycle associated with the coordinate $\chi_{13}$.}\label{3pt triang}
\end{center}
\end{figure}

In this section we apply the method developed in section \ref{Linear Problem} to the three point correlation function studied in \cite{JW2}. We use the setup of \cite{JW2}, namely the same stress-energy tensor. We aim at deriving a set of functional equations to extract the cycles used there.\\
\indent As a starting point, we introduce the WKB triangulation for this configuration from which we define the coordinates, see figure \ref{3pt triang}. From this figure, we easily derive the $\chi$-system. Since the quadrilateral is very degenerate it follows from (\ref{chi to A}) that the right hand side of the $\chi$-system is equal to 1. 
The reason is that the same auxiliary $A_{PQ}$'s appear both in numerator and denominator canceling each other.
Hence, the solution of the functional equations is simply given by the WKB asymptotics. More explicitly, the $\chi$ functions take the form
\beq\label{3pt chi}
\chi_{ac}=(-1)\exp\left(\frac{1}{2}e^{\theta}\int_{\gamma_{ac}}\omega+\frac{1}{2}e^{-\theta}\int_{\gamma_{ac}} \bar{\omega}\right)=-\frac{\mu_a \mu_c}{\mu_b}
\eeq
where $a$, $b$ and $c$ are distinct.\footnote{This result also follows directly from the definition of the coordinates in terms of the small solutions, $\chi_{ac}=-\frac{( s_c \wedge s_b )(  s_a \wedge M^{-1}_a s_b)}{(  M^{-1}_a s_b \wedge s_c)( s_b \wedge s_a )}$ for distinct $a,b$ and $c$; all the inner-products cancel and one is left with only the monodromy factors in \eqref{3pt chi}.}
The spikes must be in pointing in opposite direction as follows from the discussion of appendix \ref{gamma props}. This is the the origin of the $(-1)$ prefactor in \eqref{3pt chi}.  The cycles of $\omega$ are given in terms of the dimensions of the operators, 
\beq
\int_{\gamma_{ac}} \omega=i \pi  (-\Delta_a-\Delta_c+\Delta_b)
\eeq
Having the solutions of the functional equations, we can easily find the auxiliary quantities $A_{PQ}$ using the rules of section \ref{chi system}.  The determination of the $\eta$-cycles is also straightforward. To compare with the result in \cite{JW2} let us set 
$\Delta_1=\Delta_2=\Delta$ and $\Delta_3=\Delta_{\infty}$. We use expression \ref{extracting eta 1} to compute the cycles, and we get
\beqa
\int_{-1}^{1}\eta &=&\int_{\mathbb{R}}\frac{d \theta}{\pi} 
e^{-\theta}\log(1+A^{-}_{12})=h(2\Delta-\Delta_{\infty})+h(2\Delta+\Delta_{\infty})-2h(2 \Delta)\\
\int_{1}^{\infty}\eta&=& \int_{\mathbb{R}}\frac{d\theta}{\pi}e^{-\theta}\log(1+A^{-}_{23})=h(\Delta_{\infty})+h(2\Delta+\Delta_{\infty})-h(2\Delta)-h(2\Delta_{\infty}) 
\eeqa
where we define
\beq
h(x)=\int_{\mathbb{R}}\frac{d\theta}{\pi}\cosh\theta\log\left(1-e^{-x \pi \cosh\theta}\right)\,.
\eeq
This is precisely the result obtained in \cite{JW2}.
A last comment about the expression for the area in the three point function. It is easy to show using the same type of manipulation of the four point function case that the area can be expressed in terms of elements of the WKB triangulation as
\beq
A_{fin}=\frac{\pi}{6}-i\sum_{E\in \mathcal{T}}\omega_{E}\eta_{E}
\eeq
where the sum is over the edges of the triangulation of figure \ref{3pt triang}.  As in the case of the four point function, we define $\eta_{E_{ab}}$ as the $\eta$-cycle that passes along edge $E_{ab}$ from $P_a$ to $P_b$ and $\omega_{E_{ab}}$ as the $\omega$-cycle that intersects edge $E_{ab}$.



\begin{thebibliography}{99}

\bibitem{adscft1}
  J.~M.~Maldacena,
  ``The Large N limit of superconformal field theories and supergravity,''
  Adv.\ Theor.\ Math.\ Phys.\  {\bf 2}, 231 (1998)
  [Int.\ J.\ Theor.\ Phys.\  {\bf 38}, 1113 (1999)]
  [arXiv:hep-th/9711200].


\bibitem{tailor2} 
  J.~Escobedo, N.~Gromov, A.~Sever and P.~Vieira,
  ``Tailoring Three-Point Functions and Integrability II. Weak/strong coupling match,''
  JHEP {\bf 1109}, 029 (2011)
  [arXiv:1104.5501 [hep-th]].




\bibitem{CE} 
  J.~Caetano and J.~Escobedo,
  ``On four-point functions and integrability in N=4 SYM: from weak to strong coupling,''
  JHEP {\bf 1109}, 080 (2011)
  [arXiv:1107.5580 [hep-th]].




\bibitem{tailor3} 
  N.~Gromov, A.~Sever and P.~Vieira,
  ``Tailoring Three-Point Functions and Integrability III. Classical Tunneling,''
  JHEP {\bf 1207}, 044 (2012)
  [arXiv:1111.2349 [hep-th]].




\bibitem{quantumint} 
  N.~Gromov and P.~Vieira,
  ``Quantum Integrability for Three-Point Functions,''
  arXiv:1202.4103 [hep-th].




\bibitem{KOSTOV} 
  I.~Kostov,
  ``Three-point function of semiclassical states at weak coupling,''
  arXiv:1205.4412 [hep-th].




\bibitem{JW1} 
  R.~A.~Janik, P.~Surowka and A.~Wereszczynski,
  ``On correlation functions of operators dual to classical spinning string states,''
  JHEP {\bf 1005}, 030 (2010)
  [arXiv:1002.4613 [hep-th]].




\bibitem{JW2} 
  R.~A.~Janik and A.~Wereszczynski,
  ``Correlation functions of three heavy operators: The AdS contribution,''
  JHEP {\bf 1112}, 095 (2011)
  [arXiv:1109.6262 [hep-th]].




\bibitem{KK1} 
  Y.~Kazama and S.~Komatsu,
  ``On holographic three point functions for GKP strings from integrability,''
  JHEP {\bf 1201}, 110 (2012)
  [Erratum-ibid.\  {\bf 1206}, 150 (2012)]
  [arXiv:1110.3949 [hep-th]].




\bibitem{AMSV} 
  L.~F.~Alday, J.~Maldacena, A.~Sever and P.~Vieira,
  ``Y-system for Scattering Amplitudes,''
  J.\ Phys.\ A A {\bf 43}, 485401 (2010)
  [arXiv:1002.2459 [hep-th]].




\bibitem{KK2} 
  Y.~Kazama and S.~Komatsu,
  ``Wave functions and correlation functions for GKP strings from integrability,''
  arXiv:1205.6060 [hep-th].




\bibitem{GMN} 
  D.~Gaiotto, G.~W.~Moore and A.~Neitzke,
  ``Wall-crossing, Hitchin Systems, and the WKB Approximation,''
  arXiv:0907.3987 [hep-th].




\bibitem{POLYAKOV} 
  A.~M.~Polyakov,
  ``Gauge fields and space-time,''
  Int.\ J.\ Mod.\ Phys.\ A {\bf 17S1}, 119 (2002)
  [hep-th/0110196].




\bibitem{TSEYTLINVERTEX} 
  A.~A.~Tseytlin,
  ``On semiclassical approximation and spinning string vertex operators in AdS(5) x S**5,''
  Nucl.\ Phys.\ B {\bf 664}, 247 (2003)
  [hep-th/0304139].




\bibitem{GUBSER2002} 
  S.~S.~Gubser, I.~R.~Klebanov and A.~M.~Polyakov,
  ``A Semiclassical limit of the gauge / string correspondence,''
  Nucl.\ Phys.\ B {\bf 636}, 99 (2002)
  [hep-th/0204051].




\bibitem{RASTELLI} 
  E.~D'Hoker, D.~Z.~Freedman, S.~D.~Mathur, A.~Matusis and L.~Rastelli,
  ``Extremal correlators in the AdS / CFT correspondence,''
  In *Shifman, M.A. (ed.): The many faces of the superworld* 332-360
  [hep-th/9908160].




\bibitem{SOKATCHEVPROOF} 
  B.~Eden, P.~S.~Howe, C.~Schubert, E.~Sokatchev and P.~C.~West,
  ``Extremal correlators in four-dimensional SCFT,''
  Phys.\ Lett.\ B {\bf 472}, 323 (2000)
  [hep-th/9910150].




\bibitem{COSTA1} 
  M.~S.~Costa, J.~Penedones, D.~Poland and S.~Rychkov,
  ``Spinning Conformal Blocks,''
  JHEP {\bf 1111}, 154 (2011)
  [arXiv:1109.6321 [hep-th]].
  



\bibitem{COSTA2} 
  M.~S.~Costa, J.~Penedones, D.~Poland and S.~Rychkov,
  ``Spinning Conformal Correlators,''
  JHEP {\bf 1111}, 071 (2011)
  [arXiv:1107.3554 [hep-th]].




\bibitem{DUFFIN} 
  D.~Simmons-Duffin,
  ``Projectors, Shadows, and Conformal Blocks,''
  arXiv:1204.3894 [hep-th].




\bibitem{DOLAN} 
  F.~A.~Dolan and H.~Osborn,
  ``Conformal partial waves and the operator product expansion,''
  Nucl.\ Phys.\ B {\bf 678}, 491 (2004)
  [hep-th/0309180].




\bibitem{nullope1} 
  L.~F.~Alday, D.~Gaiotto, J.~Maldacena, A.~Sever and P.~Vieira,
  ``An Operator Product Expansion for Polygonal null Wilson Loops,''
  JHEP {\bf 1104}, 088 (2011)
  [arXiv:1006.2788 [hep-th]].


\bibitem{nullope2} 
  D.~Gaiotto, J.~Maldacena, A.~Sever and P.~Vieira,
  ``Bootstrapping Null Polygon Wilson Loops,''
  JHEP {\bf 1103}, 092 (2011)
  [arXiv:1010.5009 [hep-th]].




\bibitem{nullope3} 
  D.~Gaiotto, J.~Maldacena, A.~Sever and P.~Vieira,
  ``Pulling the straps of polygons,''
  JHEP {\bf 1112}, 011 (2011)
  [arXiv:1102.0062 [hep-th]].




\bibitem{WLDUALITY1} 
  L.~F.~Alday, B.~Eden, G.~P.~Korchemsky, J.~Maldacena and E.~Sokatchev,
  ``From correlation functions to Wilson loops,''
  JHEP {\bf 1109}, 123 (2011)
  [arXiv:1007.3243 [hep-th]].

\bibitem{WLDUALITY2} 
  B.~Eden, G.~P.~Korchemsky and E.~Sokatchev,
  ``From correlation functions to scattering amplitudes,''
  JHEP {\bf 1112}, 002 (2011)
  [arXiv:1007.3246 [hep-th]].

\bibitem{WLDUALITY3} 
  B.~Eden, G.~P.~Korchemsky and E.~Sokatchev,
  ``More on the duality correlators/amplitudes,''
  Phys.\ Lett.\ B {\bf 709}, 247 (2012)
  [arXiv:1009.2488 [hep-th]].

\bibitem{WLDUALITY4} 
  A.~V.~Belitsky, G.~P.~Korchemsky and E.~Sokatchev,
  ``Are scattering amplitudes dual to super Wilson loops?,''
  Nucl.\ Phys.\ B {\bf 855}, 333 (2012)
  [arXiv:1103.3008 [hep-th]].

\bibitem{WLDUALITY5} 
  B.~Eden, P.~Heslop, G.~P.~Korchemsky and E.~Sokatchev,
  ``The super-correlator/super-amplitude duality: Part I,''
  arXiv:1103.3714 [hep-th].

\bibitem{WLDUALITY6} 
  B.~Eden, P.~Heslop, G.~P.~Korchemsky and E.~Sokatchev,
  ``The super-correlator/super-amplitude duality: Part II,''
  arXiv:1103.4353 [hep-th].


\bibitem{TSEYTLINSPHERE} 
  E.~I.~Buchbinder and A.~A.~Tseytlin,
  ``Semiclassical correlators of three states with large $S^5$ charges in string theory in $AdS_5 x S^5$,''
  Phys.\ Rev.\ D {\bf 85}, 026001 (2012)
  [arXiv:1110.5621 [hep-th]].


\end{thebibliography}
\end{document}